\documentclass[twocolumn,showpacs,amsmath,amssymb,prd,floatfix,preprintnumbers]{revtex4-2}
\usepackage[dvipdfmx]{graphicx}
\usepackage{bm}
\usepackage{amsmath}
\usepackage{color}
\usepackage[dvipsnames]{xcolor}
\usepackage{hyperref}
\usepackage{comment}
\usepackage{multirow}
\usepackage{url}

\hypersetup{
  bookmarksnumbered=true,
  colorlinks=true,
  linkcolor=red,
  citecolor=orange,
  urlcolor=cyan
}

\newcommand{\ykrv}[1]{{\textcolor{black}{#1}}}

\newcommand{\bk}{{\bf k}}

\newcommand{\bn}{{\bf n}}

\def\avrg#1{\left\langle #1 \right\rangle}

\newcommand{\simgt}{\lower.5ex\hbox{$\; \buildrel > \over \sim \;$}}
\newcommand{\simlt}{\lower.5ex\hbox{$\; \buildrel < \over \sim \;$}}

\begin{document}

\title[]{Full-shape cosmology analysis of SDSS-III BOSS galaxy power spectrum 
using emulator-based halo model: a $5\%$ determination of $\sigma_8$}

\author{Yosuke~Kobayashi${}^{1,2}$}
\email{yosukekobayashi@email.arizona.edu}
\author{Takahiro~Nishimichi${}^{3,2}$}
\author{Masahiro~Takada${}^{2}$}
\email{masahiro.takada@ipmu.jp}
\author{Hironao~Miyatake${}^{4,2}$}
\affiliation{
${}^{1}$ Department of Astronomy/Steward Observatory, University of Arizona, 933 North Cherry Avenue, Tucson, AZ 85721-0065, USA\\
${}^{2}$ Kavli Institute for the Physics and Mathematics of the Universe
(WPI), The University of Tokyo Institutes for Advanced Study (UTIAS),
The University of Tokyo, Chiba 277-8583, Japan\\
${}^{3}$ Center for Gravitational Physics, Yukawa Institute for Theoretical Physics, Kyoto University, Kyoto 606-8502, Japan\\
${}^{4}$ Kobayashi-Maskawa Institute for the Origin of Particles and the Universe (KMI),
Nagoya University, Nagoya, 464-8602, Japan
}

\begin{abstract}
We present the results obtained from the full-shape cosmology analysis of the redshift-space power spectra for 4 galaxy samples of the SDSS-III BOSS DR12 galaxy catalog over $0.2 < z < 0.75$. 
For the theoretical template, we use an emulator that was built from an ensemble set of $N$-body simulations, which enables fast and accurate computation of the redshift-space power spectrum of ``halos''. 
Combining with the halo occupation distribution to model the galaxy-halo connection, we can compute the redshift-space power spectrum of BOSS-like galaxies in less than a CPU second, for an input model under flat $\Lambda$CDM cosmology.
In our cosmology inference, we use the monopole, quadrupole and hexadecapole moments of the redshift-space power spectrum and include 7 nuisance parameters, with broad priors, to model uncertainties in the galaxy-halo connection for each galaxy sample, but do not use any information on the abundance of galaxies. 
We demonstrate a validation of our analysis pipeline using the mock catalogs of BOSS-like galaxies, generated using different recipes of the galaxy-halo connection and including the assembly bias effect.
Assuming weak priors on cosmological parameters, except for the BBN prior on $\Omega_\mathrm{b}h^2$ and the CMB prior on $n_\mathrm{s}$, we show that our model well reproduces the BOSS power spectra. 
Including the power spectrum information up to $k_\mathrm{max}=0.25\,h\,\mathrm{Mpc}^{-1}$, we find 
\ykrv{
$\Omega_\mathrm{m}=0.301^{+0.012}_{-0.011}$, $H_0=68.2 \pm 1.4~\mathrm{km\,s}^{-1}\mathrm{Mpc}^{-1}$, and $\sigma_8=0.786^{+0.036}_{-0.037}$,} 
for the mode and 68\% credible interval, after marginalization over galaxy-halo connection parameters. 
We find little improvement in the cosmological parameters beyond a maximum wavelength $k_\mathrm{max} \simeq 0.2\,h\,\mathrm{Mpc}^{-1}$ due to the shot noise domination and marginalization of the galaxy-halo connection parameters.
\ykrv{Our results are consistent with the \textit{Planck} CMB results within $1\sigma$ statistical uncertainties.}
\end{abstract}

\preprint{IPMU21-0058}
\preprint{YITP-21-112}

\maketitle

\section{Introduction}
\label{sec:intro}

The three-dimensional distribution of galaxies, measured from wide-area spectroscopic surveys of galaxies, is a powerful probe of cosmology, \textit{e.g.}, for constraining cosmological parameters such as parameters characterizing the nature of dark energy and for testing gravity theory on cosmological scales
\citep{cole:2005aa,Eisenstein:2005su,okumura08,parkinson12,2017MNRAS.470.2617A,2016PASJ...68...38O,2021PhRvD.103h3533A}.
To attain the fundamental cosmology, there are various exiting, ongoing and planned galaxy redshift surveys: 
the SDSS-III Baryon Oscillation Spectroscopic Survey (BOSS; \cite{2013AJ....145...10D}), 
the SDSS-IV extended Baryon Oscillation Spectroscopic Survey (eBOSS; \cite{2016AJ....151...44D}), the Subaru Prime Focus Spectrograph (PFS; \cite{2014PASJ...66R...1T}), the Dark Energy Spectroscopic Instrument (DESI; \cite{Aghamousa:2016zmz}), the ESA \textit{Euclid} satellite mission \cite{Laureijs:2011gra}, and the NASA 
Nancy Grace Roman Space Telescope \cite{Gehrels:2014spa}.

The galaxy distribution observed by spectroscopic surveys is modulated by
the Doppler effect due to the line-of-sight peculiar velocities of galaxies, and exhibits characteristic anisotropies,
called the redshift-space distortion (RSD) \cite{kaiser87,Hamilton92,Hamilton_1998}.
The RSD effect is useful to improve cosmological constraints by 
breaking degeneracies between the cosmological parameters and uncertainties in galaxy bias relative to the underlying matter distribution \citep{PhysRevD.101.023510}.
In addition, since the RSD effect is a gravitational effect, it 
can be used, if precisely measured, to probe the strength of gravitational field in large-scale structure, which can be in
turn used to test gravity theory on cosmological scales \citep{2007PhRvL..99n1302Z}.

In order to exploit the full information from galaxy redshift surveys, we need a sufficiently accurate theoretical template that enables a high-fidelity comparison with the measured clustering statistics of galaxies to obtain a robust estimation of cosmological parameters.
The linear theory of cosmological fluctuations, which has been in a remarkable success in CMB analyses, ceases to be accurate at $k \gtrsim 0.1\,h{\rm Mpc}^{-1}$ due to nonlinear effects of structure formation \citep{PhysRevD.101.023510}.
The standard approach to tackle this difficulty has been analytic prescriptions based on the perturbation theory \ykrv{(PT)} of large-scale structure \cite{bernardeau02,Desjacques18}. 
This approach describes the distribution of galaxies in terms of a series expansion of both the matter density and velocity fields with a set of free coefficients/terms including bias parameters, under the single-stream approximation \citep{taruya10,nishimichi11}.
A further refined model consistently separating short-scale physics including the galaxy bias from large-scale dynamics of interest,
so-called Effective Field Theory of Large-Scale Structure (EFTofLSS), has also been developed \citep{baumann12}.
These models have been applied to actual datasets to obtain cosmological constraints \citep{2014MNRAS.444..476R,10.1093/mnras/stu1051,Beutler:2016arn,2017MNRAS.470.2617A,Ivanov_2020,d_Amico_2020,2021arXiv211005530C}.
While these \ykrv{PT-based} templates give useful predictions at linear and quasi-nonlinear scales up to $k \sim 0.2\,h\,\mathrm{Mpc}^{-1}$, an application of these models to  even smaller scales is still disturbed by even higher-order contributions of both the density and velocity fields as well as non-perturbative effects arising from the dynamics beyond shell crossing, \textit{i.e.}, formation of galaxies (or dark matter halos) \citep[\textit{e.g.},][]{Pueblas_2009,blas14,bernardeau:2014lr,nishimichi2016,taruya2017,saga2018,Halle_2020}.
Consequently, the cosmology analysis of the galaxy power spectrum has been typically limited to the wavenumber $k \lesssim 0.15$ -- $0.2\,h\,\mathrm{Mpc}^{-1}$, depending on the redshift and the accuracy of the model required to meet the statistical precision of data
\cite{Beutler:2016arn,10.1093/mnras/stu1051,2020arXiv200308277N}.
In other words, the clustering information on the higher-$k$ scales does not seem practical for cosmology in this method, because the information is used to basically constrain higher-order bias parameters and other nuisance parameters that need to be introduced for the theoretical consistency of models.

As an alternative approach, in this paper we use a simulation-based theoretical template, the \textit{emulator}, which enables fast and accurate computation of the redshift-space power spectrum of ``halos'' in the flat $\Lambda$CDM framework, developed in our previous paper \cite{Kobayashi_2020}.
Dark matter halos are \ykrv{locations} where galaxies likely form. 
It is relatively straightforward to accurately simulate the formation and evolution processes of halos using $N$-body simulations and then have an accurate prediction of their clustering properties including the redshift-space power spectrum \citep{Cooray02,Desjacques18}.
Kobayashi {\it et al.}~\cite{Kobayashi_2020} developed an emulator by training a feed-forward neural network with a dataset of the redshift-space power spectra of halos measured from halo catalogs in 
an ensemble set of $N$-body simulations for 80 models within flat $w$CDM framework in the \textsc{Dark Quest} campaign \cite{Nishimichi_2019}.
The emulator was validated using the test dataset consisting of 20 cosmological models that are not in training, and it was shown that the power spectra are sufficiently accurate up to $k=0.6\,h\,\mathrm{Mpc}^{-1}$.
The emulator includes all the non-perturbative, nonlinear effects relevant to the formation and evolution of halos: nonlinear clustering, nonlinear RSD, nonlinear bias of halos, and halo exclusion effect. 
By combining with a halo occupation distribution (HOD) model \citep{1998ApJ...494....1J,seljak:2000uq,peacock:2000qy,scoccimarro:2001fj,2005ApJ...633..791Z,zheng09}, the emulator enables to compute the redshift-space power spectrum of galaxies in less than a CPU second on a typical recent laptop computer.
Thus, the emulator allows for cosmological parameter inference of the galaxy power spectrum in a multi-dimensional parameter space. 
Such an emulator-based method is equivalent to estimating cosmological parameters from comparison of the observed power spectrum with mock spectra from simulated galaxy catalogs generated from costly high-resolution $N$-body simulations with varying cosmological models.

Hence, the purpose of this paper is to perform a cosmology analysis of the redshift-space galaxy power spectrum measured from the public BOSS DR12 large-scale structure catalog over $0.2 < z < 0.75$ \footnote{\url{https://data.sdss.org/sas/dr12/boss/lss/}}, using the aforementioned emulator. 
In doing this, we analyze the ``full-shape'' information in the monopole, quadrupole and hexadecapole moments of the redshift-space power spectra, beyond the traditional approach to extract only geometrical information through the features of baryon acoustic oscillations (BAO) and the anisotropy originating from RSD. 
We first show a series of validation checks of our method to ensure unbiased cosmological inference beyond the accuracy assessment of the emulator at the halo power spectrum level already presented in our previous paper. 
We apply the full analysis pipeline to mock signals of the galaxy power spectrum measured from the galaxy mock catalogs and then check whether our method can recover the cosmological parameters to within the statistical errors. 
For this validation, we use the mock catalogs generated using different recipes of galaxy-halo connection from our fiducial HOD model \citep[also see][for similar analyses but with different observables]{2020arXiv200308277N} and also use mock catalogs including the assembly bias effect that is one of the most dangerous, physical systematic effects in the halo model approach. 
We then apply the analysis method to \ykrv{the BOSS galaxy power spectra} under the flat $\Lambda$CDM cosmology. 
In doing this, we employ weak priors on the cosmological parameters, except for the BBN prior on $\Omega_\mathrm{b}h^2$ \citep{Aver_2015} and the CMB prior on $n_\mathrm{s}$ \citep{planck_collaboration_2020}, employ very broad priors for the galaxy-halo connection parameters, and then estimate the cosmological parameters \ykrv{after} marginalization over uncertainties in the nuisance parameters. 
We also compare our cosmological constraints with the \ykrv{recent PT-based results \citep{philcox2022boss,Chen_2022}} and the \textit{Planck} 2018 cosmological results \citep{planck_collaboration_2020}.

The structure of this paper is as follows. 
In Sec.~\ref{sec:data_ps_covariance} we will describe \ykrv{the power spectrum data and covariance we use in this work}. 
In Sec.~\ref{sec:model_and_params} we will describe details of the emulator-based halo model as the theoretical template and then the parameters and priors used in the cosmology analysis. 
In Sec.~\ref{sec:results} we show the main results of this paper: the cosmological parameters for the flat $\Lambda$CDM model. 
In Sec.~\ref{sec:discussion} we give discussion on possible residual systematic effects in the data and the theoretical template. 
Sec.~\ref{sec:conclusion} is devoted to the conclusion. 

\section{Data}
\label{sec:data_ps_covariance}

\ykrv{
For the galaxy power spectrum data and the covariance matrix, we use the updated measurement of the BOSS DR12 power spectrum recently provided in Ref.~\cite{Beutler_McDonald_2021}, which are publicly available from \url{https://fbeutler.github.io/hub/deconv_paper.html}.
We use four data chunks, named ``NGC~$z1$'', ``SGC~$z1$'', ``NGC~$z3$'', and ``SGC~$z3$'', which mean the measurements for different galaxy samples in two non-overlapping redshift bins (``$z1$'' and ``$z3$'') both for the Northern and Southern Galactic caps (``NGC'' and ``SGC''). 
``$z1$'' (``$z3$''), which we hereafter call ``low-$z$'' (``high-$z$''), corresponds to the range $0.2 < z < 0.5$ ($0.5 < z < 0.75$) with its effective redshift $z_\mathrm{eff} = 0.38$ (0.61).
We compute the theoretical model of the power spectrum for each redshift bin at its effective redshift.
}

\ykrv{
The estimator adopted in Ref.~\cite{Beutler_McDonald_2021} gives the power-spectrum multipoles convolved with
the survey window function, which can be expressed by a matrix-vector multiplication. 
Therefore, we construct a consistent model prediction of the window-convolved power spectrum as
\begin{align}
\label{eq:window_convolution}
    \mathbf{P}^\mathrm{conv} = \mathbf{W} \mathbf{M} \mathbf{P}^\mathrm{true,flat-sky},
\end{align}
where $\mathbf{P}^\mathrm{conv}$ is the vector of the convolved power spectrum multipoles, $\mathbf{P}^\mathrm{true,flat-sky} = \{P_0, P_2, P_4\}$ is that of the true (unconvolved) power spectrum multipoles computed in the global plane-parallel approximation, where the monopole, quadrupole, and hexadecapole moments are concatenated. 
$\mathbf{M}$ is the matrix that induces the odd multipoles (dipole and octopole) arising from the wide-angle effect. 
$\mathbf{W}$ is the matrix that consists of Fourier-space multipole moments of the survey window function (see Ref.~\cite{Beutler_McDonald_2021} for details). 
Due to the correction for the wide-angle effect through the matrix $\mathbf{M}$, the dipole and octopole moments (the real parts, since they are purely imaginary) are induced by the leakage from the standard even multipole moments, but we do not use these odd multipoles in this work. 
We compute the power spectrum monopole,  quadrupole, and hexadecapole using the emulator-based theoretical model we describe in the next section and substitute them to  $\mathbf{P}^\mathrm{true,flat-sky}$ in Eq.~(\ref{eq:window_convolution}). 
The matrices, $\mathbf{W}$ and $\mathbf{M}$, are also made publicly available on the same website.
} 

\ykrv{
In the measurement of the convolved power spectrum multipoles, Ref.~\cite{Beutler_McDonald_2021} assumes the flat-geometry $\Lambda$CDM model with $\Omega_\mathrm{m} = 0.31$, which is the same as the model assumed in the original measurement of the BOSS DR12 public spectra for their cosmology analysis \cite{Beutler:2016arn}. 
They measure the power spectrum in a cubic box by a fast Fourier transform (FFT) with the side length $L = 3500\,h^{-1}\,\mathrm{Mpc}$ and Nyquist wavenumber $k_\mathrm{Ny} = 0.628\,h\,\mathrm{Mpc}^{-1}$.
Each Fourier mode is divided into wavenumber bins in the range $k \in (0, 0.4)\,h\,\mathrm{Mpc}^{-1}$ with width $\Delta k = 0.01\,h\,\mathrm{Mpc}^{-1}$. 
}

\ykrv{
As for the power spectrum covariance matrix, we use the matrix provided by Ref.~\cite{Beutler_McDonald_2021}, which was estimated
from 2048 realizations of the \textsc{MultiDark-Patchy} (hereafter \textsc{Patchy}) mock catalogs \cite{Kitaura_2016,Rodriguez-Torres_2016}.
These mock catalogs were generated using an approximate $N$-body solver which combines the Lagrangian perturbation theory, a small-scale halo collapse model, and a semi-analytical galaxy biasing scheme, augmented by calibration to a reference large-volume $N$-body simulation sample selected from the \textsc{BigMultiDark} simulations \cite{Klypin_2016}. 
To correct the biased estimate of the inverse covariance, we multiply the so-called Hartlap factor \cite{Hartlap_2006} to the inverse of the estimated covariance:
\begin{align}
    f_\mathrm{Hartlap} = \frac{N_\mathrm{r}-n_\mathrm{bin}-2}{N_\mathrm{r}-1},
\end{align}
where $n_\mathrm{bin}$ is the number of bins we use in the parameter inference. 
For instance, in the case that we use the monopole, quadrupole and hexadecapole moments up to $k_\mathrm{max} = 0.25 \, h\, \mathrm{Mpc}^{-1}$, the number of bins is $n_\mathrm{bin} = 25 \times 3 = 75$ for each galaxy sample, yielding $f_\mathrm{Hartlap} = 0.9629$. 
}

\section{Theoretical model and parameters}
\label{sec:model_and_params}

\subsection{Theoretical model}
\label{sec:model}

To investigate the cosmological parameter constraint from the BOSS galaxy power spectrum, we use the theoretical template computed using the emulator for the redshift-space power spectrum of halos combined with the HOD model, developed in our previous work \citep{Kobayashi_2020}. 
We below give a brief description of our theoretical template, and see Ref.~\cite{Kobayashi_2020} \citep[also][]{PhysRevD.101.023510} for further details.

We employ the five-parameter HOD model in Ref.~\cite{2005ApJ...633..791Z}, which splits the 
galaxies into central and satellite galaxies.
The mean halo occupation number of central and satellite galaxies within host halos with mass $M$ are given as
\begin{align}
\label{eq:Nc}
  \avrg{N_\mathrm{c}}\! (M) = \frac{1}{2} \left[1 + \mathrm{erf} \left( \frac{\log M - \log M_\mathrm{min}}{\sigma_{\log M}} \right) \right],
\end{align}
and
\begin{align}
\label{eq:Ns}
  \avrg{N_\mathrm{s}}\! (M) &= \avrg{N_\mathrm{c}}\! (M) \lambda_\mathrm{s} (M) \nonumber\\
   &\equiv \avrg{N_\mathrm{c}}\! (M) \left(\frac{M- \kappa M_\mathrm{min}}{M_1} \right)^{\alpha_\mathrm{sat}}, 
\end{align}
respectively, where $\mathrm{erf}(x)$ is the error function and the logarithms in Eq.~(\ref{eq:Nc}) are base 10. 
Note that, in our model we adopt $M \equiv M_\mathrm{200} = (4\pi/3) 200 \bar{\rho}_\mathrm{m0} R_{200}^3$ as the halo mass definition, where $\bar{\rho}_\mathrm{m0}$ is the mean comoving mass density in the universe and $R_{200}$ is the spherical comoving radius within which the mean mass density is 200 times $\bar{\rho}_\mathrm{m0}$.
Here $\{\log M_\mathrm{min}, \sigma^2_{\log M},\log M_1, \alpha_\mathrm{sat}, \kappa\}$ are model parameters.
The probability distribution of galaxies given the mean number is assumed to be Bernoulli for centrals and Poisson for satellites given that the halo of interest has a central halo.
This HOD model is also used in the cosmology analysis of
the galaxy-galaxy weak lensing and projected galaxy correlation function measured from the Hyper Suprime-Cam Year1 catalog and the BOSS DR11 catalog \cite{2021arXiv211102419M}. 

Using the above HOD model, our full model of the redshift-space galaxy power spectrum is given by the sum of the one- and two-halo terms,
\begin{align}
    P_\mathrm{gg}(\bk) = P^\mathrm{1h}_\mathrm{gg}(\bk) + P^\mathrm{2h}_\mathrm{gg}(\bk),
\end{align}
where
\begin{align}
\label{eq:PS1h}
    P^\mathrm{1h}_\mathrm{gg}(\bk) &= \frac{1}{\bar{n}_\mathrm{g}^2} \int\!\!\mathrm{d}M~{\frac{\mathrm{d}n}{\mathrm{d}M}(M)} \avrg{N_\mathrm{c}}\!(M) \nonumber \\
    &\hspace{1em} \times \left[ 2\lambda_\mathrm{s}(M) \,\tilde{{\cal H}}(\bk; M) + \lambda_\mathrm{s}(M)^2 \,\tilde{{\cal H}}(\bk; M)^2
\right],
\end{align}
and
\begin{align}
\label{eq:PS2h}
&P^\mathrm{2h}_\mathrm{gg}(\bk) \nonumber \\
&=
\frac{1}{\bar{n}_\mathrm{g}^2}
\int\!\mathrm{d}M_1 {\frac{\mathrm{d}n}{\mathrm{d}M}(M_1)}\left[\avrg{N_\mathrm{c}}\!(M_1)+\avrg{N_\mathrm{s}}\!(M_1)\,
\tilde{{\cal H}}(\bk; M_1)\right]\nonumber\\
&\hspace{1em} \times \int\!\mathrm{d}M_2{\frac{\mathrm{d}n}{\mathrm{d}M}(M_2)}\left[\avrg{N_\mathrm{c}}\!(M_2)
+\avrg{N_\mathrm{s}}\!(M_2)\,\tilde{{\cal H}}(\bk; M_2)\right] \nonumber \\
&\hspace{1em} \times {P_\mathrm{hh}(\bk; M_1, M_2)},
\end{align}
\ykrv{where $P_\mathrm{hh}(\bk; M_1, M_2)$ is the redshift-space halo power spectrum, for which we use our emulator \cite{Kobayashi_2020} that includes the RSD effects in both linear and nonlinear regimes for an input set of model 
parameters (cosmological parameters, redshift and halo masses).}
These power spectra depend on the two-dimensional wavevector, $\bk=(k_\parallel,k_\perp)$ or $k(\mu,\sqrt{1-\mu^2})$, due to the redshift-space distortion effect, where $\mu$ is the cosine angle between the wavevector $\bk$ and the line-of-sight direction $\hat{\bn}$.
In the above formulae, $\mathrm{d}n/\mathrm{d}M$ is the halo mass function and $\bar{n}_\mathrm{g}$ is the global mean number density of galaxies defined as
\begin{align}
\bar{n}_\mathrm{g}=\int\!\!\mathrm{d}M~{\frac{\mathrm{d}n}{\mathrm{d}M}(M)} \left[\avrg{N_\mathrm{c}}\!(M)+\avrg{N_\mathrm{s}}\!(M)\right].
\label{eq:ng_hod}
\end{align}
For satellite galaxies, we model the intra-halo profile in redshift space as the multiplication of the Navarro-Frenk-White (NFW) profile \cite{NFW} and the velocity distribution \cite{hikage12a,hikage:2013kx,PhysRevD.101.023510}:
\begin{align}
\label{eq:sspace_profile_fourier}
  \tilde{\mathcal{H}}(\bk;M) = \tilde{u}_\mathrm{NFW}(k;M)\,\tilde{\mathcal{F}}(k_\parallel;M),
\end{align}
where $\tilde{u}_\mathrm{NFW}(k;M)$ is the Fourier transform of the NFW density profile normalized by halo mass $M$.
To specify the NFW profile, we assume the median concentration-mass relation $c(M_\mathrm{200})$ following the model in
Refs.~\cite{2015ApJ...799..108D,Diemer_2019}.
$\tilde{\mathcal{F}}(k_\parallel;M)$ is the Fourier transform of the Gaussian velocity distribution whose velocity dispersion is given as \begin{align}
    \mathcal{F}(\Delta r_\parallel; M) = \frac{1}{\sqrt{2\pi \sigma_\mathrm{v}^2(M)}} \exp \left[ - \frac{\Delta r_\parallel^2}{2\sigma_\mathrm{v}^2(M)} \right].
    \label{eq:f_vel}
\end{align}
The velocity dispersion, $\sigma_{\rm v}^2(M)$, is given by
\begin{align}
\label{eq:velocity_dispersion}
   \sigma_\mathrm{v}^2(M) = \frac{1}{a^2H^2} \frac{GM}{2 a R_\mathrm{200}(M)},
\end{align}
where $G$ is the gravitational constant, $a$ is the scale factor, $H$ is the Hubble parameter at scale factor $a$, and $R_\mathrm{200}(M)$ is the halo radius $R_\mathrm{200}$ for halos with mass $M$.
The factor $1/ (aH)^2 $ is to convert the velocity to the redshift-space displacement.
The above velocity distribution models the fingers-of-God (FoG) effect due to virial motions of satellite galaxies in their host halos. 
In order to further include uncertainties in the FoG effect, we will introduce a nuisance parameter to model the uncertainty, $\sigma_{{\rm v}}(M) \rightarrow c_{\rm vel}\sigma_{{\rm v}}(M)$, and treat $c_{\rm vel}$ as a model parameter in the cosmology analysis. 

\ykrv{As described above,} we use our emulator \cite{Kobayashi_2020} to compute the redshift-space halo power spectrum $P_\mathrm{hh}(\bk; M_1, M_2)$ in the two-halo term (Eq.~\ref{eq:PS2h}) for an input model in the flat $\Lambda$CDM cosmology.
The emulator outputs $P_{\rm hh}(\bk;M_1,M_2)$ given as a function of the two-dimensional wavevector $\bk$, rather than the multipole moments such as 
$P_{{\rm hh},\ell}(k;M_1,M_2)$.
This modeling allows us to straightforwardly include the \ykrv{Alcock-Paczy\'{n}ski effect (see the next subsection)}, even in the presence of the FoG effect.  
We also use the publicly available \texttt{Dark Emulator} (\url{https://dark-emulator.readthedocs.io/en/latest/}) \cite{Nishimichi_2019} to compute the halo mass function, and the \texttt{Colossus} code (\url{http://www.benediktdiemer.com/code/colossus/}) \cite{Diemer_2018} to compute the halo concentration-mass relation.
In the theoretical model the redshift-space halo power spectrum in the two-halo term, $P_\mathrm{hh}(\bk; M_1,M_2)$, carries cosmological information, and other functions such as the HOD and the distribution of galaxies in their host halos are all nuisance and are needed to account for uncertainties in the galaxy-halo connection. 

\subsection{Alcock-Paczy\'{n}ski effect}
\label{sec:APeffect}

Since we need to assume a reference cosmology ($\Omega_\mathrm{m} = 0.31$) in the measurement of the power spectrum, we incorporate the Alcock-Paczy\'{n}ski (AP) effect \cite{alcock79,Matsubara_1996} into the theoretical model.
Due to the difference between the true and reference cosmologies, the Fourier-space wavenumbers are transformed as
\begin{align}
  k^\mathrm{ref}_{\perp} &= \alpha_\perp k_\perp \equiv \frac{D_\mathrm{A}(z)}{D^\mathrm{ref}_\mathrm{A}(z)} k_\perp, \nonumber\\
  k^\mathrm{ref}_{\parallel} &= \alpha_\parallel k_\parallel \equiv \frac{H^\mathrm{ref}(z)}{H(z)} k_\parallel,
\end{align}
where $D_\mathrm{A}(z)$ and $H(z)$ are the angular diameter distance and Hubble parameter at the effective redshift $z$.
Quantities with superscript ``ref'' denote those for the reference cosmology. 
By this coordinate transformation, the power spectrum multipole is transformed as 
\begin{align}
P^\mathrm{ref}_{\mathrm{gg},\ell}(k^\mathrm{ref}) = \frac{2\ell+1}{2\alpha^2_\perp\alpha_\parallel } 
\int_{-1}^1\! \mathrm{d}\mu^\mathrm{ref}~P_\mathrm{gg}\!(k, \mu) 
\mathcal{L}_\ell(\mu^\mathrm{ref}).
\label{eq:PS_l_AP}
\end{align}
where $k$ and $\mu$ in the argument of $P^\mathrm{ref}_{\mathrm{gg},\ell}$ are given 
in terms of $k^\mathrm{ref}$ and $\mu^\mathrm{ref}$; $k=k(k^\mathrm{ref},\mu^\mathrm{ref})$
and $\mu=\mu(\mu^\mathrm{ref})$, given as
\begin{align}
k(k^\mathrm{ref},\mu^\mathrm{ref}) &= k^\mathrm{ref} \frac{1}{\alpha_\perp} \left [1+(\mu^\mathrm{ref})^2 \left (\frac{\alpha_\perp^2}{\alpha_\parallel^2}-1 \right ) \right ]^{1/2}, \\
\mu(\mu^\mathrm{ref}) &= \mu^\mathrm{ref} \frac{\alpha_\perp}{\alpha_\parallel}  \left [1+(\mu^\mathrm{ref})^2 \left (\frac{\alpha_\perp^2}{\alpha_\parallel^2}-1 \right ) \right ]^{-1/2}.
\label{eq:kmu_AP}
\end{align}
Since the reference cosmology is generally different from the underlying true cosmology, the AP effect induces an additional anisotropy in the measured power spectrum. 
In other words, the measured anisotropy enables us to infer the true cosmology through the geometrical information.

\subsection{Model parameters and priors}

In this subsection, we describe the model parameters that we infer, as well as their prior settings.
The default setting is summarized in Table~\ref{tab:param_table}.

\begin{table}
\begin{center}
\begin{tabular}{cc} \hline \hline
    Parameter & Prior \\ \hline \hline
	\multicolumn{2}{c}{\textbf{Cosmological parameters}} \\ \hline
	$\omega_\mathrm{b}$ & $\mathcal{N}(0.02268,0.00038)$ \\
	$\omega_\mathrm{c}$ & $\mathcal{U}(0.10782,0.13178)$ \\
	$\Omega_{\Lambda}$ & $\mathcal{U}(0.54752,0.82128)$ \\
	$\ln (10^{10}A_\mathrm{s})$ & $\mathcal{U}(2.4752,3.7128)$ \\
	$n_\mathrm{s}$ & $\mathcal{N}(0.9649,0.0042)$ \\ \hline \hline
	\multicolumn{2}{c}{\textbf{HOD parameters}} \\ \hline 
	$\log M_\mathrm{min}$ & $\mathcal{U}(12.0,15.0)$ \\
	$\sigma_{\log M}^2$ & $\mathcal{U}(0.0001,2.0)$ \\
	$\log M_1$ & $\mathcal{U}(12.0,16.0)$ \\
	$\alpha_\mathrm{sat}$ & $\mathcal{U}(0.01,5.0)$ \\ 
	$\kappa$ & $\mathcal{U}(0.01,5.0)$ \\ \hline \hline
	\multicolumn{2}{c}{\textbf{Other nuisance parameters}} \\ \hline 
	$c_\mathrm{vel}$ & $\mathcal{U}(0.01,10.0)$ \\
	$P_\mathrm{shot}$ & $\mathcal{U}(-10^4,10^4)\,h^{-3}\,\mathrm{Mpc}^3$ \\ 
	\hline \hline
	\multicolumn{2}{c}{\textbf{Derived parameters}} \\ \hline 
	$\Omega_\mathrm{m}$ & $ - $ \\
	$H_0$ & $ - $ \\
	$\sigma_8$ & $ - $ \\ 
	\hline \hline
\end{tabular}
\caption{
Model parameters and priors used in our cosmology analysis in flat $\Lambda$CDM cosmology. 
$\mathcal{N}(\mu,\sigma)$ denotes the Gaussian distribution with mean $\mu$ and standard deviation $\sigma$.
$\mathcal{U}(a,b)$ denotes the uniform distribution between the minimum value $a$ and the maximum value $b$.
The Gaussian priors on $\omega_\mathrm{b}$ and $n_\mathrm{s}$ are based on the BBN and \textit{Planck} CMB constraints, respectively (see text for details).
The flat priors on other cosmological parameters are set to be within the parameter ranges on which the emulator is supported.
}
\label{tab:param_table}
\end{center}
\end{table}

\subsubsection{Cosmological parameters}

Since we want to infer the cosmological parameters within the flat-geometry $\Lambda$CDM framework, we sample all of the five parameters:
\begin{align}
    \mathbf{p}_\mathrm{cosmo} = \left\{ \omega_\mathrm{b}, \omega_\mathrm{c}, \Omega_{\Lambda}, \ln (10^{10} A_\mathrm{s}), n_\mathrm{s} \right\},
\end{align}
where $\omega_\mathrm{b} = \Omega_\mathrm{b} h^2$ and $\omega_\mathrm{c} = \Omega_\mathrm{c} h^2$ are the physical energy density parameters of baryon and cold dark matter, $\Omega_{\Lambda}$ is the energy density parameter of the cosmological constant, and $A_\mathrm{s}$ and $n_\mathrm{s}$ are the amplitude (at the pivot scale $k_\mathrm{pivot} = 0.05\,\mathrm{Mpc}^{-1}$) and the spectral tilt of the power spectrum of primordial curvature perturbations.
For $\omega_\mathrm{b}$ and $n_\mathrm{s}$, we impose priors that are inferred from other cosmological probes.
More specifically, we adopt a Gaussian prior on $\omega_\mathrm{b}$ from the primordial deuterium and helium abundance data compared with the standard Big Bang nucleosynthesis (BBN) model \cite{Aver_2015,Cooke_2018,Sch_neberg_2019}.
On the other hand, we impose a Gaussian prior on $n_\mathrm{s}$ given in Table~1 of Ref.~\cite{planck_collaboration_2020} for the \textit{Planck} 2018 ``TT, TE, EE+lowE+lensing''.
We treat $\Omega_\mathrm{m} (= 1-\Omega_\Lambda)$, the total matter energy density, $H_0 = 100 h~\mathrm{km~s}^{-1}\mathrm{Mpc}^{-1}$, the Hubble constant, and $\sigma_8$, the standard deviation of linear matter perturbations at $z=0$ averaged within a sphere with comoving radius $8\,h^{-1}\,\mathrm{Mpc}$, as the derived parameters. 
We fix the density parameter of neutrinos to $\omega_\nu=0.00064$, corresponding to 0.06~eV for the total mass of the three mass eigenstates, which is used when computing the linear matter power spectrum used to set up the initial conditions of cosmological simulations that are used for the \textsc{Dark Emulator} development 
\cite{Nishimichi_2019}.

\subsubsection{Nuisance parameters}

As we described in Sec.~\ref{sec:model}, we employ five parameters to specify the HOD model for each galaxy sample.
In addition, we include two additional nuisance parameters.
\begin{itemize}
    \item $c_\mathrm{vel}$ --- The multiplicative coefficient on the velocity dispersion of galaxies relative to the halo center. 
    It regulates the uncertainty on the strength of the FoG effect (see around Eq.~\ref{eq:f_vel}).
    \item $P_\mathrm{shot}$ --- The residual shot noise contribution apart from the simple Poisson shot noise. 
    We add $P_\mathrm{shot}$ to the galaxy power spectrum $P_\mathrm{gg}(k,\mu)$, and hence it is relevant only to the monopole moment.
\end{itemize}
We employ a flat prior over the range $[-10^4,10^4]\,(h^{-1}\,\mathrm{Mpc})^3$ for the latter. 
The mean number density of galaxies for each sample is a factor of a few times $10^{-4}(h^{-1}\mathrm{Mpc})^3$, so the prior range is sufficiently wide.
Thus, we have seven nuisance parameters on the galaxy-halo connection:
\begin{align}
    \mathbf{p}_\mathrm{galaxy} = \{ \log M_\mathrm{min}, \sigma_{\log M}^2, \log M_1, \alpha_\mathrm{sat}, \kappa, c_\mathrm{vel}, P_\mathrm{shot} \},
\end{align}
for each of the 4 galaxy samples. 
Hence, the total number of parameters is $5 + 7 \times 4 = 33$ within the flat $\Lambda$CDM framework. 
As can be found from Table~\ref{tab:param_table}, we employ a broad prior range for each of the galaxy-halo connection parameters.
For example, the range of $M_\mathrm{min}$, which is one of the parameters that are sensitive to the linear bias of the galaxy sample, corresponds to halos that have $b_1\simeq [1.2,10.2]$ at $z=0.5$ for the {\it Planck} cosmology. 
Thus our approach can be considered conservative in parameter inference.

\subsection{Parameter inference}

We employ the Bayesian inference
to derive the parameter posterior distribution:
\begin{align}
    p_\mathrm{post}(\mathbf{p}|\mathcal{D}) \propto \mathcal{L}(\mathcal{D}|\mathbf{p}) p_\mathrm{prior}(\mathbf{p}),
\end{align}
where $p_\mathrm{prior}$ and $p_\mathrm{post}$ are the prior and posterior distributions of model parameters, and $\mathcal{L}(\mathcal{D}|\mathbf{p})$ is the likelihood function of the observational data $\mathcal{D}$ given parameters $\mathbf{p}$.
Using the power spectrum data vector and the covariance matrix, we compute the log-likelihood function:
\begin{align}
&\ln \mathcal{L}(\mathcal{D}|\mathbf{p}) = - \frac{1}{2} \sum_{\rm samp}
\sum_{\ell,\ell'} \sum_{i,j}^{k_\mathrm{max}} \left[ {P}^\mathcal{D}_{\ell}(k_i) - {P}_{\ell}(k_i;\mathbf{p}) \right] \nonumber\\
&\,\,\, \times \mathrm{Cov}^{-1}\left[{P}_\ell(k_i),{P}_{\ell'} (k_j)\right] \left[ {P}^\mathcal{D}_{\ell'}(k_j) - {P}_{\ell'}(k_j;\mathbf{p}) \right],
\label{eq:loglike}
\end{align}
where we assume the Gaussian likelihood and omit the normalization factor.
${P}^\mathcal{D}_{\ell}(k_i)$ denotes the data of the $\ell$-th multipole moment of the power spectrum in the $i$-th wavenumber bin, ${P}_{\ell}(k_i;\mathbf{p})$ is its theoretical model prediction, and $\mathbf{p}$ is the model parameters.
We include the power spectrum information over $k_\mathrm{min} \le k \le k_\mathrm{max}$, and we employ $k_\mathrm{min}=0.005\,h\,\mathrm{Mpc}^{-1}$ and $k_\mathrm{max}=0.25\,h\,\mathrm{Mpc}^{-1}$ as our fiducial choices, respectively. 
We will below give a validation of the choice of $k_\mathrm{max}=0.25\,h\,\mathrm{Mpc}^{-1}$ and discuss how different choices of $k_\mathrm{min}$ or $k_\mathrm{max}$ change the cosmological results.
The summation $\sum_{\rm samp}$ denotes the summation over galaxy samples when the power 
spectrum information for different galaxy samples are combined.

For the parameter sampling, we employ the Markov-chain Monte Carlo (MCMC) sampling of the standard Metropolis algorithm \cite{metropolis_1949}, or the nested sampling algorithm \texttt{MultiNest} \cite{Feroz_2009} implemented in the public Python package \texttt{PyMultiNest} \cite{Buchner_2014}.
In the MCMC sampling, we monitor the convergence for the cosmological parameters by using a method in Ref.~\cite{Vehtari_2021}, which is an improved variant of the Gelman-Rubin diagnostic \cite{Gelman_Rubin_1992,Brooks_Gelman_1998}.
More specifically, we apply the rank-normalization \cite{Vehtari_2021} to the MCMC chains and measure the Gelman-Rubin statistic $\hat{R}$ of the chains split in half (so-called the split-$\hat{R}$ \cite{Gelman_2013}), after discarding 1000 points at the beginning of each chain as the burn-in phase.
We run the MCMCs until the criteria $\hat{R} < 1.05$ for the cosmological parameters are met. 
\ykrv{We use \texttt{GetDist} \cite{Lewis:2019xzd} package to draw triangle plots of parameter posteriors.}

We should note that, throughout this paper, we \textit{do not} include the abundance of galaxies (the mean number density) nor the BAO information after reconstruction \citep[\textit{e.g.}, see][for such a study]{2020JCAP...05..032P} in the data vector in the parameter inference.

\begin{figure*}
\centering
    \includegraphics[width=0.95\columnwidth]{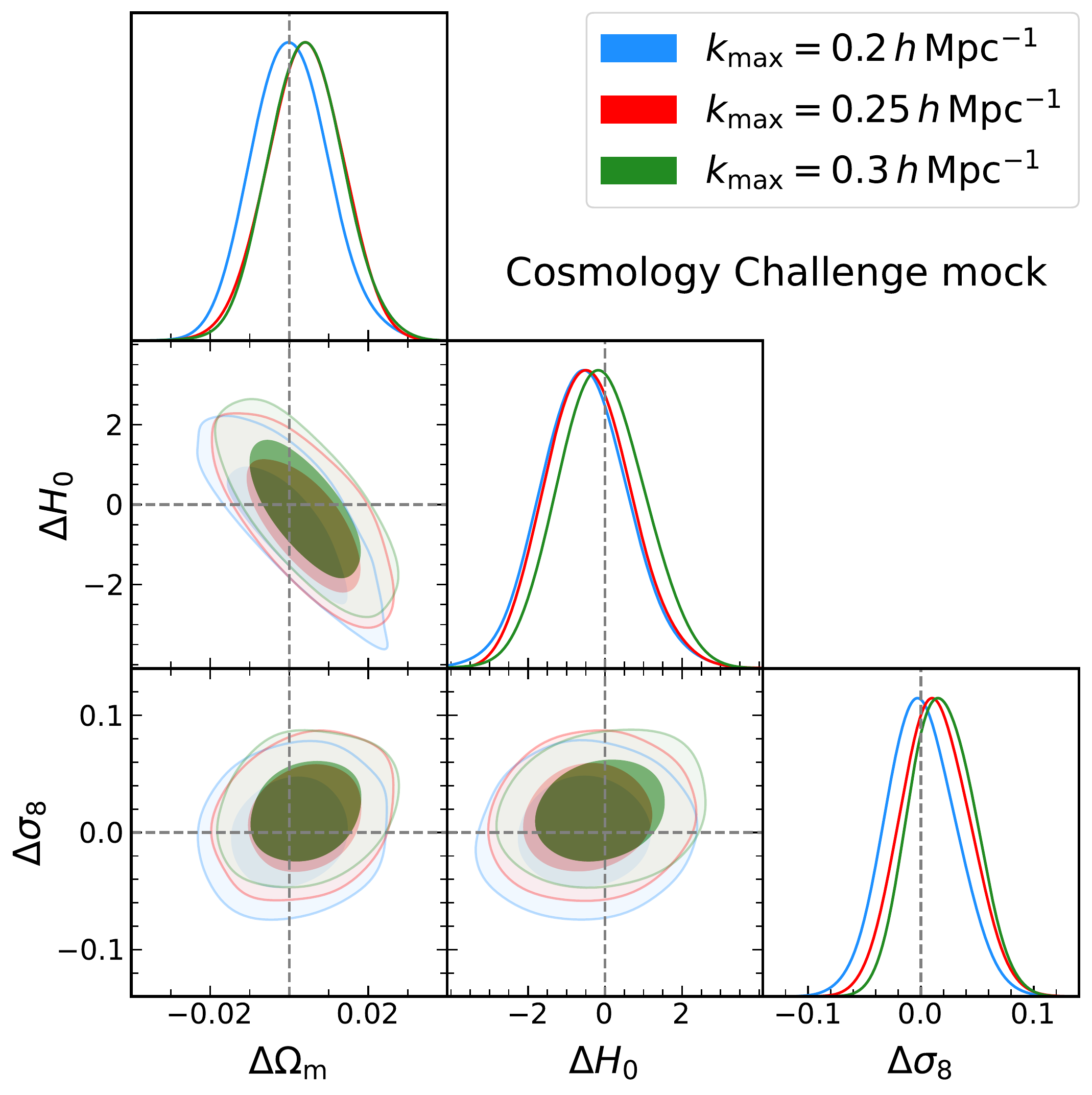}
    \includegraphics[width=0.95\columnwidth]{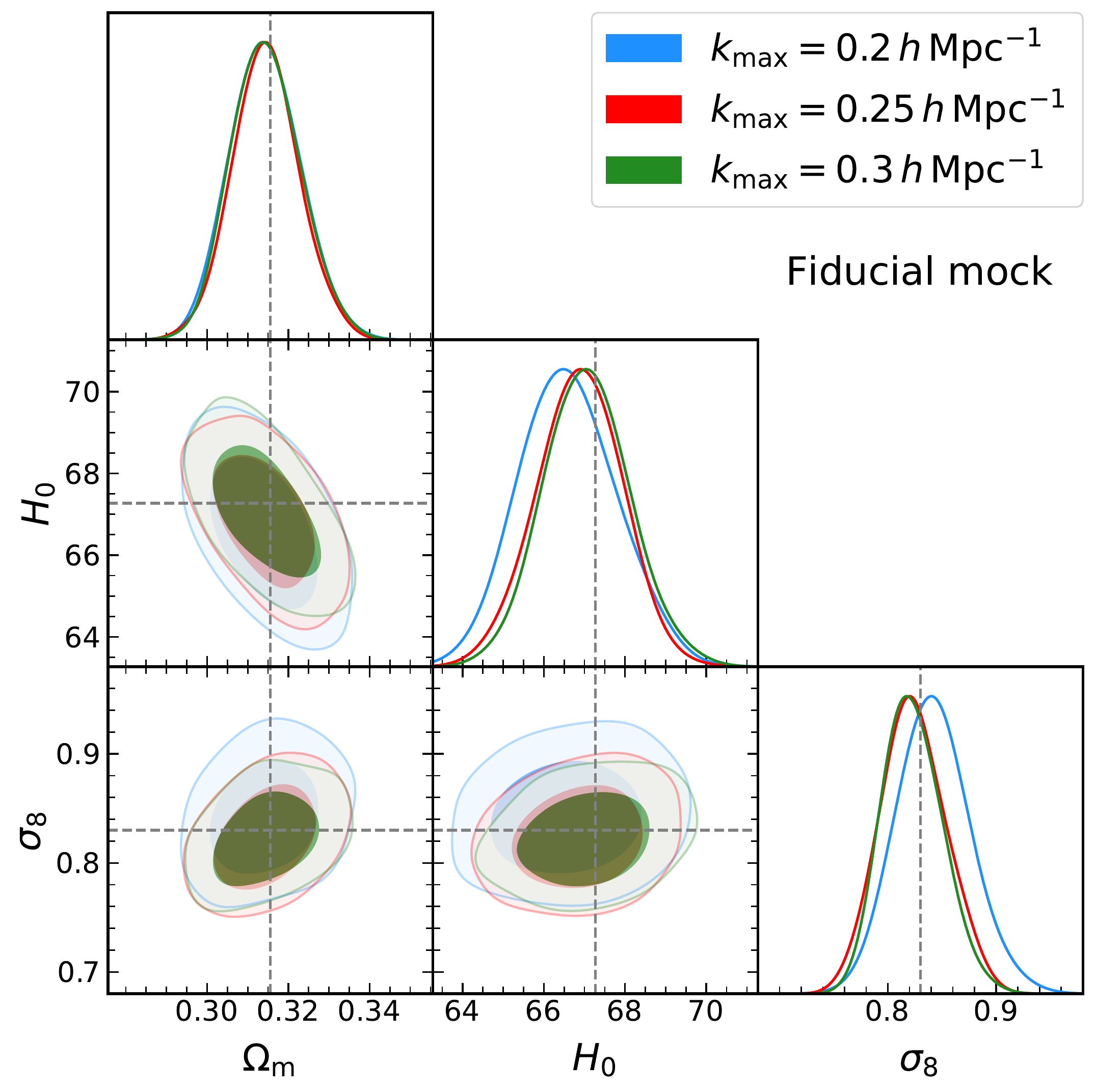}
\caption{
Results of validation tests of our cosmology analysis pipeline: we apply the analysis pipeline to the mock signals for the 4 BOSS-like galaxy samples using the covariance matrix of the actual BOSS spectra.
We include the monopole, quadrupole and hexadecapole moments of the redshift-space power spectrum for each of the 4 galaxy samples over $0.005\,h\,\mathrm{Mpc}^{-1} \le k \le k_\mathrm{max}$, and we show the results for $k_\mathrm{max}=0.20,0.25$ or 0.30$\,h\,\mathrm{Mpc}^{-1}$, respectively. 
The contours show the posterior distributions of $\Omega_\mathrm{m}, H_0,$ and $\sigma_8$ including marginalization over uncertainties in other parameters including the galaxy-halo connection parameters. 
The left plot shows the results for the galaxy mocks that are generated using a different recipe of the galaxy-halo connection from our fiducial HOD model: the mock galaxy catalogs used in the Cosmology Challenges in Ref.~\cite{2020arXiv200308277N} (see text for details). Since we would like to blind the true values of the cosmological parameters, we show the results in terms of the parameter difference such as $\Delta\Omega_\mathrm{m}=\Omega_\mathrm{m}-\Omega_\mathrm{m, true}$.
The right plot shows the results for the mock catalogs generated using the same form of HOD model as in our HOD model.
The dashed lines in the right panel are the true values of the cosmological parameters used in the mock catalogs, \textit{i.e.}, those for the {\it Planck} cosmology. 
}
\label{fig:mock_z02_NS_p024_kmax0.25_bbn-planck-gauss-fsigv-Pshot_kmax0.2-0.3}
\end{figure*}

\section{Results}
\label{sec:results}

We show the main results of our cosmology analysis in this section. 
Throughout this paper, we mainly focus on the constraints on three cosmological parameters $\Omega_\mathrm{m}$, $H_0$, and $\sigma_8$, which are well constrained by the redshift-space galaxy power spectrum, in flat $\Lambda$CDM model.

\subsection{Validation tests}
\label{sec:validation_maintext}

Before showing the main results, we first present validation tests of our emulator based method. 
To test the validity and usefulness of the emulator-based method, we performed various cosmology challenges: we apply our cosmology analysis pipeline to simulated mock signals of the redshift-space galaxy power spectrum to address whether the pipeline can recover the underlying true cosmological parameters used in the simulations. 
Please see Ref.~\cite{2020arXiv200308277N} for details of the procedures and aims.
Y.~K., who is one of the authors of this paper and led the actual cosmology inference analysis of the BOSS data, applied the pipeline to the mock signals for the BOSS-like galaxies used in Ref.~\cite{2020arXiv200308277N}. 
The mock galaxy catalogs were generated using a different recipe for the galaxy-halo connection based on subhalos, so it is not entirely clear whether our HOD method can recover the underlying cosmological parameters. 
For instance, the spatial and velocity structures of satellite galaxies in host halos are generally different from those in our fiducial halo model. 
In this test, Y.~K. was not informed of the cosmological parameters, and the validation test was done effectively in a blind manner. 
He submitted the results to T.~N., who is a co-author of this paper and is the main organizer and the maintainer of the challenge program, and then the ground truth cosmological parameters were revealed with the mutual agreement not to update the analysis anymore. 
The submitted results are recorded and presented on the challenge webpage (\url{https://www2.yukawa.kyoto-u.ac.jp/~takahiro.nishimichi/data/PTchallenge/}).

In Figure~\ref{fig:mock_z02_NS_p024_kmax0.25_bbn-planck-gauss-fsigv-Pshot_kmax0.2-0.3} we show the results of our validation tests. 
The left panel shows the results using the simulated signals that are generated from the mock catalogs of the above cosmology challenges of Ref.~\cite{2020arXiv200308277N}. 
To mimic the cosmology analysis of the BOSS power spectra, we use the mock catalogs at $z=0.38$ and $0.61$ to simulate the redshift-space spectra for the 4 subsamples, the NGC/SGC in the low-$z$ and high-$z$ bins, and use the same covariance matrix as that used in the following cosmology analysis of BOSS spectra (see Sec.~\ref{sec:data_ps_covariance}). 
\ykrv{These mock catalogs are based on the realizations of $N$-body simulations for the total volume of $566\,(h^{-1}\mathrm{Gpc})^3$ \cite{2020arXiv200308277N}}, which is about a hundred times that of the BOSS DR12 galaxy sample. 
The large volume of the simulations allows us to sufficiently reduce the sample variance errors in the simulated power spectra. 
Therefore, we can conduct a fairly stringent test of the systematic error due to an imperfect modeling of the power spectrum.
The figure shows that our analysis method recovers the true cosmological parameters to within the statistical errors of the BOSS galaxy spectra. 
We also stress that our emulator-based method passes the validation test even if including the power spectrum information up to $k_\mathrm{max}=0.3\,h\,\mathrm{Mpc}^{-1}$, where the perturbation theory breaks down.
Here the constraints on $\Omega_{\rm m}$ and $H_0$ are mainly from the BAO features and partly from the power spectrum shape via the AP effect.
On the other hand, the constraint on $\sigma_8$ is from the power spectrum amplitude, after the degeneracies with galaxy bias uncertainty (uncertainties in the galaxy-halo connection in our model) are lifted by measurements of the RSD effect as we will discuss later (Sec.~\ref{sec:fixed_HOD}).
However, including the information beyond $k\simeq 0.2\,h\,\mathrm{Mpc}^{-1}$ gives little improvement in the cosmological parameters, due to the shot noise domination and the degeneracies with the galaxy-halo connection parameters. 
On the other hand, we find that the constraints on HOD parameters are improved by including the higher-$k$ information. 

As another sanity check, we also perform the validation tests using the mock galaxy catalogs \ykrv{generated in Refs.~\cite{Nishimichi_2019,PhysRevD.101.023510}. They are based on the simulations for the {\it Planck} cosmology using the same HOD prescription as that in our analysis.}
Similarly to the above test, we used the outputs of $N$-body simulations at $z=0.251$ and 0.617, adopted the same HOD model as that in Ref.~\cite{2021arXiv210100113M} to populate galaxies into halos of each realization, and then simulated the mock galaxy spectra for the 4 galaxy samples.
We used the simulations of total volume $128\,(h^{-1}\mathrm{Gpc})^3$ to generate the simulated data vector \citep[see][for details of the simulations]{PhysRevD.101.023510}, and performed the cosmology analysis using the same covariance matrix of the BOSS spectra.
The right panel of Figure~\ref{fig:mock_z02_NS_p024_kmax0.25_bbn-planck-gauss-fsigv-Pshot_kmax0.2-0.3} shows that our analysis method recovers the cosmological parameters to within the statistical errors, for all the $k_\mathrm{max}$ values. 
One might notice a slight bias in each cosmological parameter, even if both the theoretical template and the mock catalogs employ the same form of HOD model. 
We ascribe the parameter shift to the projection effect of the full posterior distribution in a multi-dimensional parameter space \citep[see][for the similar discussion]{Sugiyama_2020}.
Here we note that some of the galaxy-halo connection parameters are not necessarily recovered by the analysis as explicitly shown in Figure~\ref{fig:posterior_galaxy-halo_parameters_mockchallenge} of Appendix~\ref{sec:hod_parameters_mockchallenges}.
The figure in the appendix also shows that including the power spectrum information on the higher $k_\mathrm{max}$ gives smaller error bars for the galaxy-halo connection parameters, although the central values are biased for some of the galaxy-halo parameters. 
On the other hand, the credible intervals of the cosmological parameters are not much improved \ykrv{by including the information from 
$k_{\rm max}=0.25$ to $0.30\,h{\rm Mpc}^{-1}$,} as can be found from Figure~\ref{fig:mock_z02_NS_p024_kmax0.25_bbn-planck-gauss-fsigv-Pshot_kmax0.2-0.3}. 
Hence, from these results, we conclude that our method can robustly recover the cosmological parameters to within the statistical errors of the BOSS power spectra, after marginalization over the galaxy-halo connection parameters.
Here note that the cosmological information is extracted from the redshift-space power spectrum of halos predicted by the emulator in our method. 
In this paper, we employed $k_\mathrm{max}=0.25\,h\,\mathrm{Mpc}^{-1}$ as our fiducial choice of the maximum wavenumber. 

We will also later show the validation test of our analysis method using the mock catalogs including the assembly bias effect, which is one of the most dangerous, physical systematic effects in the halo model approach.

\begin{figure*}
\centering
\includegraphics[width=0.7\textwidth]{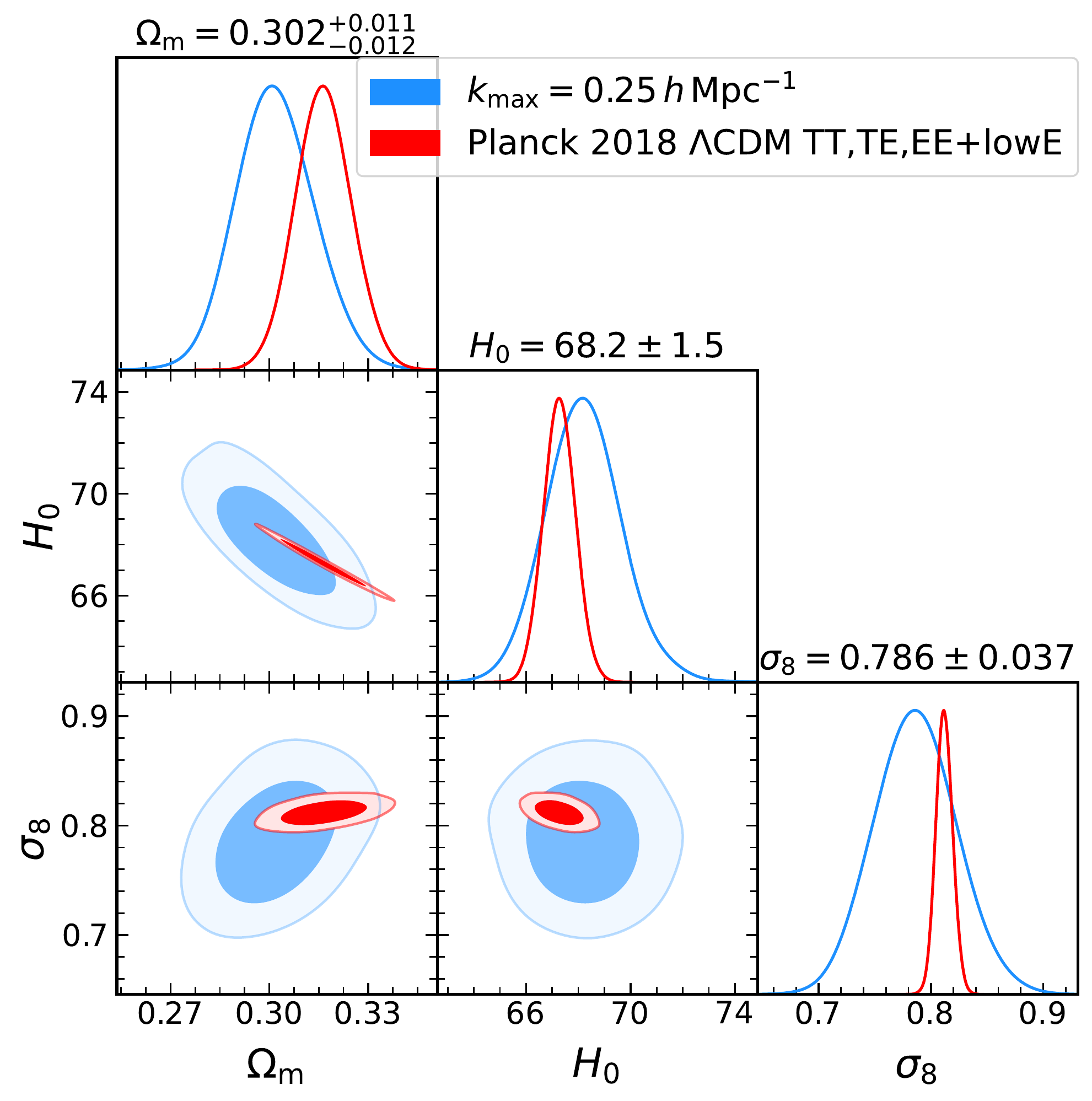}
\caption{
The posterior distributions of $\Omega_\mathrm{m}, H_0$ and $\sigma_8$, obtained from the cosmology inference including the full shape of the monopole, quadrupole and hexadecapole moments of the \ykrv{BOSS DR12 galaxy power spectra} up to $k_\mathrm{max} = 0.25\,h\,\mathrm{Mpc}^{-1}$, for the flat $\Lambda$CDM model. 
For the theoretical templates we use the emulator-based halo model, and the posteriors include marginalization over uncertainties in other cosmological parameters and the nuisance parameters including the galaxy-halo connection parameters. 
For comparison, the red contours show the results from the \textit{Planck} 2018 ``TT, TE, EE+lowE'' analysis.
}
\label{fig:real_z02_NS_p024_bbn-planck-gauss-fsigv-Pshot_kmax0.25}
\end{figure*}
\begin{table*}
    \centering
    \begin{tabular}{l|ccc|c} \hline \hline
        Parameter & MAP & Median & 68\% CI & Planck 68\% CI \\ \hline
        $\ln (10^{10} A_\mathrm{s})$ & \ykrv{2.93} & \ykrv{3.01} & \ykrv{$3.01_{-0.087}^{+0.089}$} & $3.045 \pm 0.016$ \\ \hline
        $\Omega_\mathrm{m}$ & \ykrv{0.300} & \ykrv{0.302} & \ykrv{$0.301_{-0.011}^{+0.012}$} & $0.3166 \pm 0.0084$ \\
        $H_0\,[\mathrm{km}\,\mathrm{s}^{-1}\,\mathrm{Mpc}^{-1}]$ & \ykrv{68.3} & \ykrv{68.2} & \ykrv{$68.2_{-1.4}^{+1.4}$} & $67.27 \pm 0.60$ \\
        $\sigma_8$ & \ykrv{0.754} & \ykrv{0.786} & \ykrv{$0.786_{-0.037}^{+0.036}$} & $0.8120 \pm 0.0073$ \\ \hline
        $S_8 \equiv \sigma_8 (\Omega_\mathrm{m}/0.3)^{0.5}$ & \ykrv{0.754} & \ykrv{0.787} & \ykrv{$0.784_{-0.042}^{+0.048}$} & $0.834 \pm 0.016$ \\
        $f\sigma_8(z_{\rm eff}=0.38)$ & \ykrv{0.474} & \ykrv{0.471} & \ykrv{$0.467_{-0.028}^{+0.035}$} & $0.4771 \pm 0.0066$ \\
        $f\sigma_8(z_{\rm eff}=0.61)$ & \ykrv{0.434} & \ykrv{0.434} & \ykrv{$0.430_{-0.026}^{+0.034}$} & $0.4696 \pm 0.0053$ \\
        \hline \hline
    \end{tabular}
    \caption{
    The results of the cosmological parameter inference of the BOSS power spectra for $\Lambda$CDM model as in Figure~\ref{fig:real_z02_NS_p024_bbn-planck-gauss-fsigv-Pshot_kmax0.25}.
    Note that we include the BBN prior on $\omega_\mathrm{b}$ and the \textit{Planck} CMB prior on the spectral tilt $n_\mathrm{s}$ (see Table~\ref{tab:param_table}).
    For each parameter, we show the parameter value at the \textit{maximum a posteriori} (MAP), the median, and the mode with 68\% credible interval (CI) for the 1d posterior distribution of each parameter.
    The parameters other than $\ln(10^{10}A_{\rm s})$ are derived parameters (see Table~\ref{tab:param_table}).
    For comparison with constraints from other experiments, we also show the constraints on $S_8\equiv \sigma_8(\Omega_{\rm m}/0.3)^{0.5}$ and $f(z)\sigma_8(z)$, where $f$ is the linear growth rate $f \equiv \mathrm{d}\ln D/\mathrm{d}\ln a$. 
    For $f\sigma_8$, we show the results obtained from the cosmology analysis using the subsample of low-$z$ NGC+SGC or 
    high-$z$ NGC+SGC sample at the effective redshift $z_{\rm eff}=0.38$ and 0.61, respectively 
    (see the right panel of Figure~\ref{fig:real_z02_NS_p024_kmax0.25_bbn-planck-gauss-fsigv-Pshot_2chunks}).
    For comparison, we also show the 68\% credible interval from the {\it Planck} 2018 ``TT, TE, EE+lowE'' analysis, 
    taken from Table~2 of Ref.~\cite{planck_collaboration_2020}.
    }
    \label{tab:constraint_result}
\end{table*}
\begin{figure*}
\centering
\includegraphics[width=0.8\textwidth]{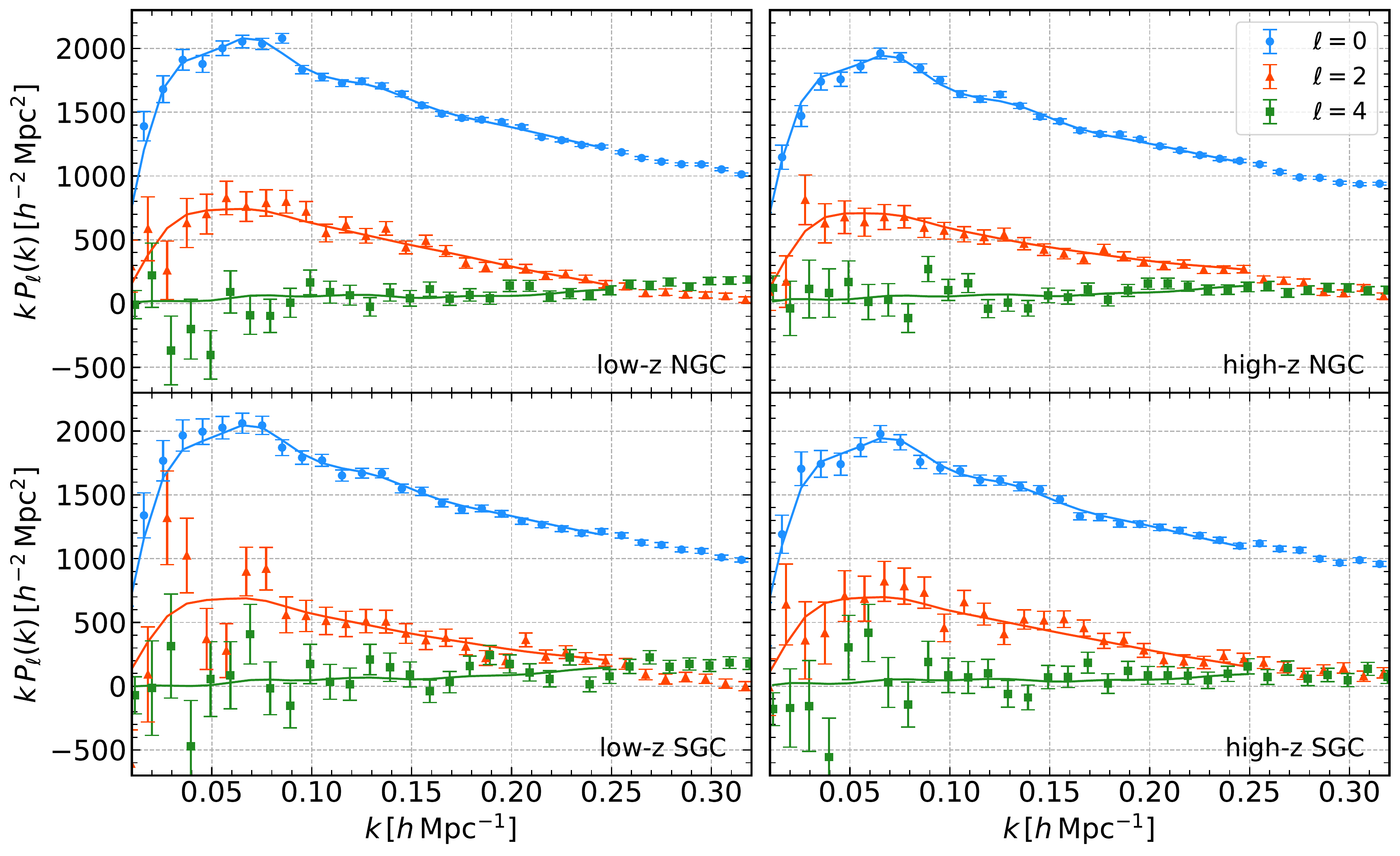}
\caption{
\ykrv{The comparison of monopole (blue), quadrupole (red), and hexadecapole (green) moments between the data (symbols with error bars) and the model predictions (solid lines) at MAP of the MCMC chains in our cosmology analysis shown in Figure~\ref{fig:real_z02_NS_p024_bbn-planck-gauss-fsigv-Pshot_kmax0.25}.
The error bars are computed from the diagonal elements of the covariance matrix, described in Section \ref{sec:data_ps_covariance}.}
}
\label{fig:boss_dr12_power_MAP}
\end{figure*}
\begin{figure*}
\centering
\includegraphics[width=0.7\textwidth]{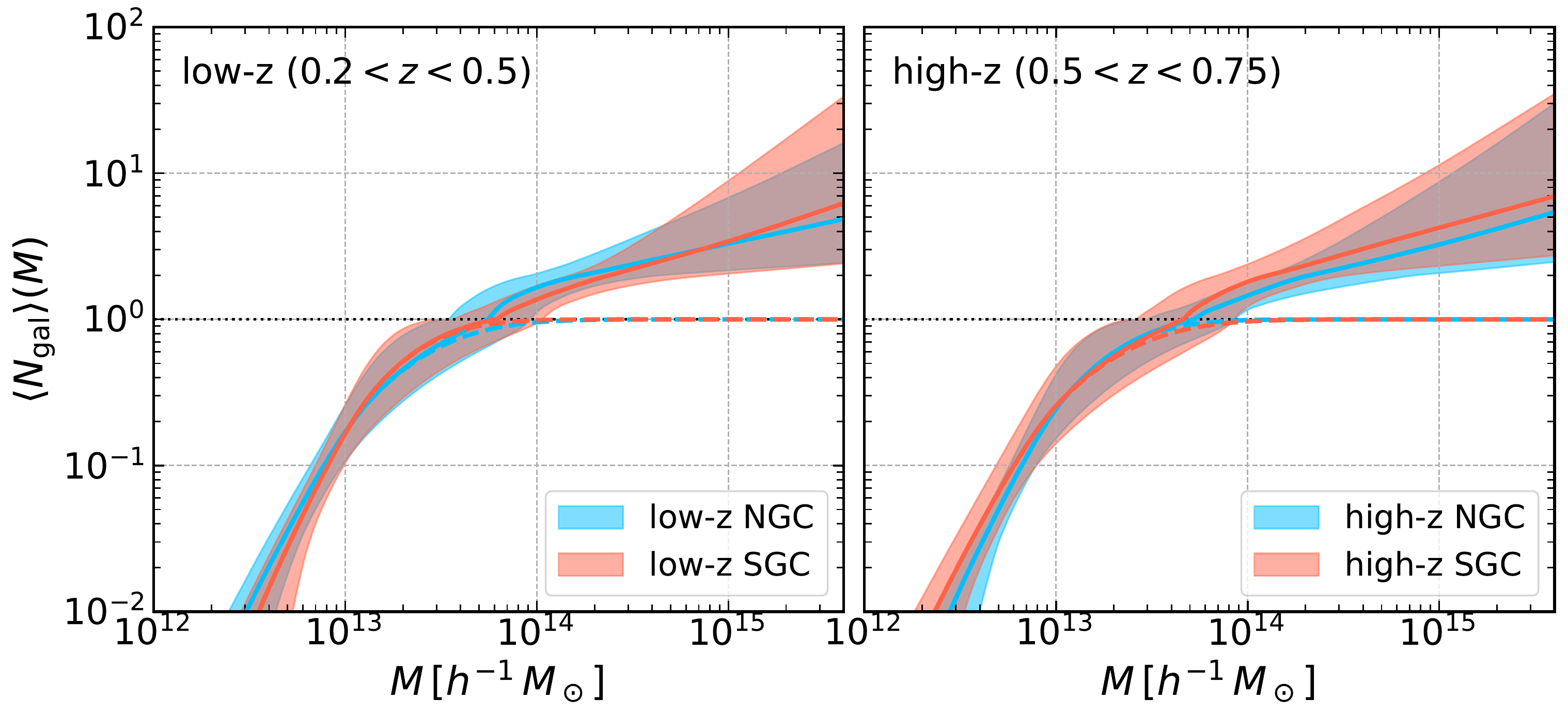}
\caption{
The median and 68\% percentile interval of the HOD functions for each of the BOSS galaxy samples (low-$z$/high-$z$ NGC/SGC samples), obtained from the MCMC chains in Figure~\ref{fig:real_z02_NS_p024_bbn-planck-gauss-fsigv-Pshot_kmax0.25}.
The solid and dashed lines are the medians of central+satellite and central-only HOD functions, respectively.
}
\label{fig:hod_bossdr12}
\end{figure*}
\subsection{Results: $\Lambda$CDM cosmology}
\label{sec:main_result}
Figure~\ref{fig:real_z02_NS_p024_bbn-planck-gauss-fsigv-Pshot_kmax0.25} shows the main results of this paper, \ykrv{the cosmological parameters obtained from the actual BOSS power spectra. The figure shows} the projected posterior distributions of $\Omega_\mathrm{m}$, $H_0$ and $\sigma_8$ for the flat $\Lambda$CDM model, after marginalizing over the other parameters such as the galaxy-halo connection parameters. 
These constraints include the BAO and full shape information of the redshift-space power spectrum for the 4 samples of the low-$z$/high-$z$ NGC and SGC samples, up to $k_\mathrm{max}=0.25\,h\,\mathrm{Mpc}^{-1}$ where the nonlinear effects such as nonlinear bias, nonlinear clustering and nonlinear RSD are properly included. 
Our method achieves precise measurements of the cosmological parameters:
\ykrv{
\begin{align}
&\Omega_\mathrm{m}=0.301^{+0.012}_{-0.011}\nonumber\\
&H_0=68.2 \pm 1.4 \nonumber\\
&\sigma_8=0.786^{+0.036}_{-0.037},
\end{align}
}
where we report the mode and the 68\% credible interval of the marginalized posterior distribution for each parameter, respectively.
Note that $H_0$ is in units of $\mathrm{km\,s}^{-1}\mathrm{Mpc}^{-1}$. 

\ykrv{Table~\ref{tab:comparison_to_PT} compares our result with those from the recent similar full-shape analyses of the power spectrum multipoles using the PT-based theory models: the EFTofLSS model \cite{philcox2022boss} and the model based on Lagrangian perturbation theory (LPT) \cite{Chen_2022}. 
They used the same power spectrum signals and window function of BOSS DR12 provided by Ref.~\cite{Beutler_McDonald_2021}, \textit{i.e.,} the same dataset as that we use. 
}
It is interesting to note that our results are consistent with \ykrv{these two analyses} to within the statistical errors, even though our halo model-based method and \ykrv{the PT-based ones} are constructed based on totally different frameworks. 
The size of the error bar in each parameter is comparable with those of the their results. 
This might be counter-intuitive, because one might think that our halo model based method is a more restrictive model than \ykrv{PT-based} models, so expect our method gives tighter constraints on the cosmological parameters. 
Probably this is due to the wide priors of the galaxy-halo connection parameters in our analysis, and we will later discuss a possible room of improvements in the cosmological parameters within our method. 

\begin{table*}
    \centering
    \ykrv{
    \begin{tabular}{l|ccc} \hline \hline
        Parameter & This work & EFTofLSS model \cite{philcox2022boss} & LPT model \cite{Chen_2022} \\ \hline
        $\Omega_\mathrm{m}$ & $0.301_{-0.011}^{+0.012}$ & $0.312_{-0.012}^{+0.011}$ & $0.305 \pm 0.01$ \\
        $H_0\,[\mathrm{km}\,\mathrm{s}^{-1}\,\mathrm{Mpc}^{-1}]$ & $68.2 \pm 1.4$ & $68.5_{-1.3}^{+1.1}$ & $68.5 \pm 1.1$ \\
        $\sigma_8$ & $0.786_{-0.037}^{+0.036}$ & $0.737_{-0.044}^{+0.040}$ & $0.738 \pm 0.048$ \\ \hline
    \end{tabular}
    }
    \caption{
    \ykrv{
    Comparison of the parameter bounds from recent ``full-shape'' analyses of the BOSS DR12 power spectrum ($68\%$ CI). 
    For the analysis of the EFTofLSS model, we quote the results using the window-convolved power spectrum (top row of Table~IV of Ref.~\cite{philcox2022boss}).
    All the results shown here are obtained from exactly the same power spectrum data provided by Ref.~\cite{Beutler_McDonald_2021}.
    }
    }
    \label{tab:comparison_to_PT}
\end{table*}
\ykrv{
We also note other works \citep{Ivanov_2020,d_Amico_2020,Zhang_2022,2021arXiv211103156S,2021PhRvD.104l1301B}
that recently performed the full-shape cosmology analysis of the BOSS power spectrum. 
However, these works used the different dataset and/or analysis method from those in this paper, so it is not easy to make an apple-to-apple comparison with our result. 
We here give a note for completeness of our discussion, stressing that the full-shape cosmology analysis is becoming possible to obtain cosmological constraints for a given cosmological framework such as the flat $\Lambda$CDM model.
}

Our results can also be compared with the \textit{Planck} 2018 CMB constraints: we overplot the \textit{Planck} 2018 cosmological constraints from the baseline likelihood of ``TT, TE, EE+lowE''; 
the MCMC chains of \textit{Planck} 2018 cosmology analysis are downloaded from the Planck Legacy Archive (\url{http://pla.esac.esa.int/pla/#cosmology}), where the neutrino mass is fixed to $0.06\,$eV as we did in our analysis.
Our results are in good agreement with the {\it Planck} results for \ykrv{all of the three cosmological parameters}.
In Table~\ref{tab:constraint_result}, we also show the parameter value at the \textit{maximum a posteriori} 
(MAP), the median, and the mode with 68\% credible interval (CI) for the 1d posterior distribution of each parameter. 
For comparison with the weak lensing survey results \citep[\textit{e.g.}][]{2019PASJ...71...43H,2021arXiv211102419M}, we also give the results for $S_8$, the parameter on which the weak lensing surveys can give most stringent constraint. 
In addition, we show the results for $f(z)\sigma_8(z)$, which is the parameter often used to characterize the constraint mainly from the RSD measurement \ykrv{on linear scales}. 
\ykrv{Note that, since we assume the flat $\Lambda$CDM cosmology, the linear growth rate $f(z)$ is determined solely by $\Omega_\mathrm{m}$ through
\begin{align}
    &f(z) = - \frac{3}{2} \Omega_\mathrm{m}(z) + \left[ \int_0^1 \frac{\mathrm{d}x}{\left\{ \Omega_\mathrm{m}(z) x^{-1} + (1-\Omega_\mathrm{m}(z)) x^2 \right\}^{\frac{3}{2}} } \right]^{-1},
\end{align}
where $\Omega_\mathrm{m}(z) = \Omega_\mathrm{m}(1+z)^3 [H_0/H(z)]^2 $ is the time-dependent matter density parameter.
It is in contrast to traditional RSD analyses such as Ref.~\cite{Beutler:2016arn}, where one performs \textit{model independent} linear-growth measurements, marginalizing over possible nonlinear corrections and focusing only on the amplitude of the apparent anisotropies arising from the linear RSD effect.
Therefore, as we have already quoted earlier in this subsection, we can determine fundamental cosmological parameters, breaking the degeneracy between $\sigma_8$ and $f$ under the assumed cosmological model.}
The table gives the constraints for $f(z)\sigma_8(z)$ that are obtained from the MCMC analyses using either of the low-$z$ NGC+SGC sample or high-$z$ NGC+SGC sample at the effective redshift $z_{\rm eff}=0.38$ or $0.61$, respectively (see below).
Our constraint is from the combined information of the BAO, the AP effect, the RSD effect and the amplitude and shape information of the power spectrum under the flat $\Lambda$CDM framework. 

Figure~\ref{fig:boss_dr12_power_MAP} shows that our model at MAP well reproduces all the measured multipoles of redshift-space power spectra. 
\ykrv{The reduced chi-square value at the MAP is $\chi^2/\mathrm{d.o.f.} \simeq 1.05$ for 267$(=300-33)$ degrees of freedom (the $p$-value $p \simeq 0.28$), implying that the MAP model gives an acceptable fit to the data.}
Taking a closer look at the figure, our model has a slightly weaker BAO feature than data. 
This tendency reflects the fact that our power spectrum emulator is based on the training data which have a larger $k$-bin width ($\Delta k = 0.02 \,h\,\mathrm{Mpc}^{-1}$) than the data ($\Delta k = 0.01\,h\,\mathrm{Mpc}^{-1}$), to suppress the statistical scatters on the training data (also see Appendix~A of Ref.~\cite{Kobayashi_2020}). 
Since the BAO feature leads to a tight constraint on $\Omega_\mathrm{m}$ through the AP effect, the sharpening of the BAO feature in the emulator could improve the cosmological constraints.

Figure~\ref{fig:hod_bossdr12} shows the mean HOD functions obtained from the MCMC chains. 
It can be found that the galaxy population inferred from the BOSS DR12 galaxy power spectra is almost confined to halos with masses, $M \gtrsim 10^{12} \, h^{-1} \, M_\odot$. 
This figure also shows that there is no remarkable difference in the HOD among the 4 galaxy samples.
Our results are qualitatively consistent with the HODs estimated in the previous works \cite{reid09,2011ApJ...728..126W,2015ApJ...806....2M}.
We again stress that the HOD constraints are obtained from the fitting of the emulator-based halo model to the redshift-space power spectrum, without employing any strong priors on the HOD parameters nor using the abundance information (the mean number density of galaxies). 
Hence, the posterior distributions of HOD are purely from the redshift-space clustering information, while the previous works took into account different clustering information such as the galaxy-galaxy weak lensing and/or the projected correlation function \ykrv{\citep[e.g.][for such study]{2021arXiv211102419M}.}

\section{Discussion}
\label{sec:discussion}

In this section, we discuss the robustness of our cosmology results: we study how different analysis methods and datasets change the inferred cosmological parameters.

\begin{figure*}
\centering
    \includegraphics[width=0.99\columnwidth]{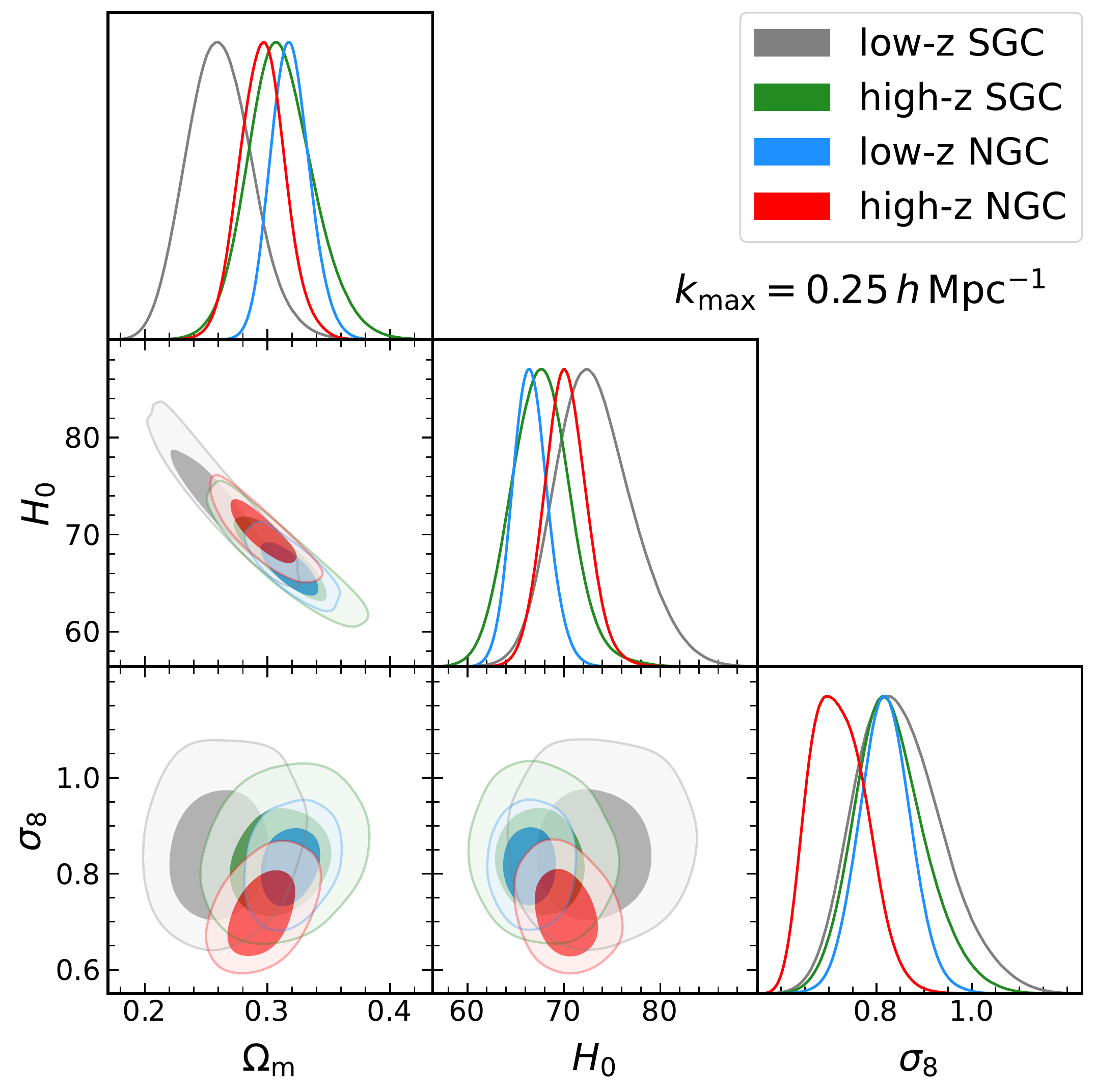}
    \includegraphics[width=0.99\columnwidth]{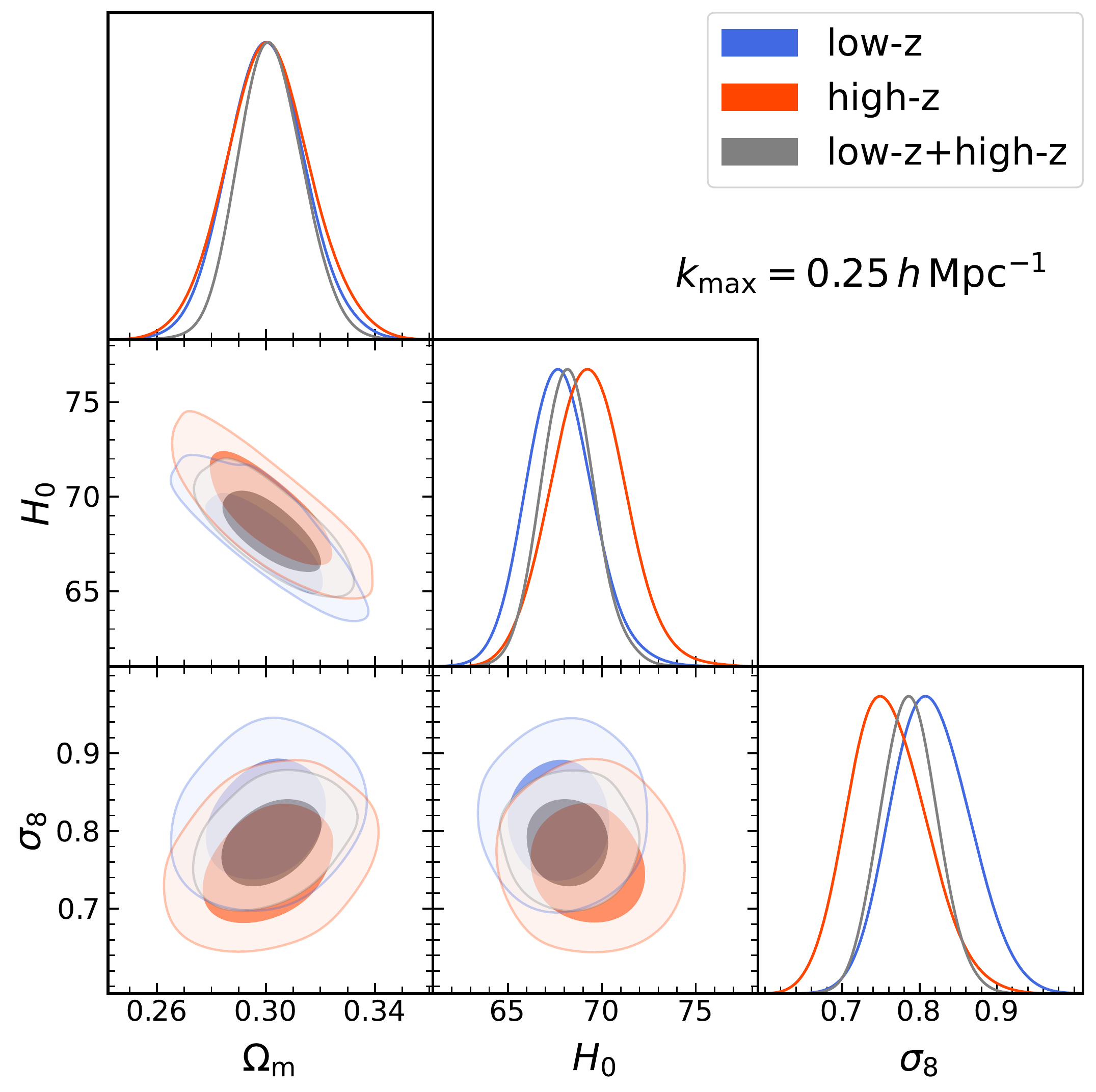}
\caption{
The posterior distributions for cosmological parameters, obtained from the different galaxy subsamples.
The left panel shows the results for 4 individual galaxy samples: 
low-$z$ NGC, low-$z$ SGC, high-$z$ NGC, and high-$z$ SGC.
The right panel shows the results for each of low-$z$ or high-$z$ NGC+SGC samples.
In the right panel, we overplot the gray contours to show the distribution for the full sample for comparison, which is the same as the blue contours in Figure~\ref{fig:real_z02_NS_p024_bbn-planck-gauss-fsigv-Pshot_kmax0.25}.
% These results can be compared with Figure~2 in Ref.~\cite{Ivanov_2020}, which shows the similar posterior distributions obtained using the EFTofLSS.
}
\label{fig:real_z02_NS_p024_kmax0.25_bbn-planck-gauss-fsigv-Pshot_2chunks}
\end{figure*}
\begin{table*}
    \centering
    \begin{tabular}{l|ccc|ccc} \hline \hline
         & \multicolumn{3}{c|}{low-$z$ NGC} & \multicolumn{3}{c}{low-$z$ SGC} \\
        Parameter & MAP & Median & 68\% CI & MAP & Median & 68\% CI \\ \hline
        $\ln (10^{10} A_\mathrm{s})$ & \ykrv{3.05} & \ykrv{3.10} & \ykrv{$3.12_{-0.14}^{+0.11}$} & \ykrv{3.17} & \ykrv{3.14} & \ykrv{$3.14_{-0.21}^{+0.20}$} \\ \hline
        $\Omega_\mathrm{m}$ & \ykrv{0.311} & \ykrv{0.318} & \ykrv{$0.317_{-0.015}^{+0.016}$} & \ykrv{0.261} & \ykrv{0.261} & \ykrv{$0.259_{-0.025}^{+0.028}$} \\
        $H_0\,[\mathrm{km}\,\mathrm{s}^{-1}\,\mathrm{Mpc}^{-1}]$ & \ykrv{66.1} & \ykrv{66.5} & \ykrv{$66.4_{-1.7}^{+1.8}$} & \ykrv{71.1} & \ykrv{73.0} & \ykrv{$72.4_{-3.6}^{+4.2}$} \\
        $\sigma_8$ & \ykrv{0.782} & \ykrv{0.817} & \ykrv{$0.816_{-0.051}^{+0.055}$} & \ykrv{0.811} & \ykrv{0.842} & \ykrv{$0.826_{-0.077}^{+0.102}$} \\ \hline
        $S_8 \equiv \sigma_8 (\Omega_\mathrm{m}/0.3)^{0.5}$ & \ykrv{0.796} & \ykrv{0.841} & \ykrv{$0.840_{-0.061}^{+0.063}$} & \ykrv{0.756} & \ykrv{0.783} & \ykrv{$0.763_{-0.084}^{+0.106}$} \\
        $f\sigma_8(z_{\rm eff}=0.38)$ & \ykrv{0.457} & \ykrv{0.481} & \ykrv{$0.481_{-0.033}^{+0.034}$} & \ykrv{0.451} & \ykrv{0.467} & \ykrv{$0.455_{-0.045}^{+0.061}$} \\
        \hline \hline
         & \multicolumn{3}{c|}{high-$z$ NGC} & \multicolumn{3}{c}{high-$z$ SGC} \\
        Parameter & MAP & Median & 68\% CI & MAP & Median & 68\% CI \\ \hline
        $\ln (10^{10} A_\mathrm{s})$ & \ykrv{2.65} & \ykrv{2.76} & \ykrv{$2.78_{-0.193}^{+0.127}$} & \ykrv{3.02} & \ykrv{3.11} & \ykrv{$3.10_{-0.17}^{+0.18}$} \\ \hline
        $\Omega_\mathrm{m}$ & \ykrv{0.298} & \ykrv{0.296} & \ykrv{$0.297_{-0.020}^{+0.017}$} & \ykrv{0.305} & \ykrv{0.310} & \ykrv{$0.307_{-0.024}^{+0.028}$} \\
        $H_0\,[\mathrm{km}\,\mathrm{s}^{-1}\,\mathrm{Mpc}^{-1}]$ & \ykrv{70.3} & \ykrv{70.1} & \ykrv{$70.1_{-2.1}^{+2.2}$} & \ykrv{67.6} & \ykrv{67.6} & \ykrv{$67.7_{-3.1}^{+2.8}$} \\
        $\sigma_8$ & \ykrv{0.685} & \ykrv{0.717} & \ykrv{$0.697_{-0.044}^{+0.074}$} & \ykrv{0.783} & \ykrv{0.824} & \ykrv{$0.814_{-0.066}^{+0.079}$} \\ \hline
        $S_8 \equiv \sigma_8 (\Omega_\mathrm{m}/0.3)^{0.5}$ & \ykrv{0.683} & \ykrv{0.711} & \ykrv{$0.692_{-0.056}^{+0.084}$} & \ykrv{0.789} & \ykrv{0.838} & \ykrv{$0.824_{-0.079}^{+0.096}$} \\
        $f\sigma_8(z_{\rm eff}=0.61)$ & \ykrv{0.393} & \ykrv{0.410} & \ykrv{$0.399_{-0.027}^{+0.043}$} & \ykrv{0.450} & \ykrv{0.475} & \ykrv{$0.470_{-0.040}^{+0.047}$}  \\
        \hline \hline
    \end{tabular}
    \caption{
    Similar to Table~\ref{tab:constraint_result}, but this table shows the results obtained from the cosmology analysis using each of the 4 galaxy samples. 
    }
    \label{tab:constraint_result_each-chunk}
\end{table*}

\subsection{Variations in the cosmological parameters for different galaxy subsamples}
\label{sec:variations_subsamples}

In Figure~\ref{fig:real_z02_NS_p024_kmax0.25_bbn-planck-gauss-fsigv-Pshot_2chunks} we study how the cosmological parameters are changed when using different subsets of the data vector:
the left panel shows the results for 4 individual galaxy samples, and the right panel shows the combined results for each of the two redshift bins, where the two galactic hemispheres, NGC and SGC, are combined.
Shifts in each parameter display a similar trend to those shown in Figure~2 of Ref.~\cite{Ivanov_2020}, even though we use a totally different theoretical template, \textit{i.e.}, the halo model based method, compared to the EFTofLSS in their paper. 
Hence, we think that the parameter shifts are likely due to the sample variances in each subsample.
For completeness of our discussion, in Appendix~\ref{sec:posterior_full_parameters} we show the 2d posterior distributions for the full parameters for each of the 4 individual galaxy samples (see Figure~\ref{fig:real_p024_kmax0.25_bbn-planck-gauss-fsigv-Pshot_each-chunk}). 
The low-$z$ SGC sample displays a sizable difference in some parameters compared to the other samples, but the difference is still within the statistical errors. 
Hence, we cannot give any definite conclusion as to that sample could have a potential observational systematics compared to the others.

\subsection{The impact of hexadecapole moments}
\label{sec:hexadecapole}
\begin{figure}
\centering
\includegraphics[width=0.45\textwidth]{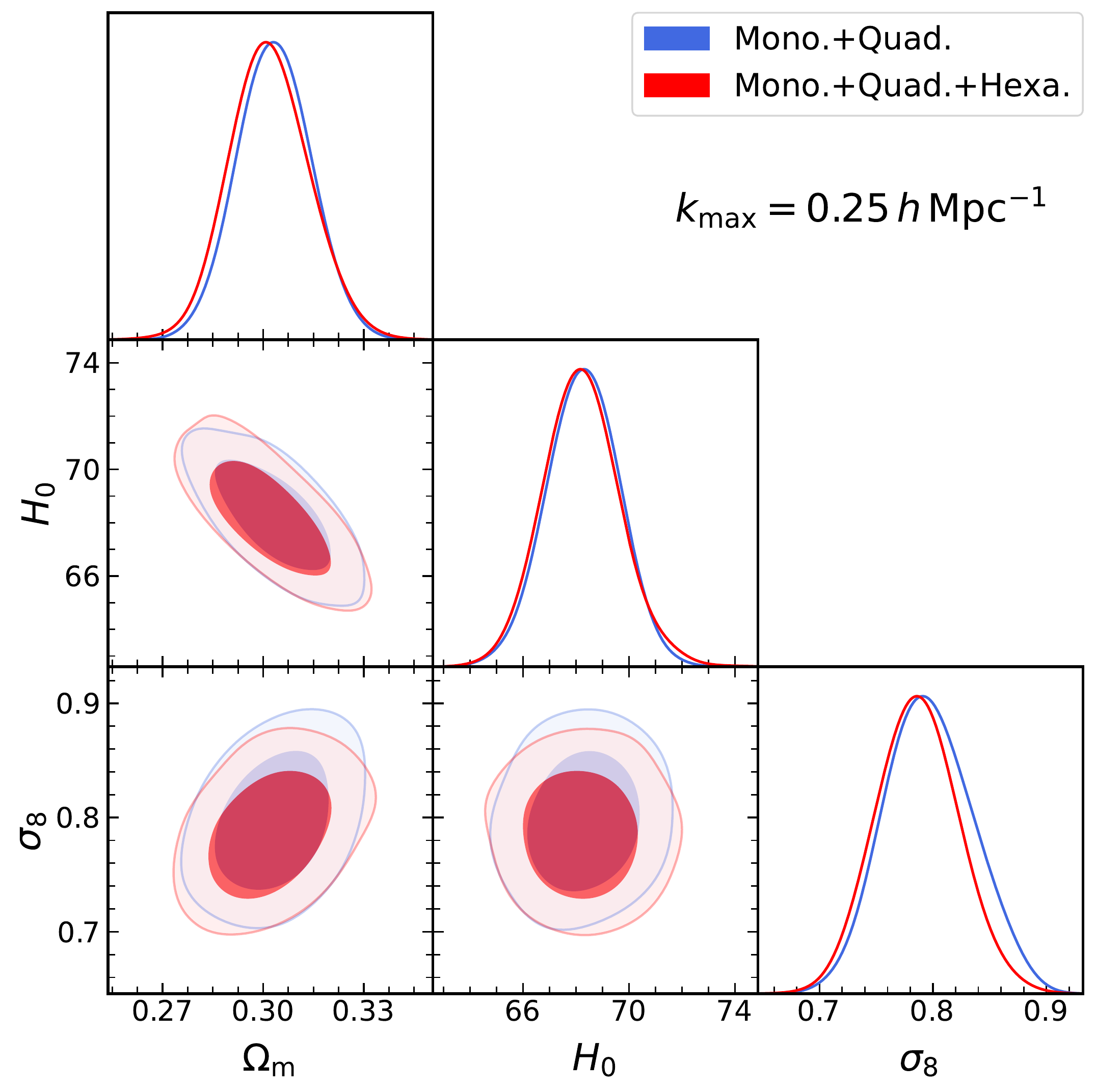}
\caption{
The posterior distributions of cosmological parameters in our fiducial cosmology analysis for the full galaxy sample, obtained if or not we include the hexadecapole moments of the redshift-space power spectrum in the parameter inference. 
The red contours are the same as those in Figure~\ref{fig:real_z02_NS_p024_bbn-planck-gauss-fsigv-Pshot_kmax0.25}.
}
\label{fig:real_z02_NS_p02-p024_bbn-planck-gauss-fsigv-Pshot}
\end{figure}
Figure~\ref{fig:real_z02_NS_p02-p024_bbn-planck-gauss-fsigv-Pshot} shows that the inclusion of the hexadecapole moment of the redshift-space power spectrum yields only a subtle improvement in the cosmological parameters, because the hexadecapole moments have lower signal-to-noise ratios than the monopole and quadrupole moments as shown in Figure~\ref{fig:boss_dr12_power_MAP}. 
However, we note that the hexadecapole indeed improves some 
of the HOD parameters, which is consistent with the finding in Ref.~\cite{hikage:2013lr}.

\subsection{The cosmological information in the different range of $k$}
\label{sec:cosmology_different_kmax}
\begin{figure}
\centering
\includegraphics[width=0.45\textwidth]{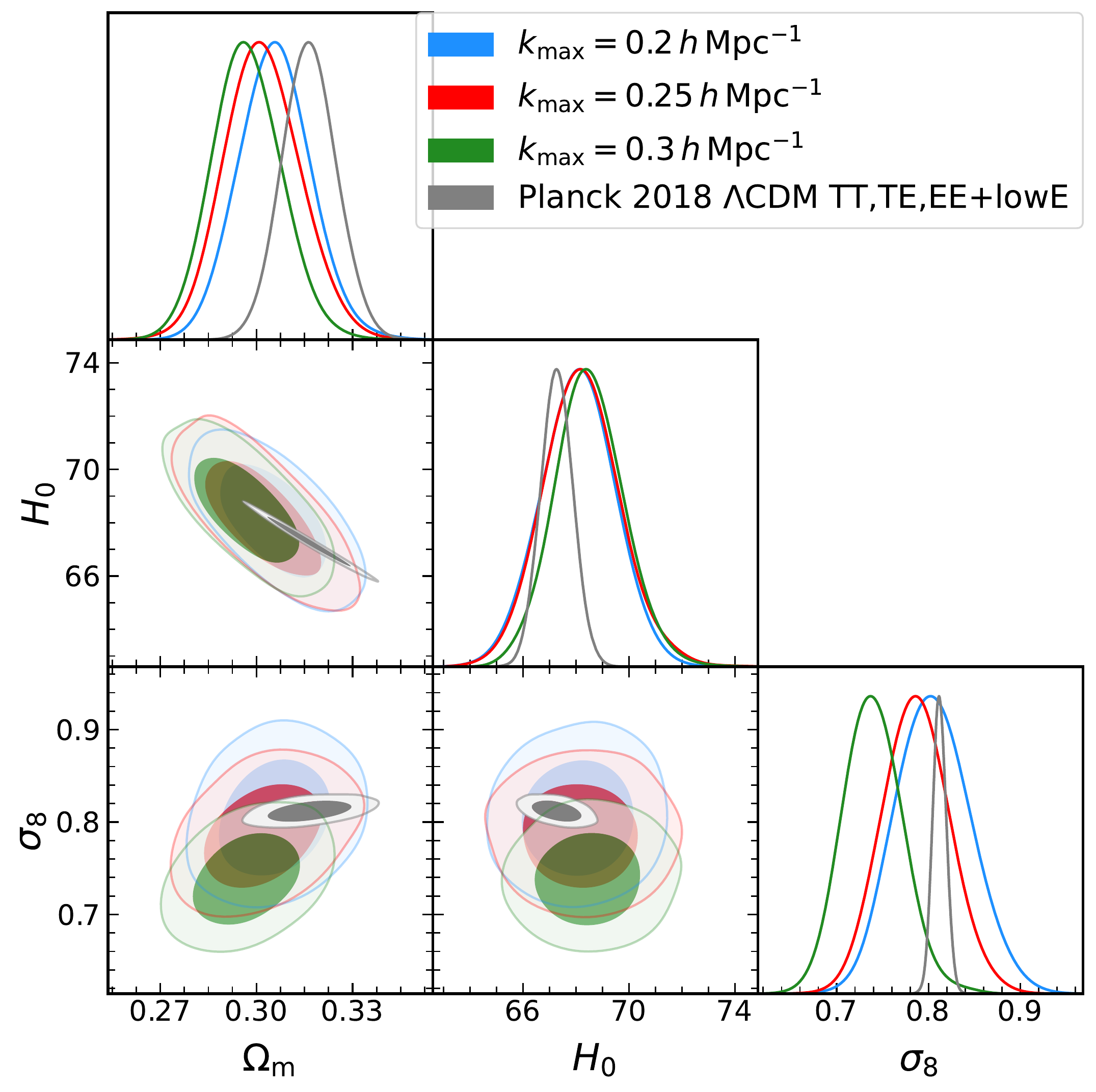}
\caption{
Comparison of the posterior distributions obtained when including the redshift-space power spectrum information up to a different $k_{\rm max}$ in the analysis. 
We show the cases of $k_\mathrm{max} = 0.2$ (blue), 0.25 (red), and 0.3 (green) $h\,\mathrm{Mpc}^{-1}$ and also show the results of the \textit{Planck} 2018 $\Lambda$CDM ``TT, TE, EE+lowE'' (gray). 
The cases of $k_\mathrm{max} = 0.25\,h\,\mathrm{Mpc}^{-1}$ and the \textit{Planck} are identical to those in Figure~\ref{fig:real_z02_NS_p024_bbn-planck-gauss-fsigv-Pshot_kmax0.25}.
}
\label{fig:real_z02_NS_p024_bbn-planck-gauss-fsigv-Pshot_kmax0.2-0.3}
\end{figure}
\begin{figure}
\centering
    \includegraphics[width=1.05\columnwidth]{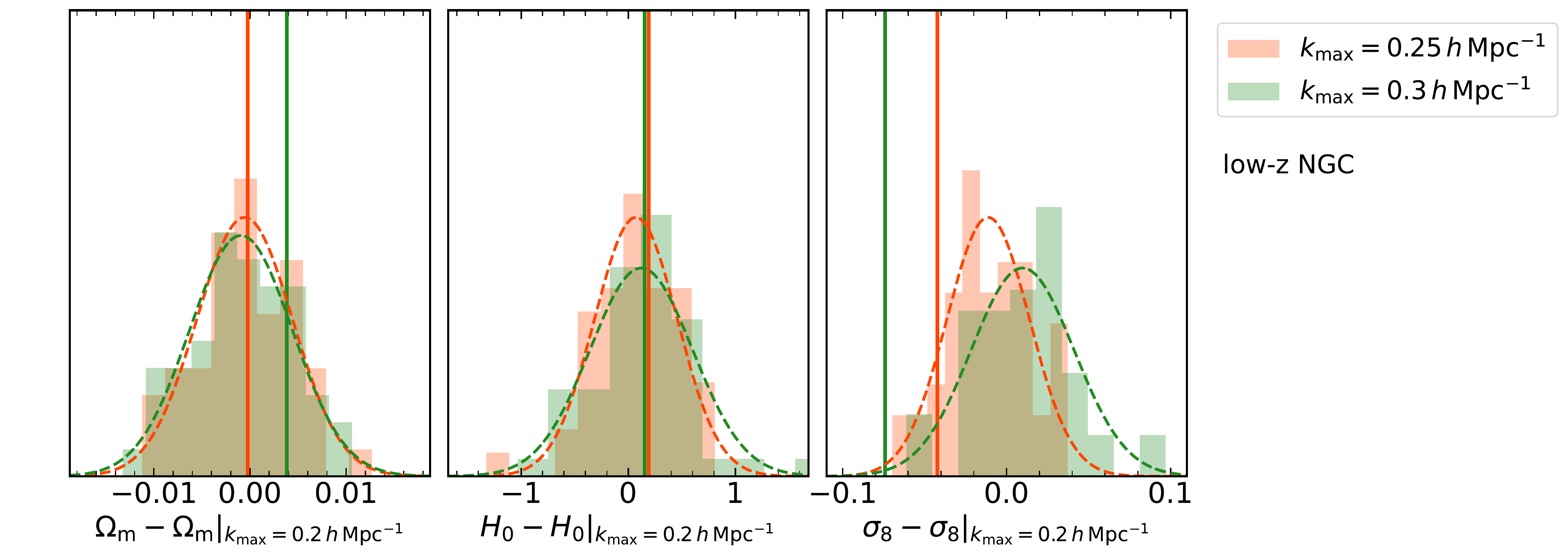}
    \includegraphics[width=1.05\columnwidth]{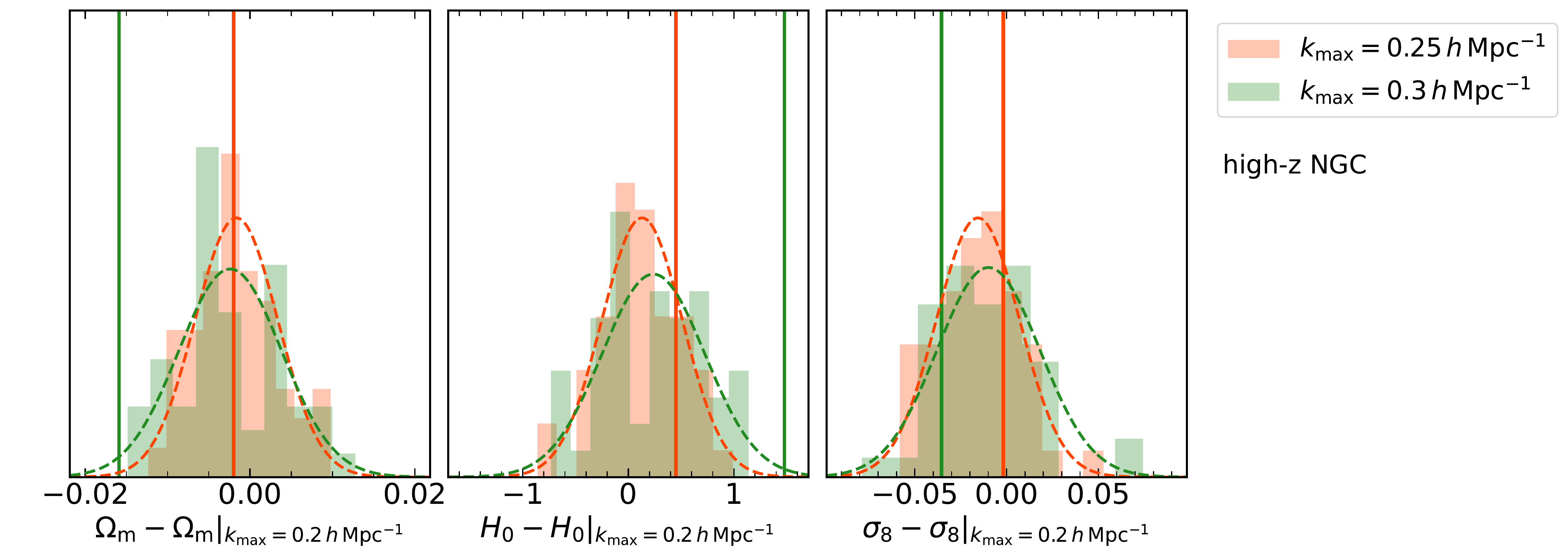}
\caption{
The distribution of scatters in each cosmological parameter, obtained from the cosmology analysis of noisy mock power spectra for the low-$z$ and high-$z$ NGC samples, using the different $k_\mathrm{max}$ cuts; 
the $x$-axis shows a shift in the \ykrv{mode value in the 1d posterior} of each parameter at $k_\mathrm{max}=0.25$ or 0.3\,$h\,\mathrm{Mpc}^{-1}$, compared to that at 
$k_\mathrm{max}=0.2\,h\,\mathrm{Mpc}^{-1}$. 
We used the 50 realizations of the noisy mock spectra, and the histogram in each panel displays the distribution of parameter shifts. 
The vertical red and green lines denote the shift found from the cosmology analyses of the real BOSS data, shown in Figure~\ref{fig:real_z02_NS_p024_bbn-planck-gauss-fsigv-Pshot_kmax0.2-0.3}.
\ykrv{The dashed curves denote the Gaussian distributions specified by the mean and variance of the parameter shifts among the 50 realizations.}
}
\label{fig:noisy_mock_hist}
\end{figure}
One advantage of our approach, \textit{e.g.}, compared to the \ykrv{PT-based} method such as the EFTofLSS, is that ours enables to compare the model predictions with the measurements up to higher $k_\mathrm{max}$. 
As described in our previous paper \cite{Kobayashi_2020}, our emulator is designed to give an accurate prediction of the redshift-space power spectrum up to $k=0.6\,h\,\mathrm{Mpc}^{-1}$.
However the power spectrum measurements at $k \gtrsim 0.2\,h\,\mathrm{Mpc}^{-1}$ are in the shot noise dominated regime, and therefore it is not clear whether the cosmological parameters are improved even if we include the data points at higher wavenumbers. 
In Figure~\ref{fig:real_z02_NS_p024_bbn-planck-gauss-fsigv-Pshot_kmax0.2-0.3} we study how the cosmological parameters are changed for different choices of $k_\mathrm{max}$; 
we consider $k_\mathrm{max}$ $=0.2$, 0.25 or $0.3\,h\,\mathrm{Mpc}^{-1}$, respectively.
The figure shows that the size of the credible intervals is not largely changed for the different $k_\mathrm{max}$ values, confirming the shot noise domination in the power spectrum measurements at $k \gtrsim 0.2\,h\,\mathrm{Mpc}^{-1}$. 
We confirmed that the statistical errors of the galaxy-halo connection parameters such as the residual shot noise and the HOD parameters are indeed improved by including the information on the higher $k$.
A closer look reveals that the results for $k_\mathrm{max}=0.2$ and $0.25\,h\,\mathrm{Mpc}^{-1}$ are consistent with each other. 
However, the result for $k_\mathrm{max}=0.3\,h\,\mathrm{Mpc}^{-1}$ shows a sizable shift in $\sigma_8$. 
For the validation tests using the \textit{noiseless} mock signals in Figure~\ref{fig:mock_z02_NS_p024_kmax0.25_bbn-planck-gauss-fsigv-Pshot_kmax0.2-0.3}, we did not find this level of shift in the cosmological parameters. 

As a further test of this shift, we use 50 realizations of \textit{noisy} signals that are generated by adding random noise realizations drawn from the BOSS covariance matrix to the noiseless mock signals (the mock signals in the right panel of Figure~\ref{fig:mock_z02_NS_p024_kmax0.25_bbn-planck-gauss-fsigv-Pshot_kmax0.2-0.3}).
Then we applied the same cosmology analysis pipeline to each of the mock signals to estimate the cosmological parameters. Figure~\ref{fig:noisy_mock_hist} shows the distributions of shifts in the cosmological parameters at $k_\mathrm{max}=0.25$ or $0.30\,h\,\mathrm{Mpc}^{-1}$ with respect to those at $k_\mathrm{max}=0.20\,h\,\mathrm{Mpc}^{-1}$, for the low-$z$ and high-$z$ NGC samples which give the dominant contributions to the cosmological constraints of our full analysis. 
The figure shows that there is a reasonable chance to have the parameter shifts for $k_\mathrm{max}=0.25\,h\,\mathrm{Mpc}^{-1}$ seen from the actual BOSS data. 
However, the shifts in some parameters for $k_\mathrm{max}=0.3\,h\,\mathrm{Mpc}^{-1}$ are at the tail of the distribution of the mock results, indicating a possible hint in the systematic effects at $k \gtrsim 0.25\,h\,\mathrm{Mpc}^{-1}$, \textit{e.g.}, a limitation of the halo model approach at such high $k$ scales or a residual systematic error in the power spectrum data. 
Hence our fiducial choice of $k_\mathrm{max}=0.25\,h\,\mathrm{Mpc}^{-1}$ seems reasonable against possible systematic effects. 

In Appendix~\ref{sec:fibercollision_kmin} we also study possible effects of the fiber collision and the minimum wavenumber $k_\mathrm{min}$ on the cosmological results. 
Here $k_\mathrm{min}$ is the minimum wavenumber in that we include the power spectrum information over $k_\mathrm{min} \le k \le k_\mathrm{max}$ in the cosmology analysis. 
A brief summary is that these effects do not appear to cause any major systematic effect in our cosmological results.

\subsection{A further test of emulation accuracy}
\label{sec:theory_error}

\begin{figure}
\centering
\includegraphics[width=0.45\textwidth]{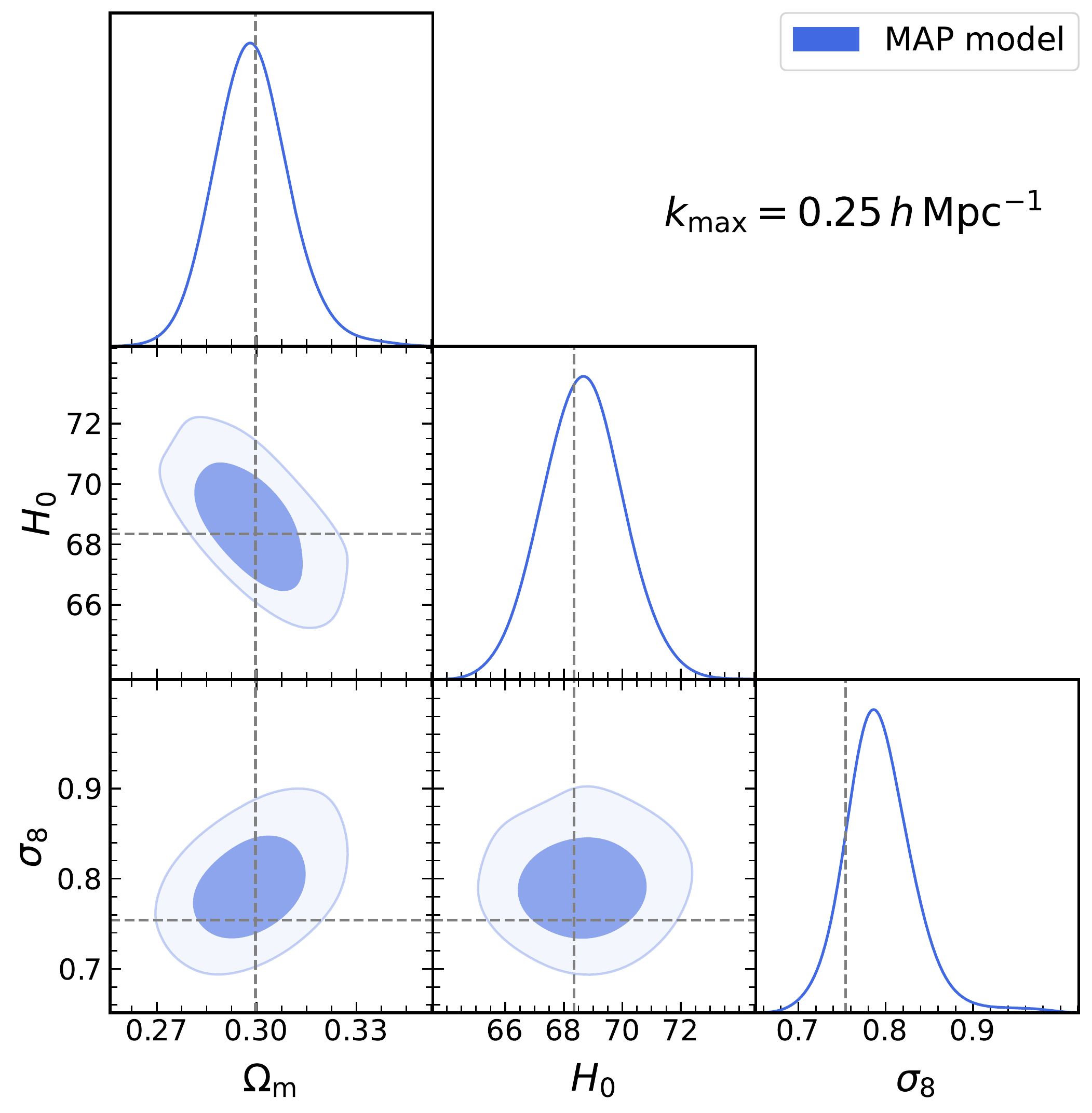}
\caption{
The posterior distributions of the cosmological parameters using the mock catalogs that are generated from the $N$-body simulations for \ykrv{a cosmological model at MAP} in Table~\ref{tab:constraint_result} (see text for details). 
\ykrv{The horizontal and vertical dashed lines show the input parameter values, {\it i.e.}, the cosmology at MAP.}
This is a similar test to Figure~\ref{fig:mock_z02_NS_p024_kmax0.25_bbn-planck-gauss-fsigv-Pshot_kmax0.2-0.3}, and the mock catalogs are generated using the same recipe of galaxy-halo connection as that in the left panel of Figure~\ref{fig:mock_z02_NS_p024_kmax0.25_bbn-planck-gauss-fsigv-Pshot_kmax0.2-0.3}.
}
\label{fig:real-map_z02_NS_p024_bbn-planck-gauss-fsigv-Pshot_kmax0.25}
\end{figure}

Now we turn to discussion on a possible uncertainty in the model predictions. 
As discussed in Refs.~\cite{Nishimichi_2019} and \cite{Kobayashi_2020}, our emulator for the redshift-space halo power spectrum is calibrated using a dataset of $N$-body simulations for the 101 flat $w$CDM models, where we used 15 realizations for the fiducial {\it Planck} cosmology and one realization for each of 100 models (more exactly among these we used the data for 80 models as training data). 
Here the 100 models are sampled using the optimal maximum-distance sliced Latin hypercube design in the 6-dimensional parameter space of $w$CDM cosmology \citep[see][for details]{Nishimichi_2019}. 
One might think that our emulator has a better accuracy around the \textit{Planck} cosmology, which is different from the MAP model or the model preferred by our cosmology analysis of the BOSS power spectra. 
\ykrv{To test this possible uncertainty, we ran a new set of $N$-body simulations for a model at the MAP cosmology in Table~\ref{tab:constraint_result}: 
the model with $\Omega_\mathrm{m}=0.300, H_0=68.3\,{\rm km~s}^{-1}{\rm Mpc}^{-1}$, $\sigma_8=0.754$, and other cosmological parameters are set to the values at MAP}.
More exactly, we ran each $N$-body simulation with box side length of 2.5$\,h^{-1}$Gpc and $3000^3$ particles, and use 5 realizations. 
The total volume is about 78$\,(h^{-1}\mathrm{Gpc})^3$, much larger than the BOSS volume ($\sim 5.7\,(h^{-1}\mathrm{Gpc})^3$), so we can sufficiently reduce the sample variance effect in the simulated power spectra. 
Then we populate galaxies into halos using the same recipe of the galaxy-halo connection used in the Cosmology Challenge paper \citep{2020arXiv200308277N}, which is different from our fiducial HOD method. 
Using the same covariance matrix as we used in the actual cosmology analysis of BOSS data, we perform the same cosmology analysis to the simulated power spectra. 
This is very similar to the validation test of our method using the simulated power spectrum used in the Cosmology Challenge \cite{2020arXiv200308277N}, but this new test gives a validation of our method \ykrv{at} the MAP model.
Also note that the new set of $N$-body simulations employs the fixed neutrino mass of $0.06\,\mathrm{eV}$ as in the simulations used for the emulator development, while the simulations in the Cosmology Challenges assume zero neutrino mass. 
Hence, this test also gives a confirmation on the subtle effect of the non-zero neutrino mass, which mainly affects the transfer function of matter density fluctuations.
Figure~\ref{fig:real-map_z02_NS_p024_bbn-planck-gauss-fsigv-Pshot_kmax0.25} shows that our method recovers each cosmological parameter accurately.

\begin{figure}
\centering
\includegraphics[width=0.45\textwidth]{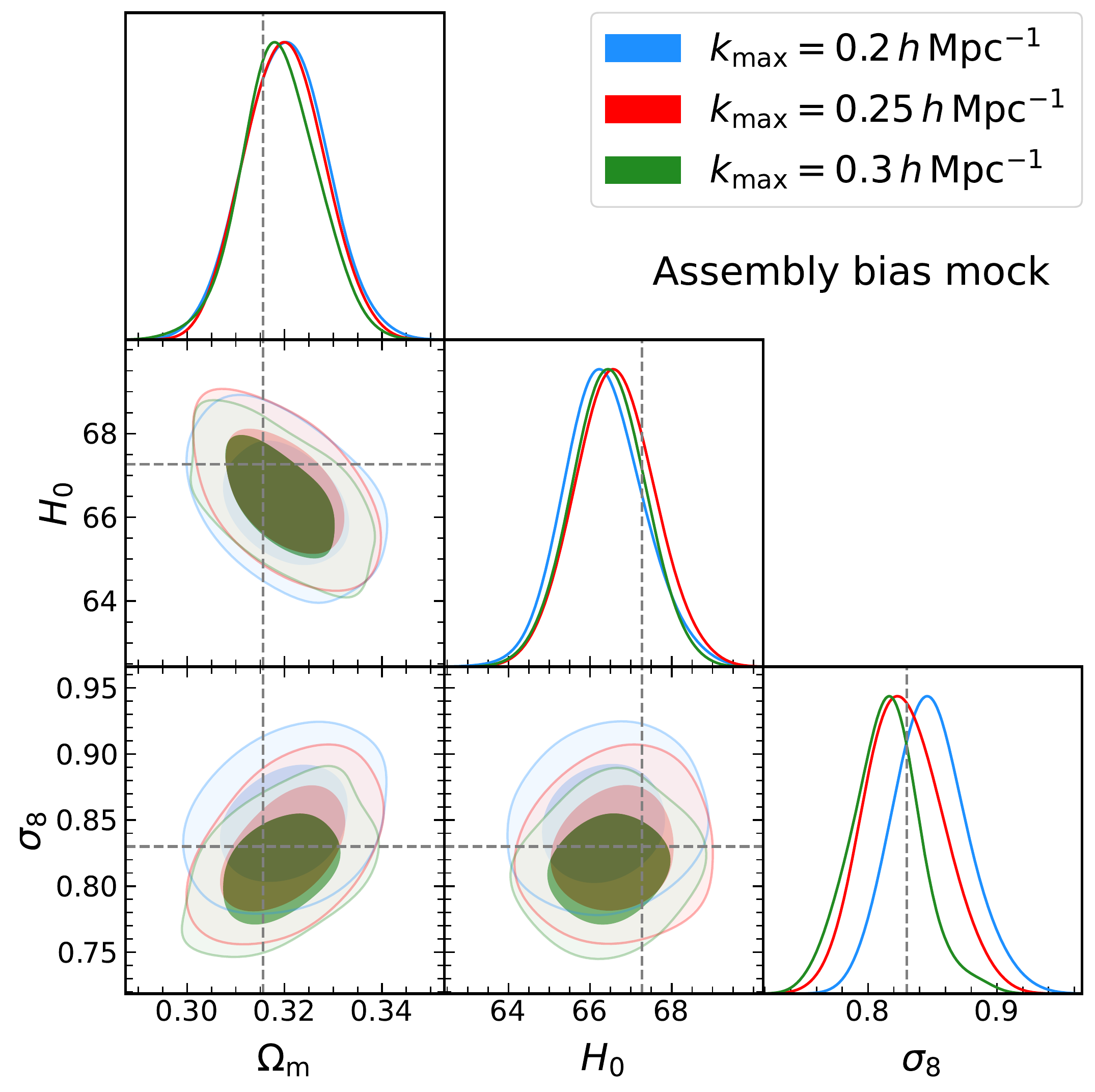}
\caption{
The posterior distributions of $\Omega_\mathrm{m}, H_0$, and $\sigma_8$, obtained when we apply our analysis pipeline to the mock power spectrum signals that are generated from the mock galaxy catalogs including the assembly bias effect (see text for the details). 
\ykrv{The horizontal and vertical dashed lines show the input parameter values.}
The mock galaxy catalogs have the same HOD shape as that of the mock catalogs in the right panel of Figure~\ref{fig:mock_z02_NS_p024_kmax0.25_bbn-planck-gauss-fsigv-Pshot_kmax0.2-0.3}, but were generated by populating mock galaxies preferentially into halos that have lower concentrations, in each halo mass bin. 
}
\label{fig:dr11_assembly-bias-sigma0.10_z02_NS_p024_bbn-planck-gauss-fsigv-Pshot_kmax0.2-0.3}
\end{figure}
\begin{figure*}
\centering
\includegraphics[width=0.9\textwidth]{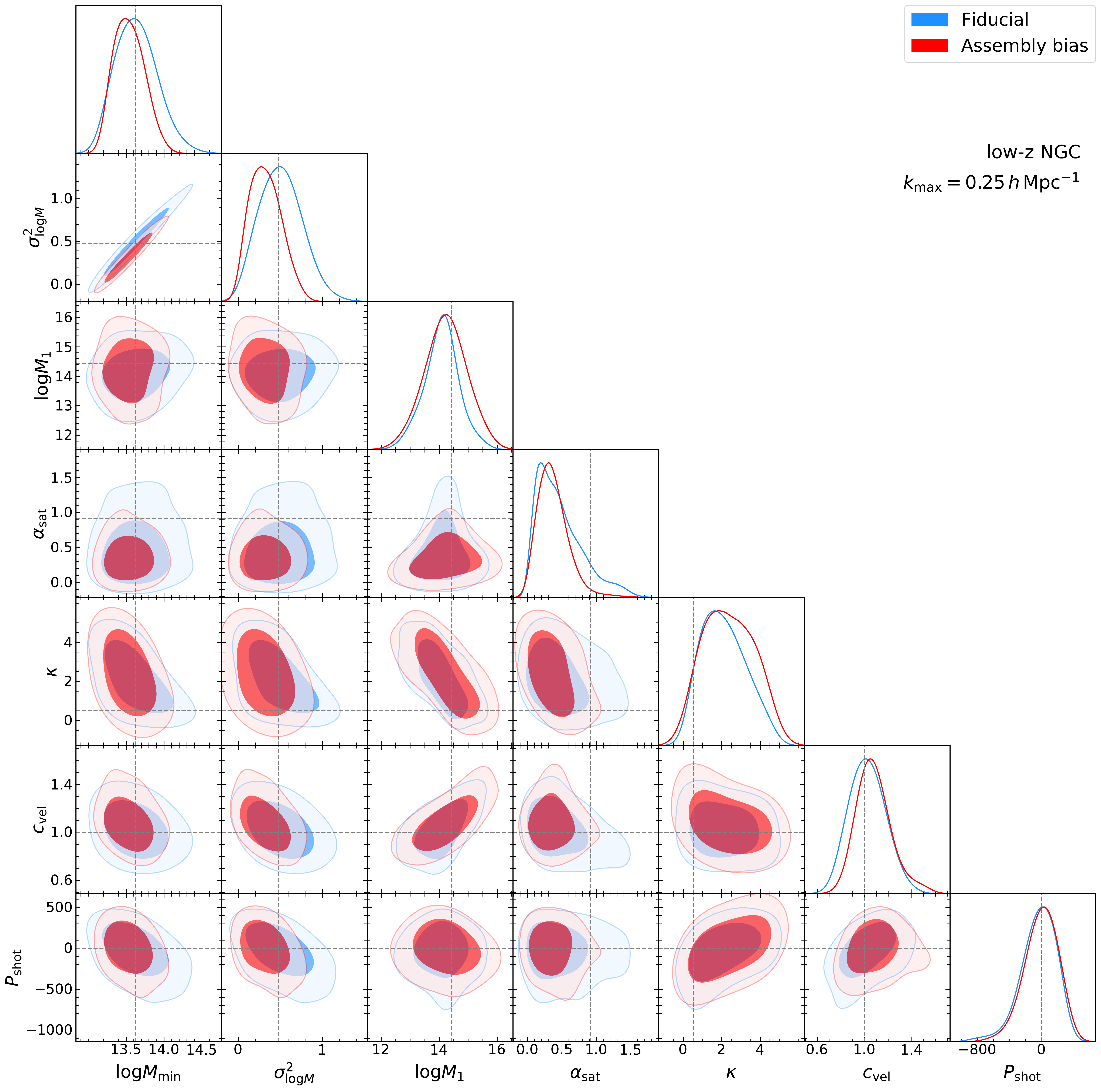}
\caption{
\ykrv{The red contours show the posterior distributions of the galaxy-halo conection parameters for the low-$z$ NGC sample, obtained for 
the same analysis using the assembly bias mock catalog as in Figure~\ref{fig:dr11_assembly-bias-sigma0.10_z02_NS_p024_bbn-planck-gauss-fsigv-Pshot_kmax0.2-0.3}.
For comparison, the blue contours show the posterior distributions obtained from the fiducial mock catalog, which are from the same analysis in the right panel of Figure~\ref{fig:mock_z02_NS_p024_kmax0.25_bbn-planck-gauss-fsigv-Pshot_kmax0.2-0.3}.
The two mock catalogs are constructed using the same galaxy-halo connection model, and the horizontal and vertical dashed lines show the input parameter values.}
}
\label{fig:assemblybias_hod_posterior}
\end{figure*}
\begin{figure}
\centering
\includegraphics[width=0.48\textwidth]{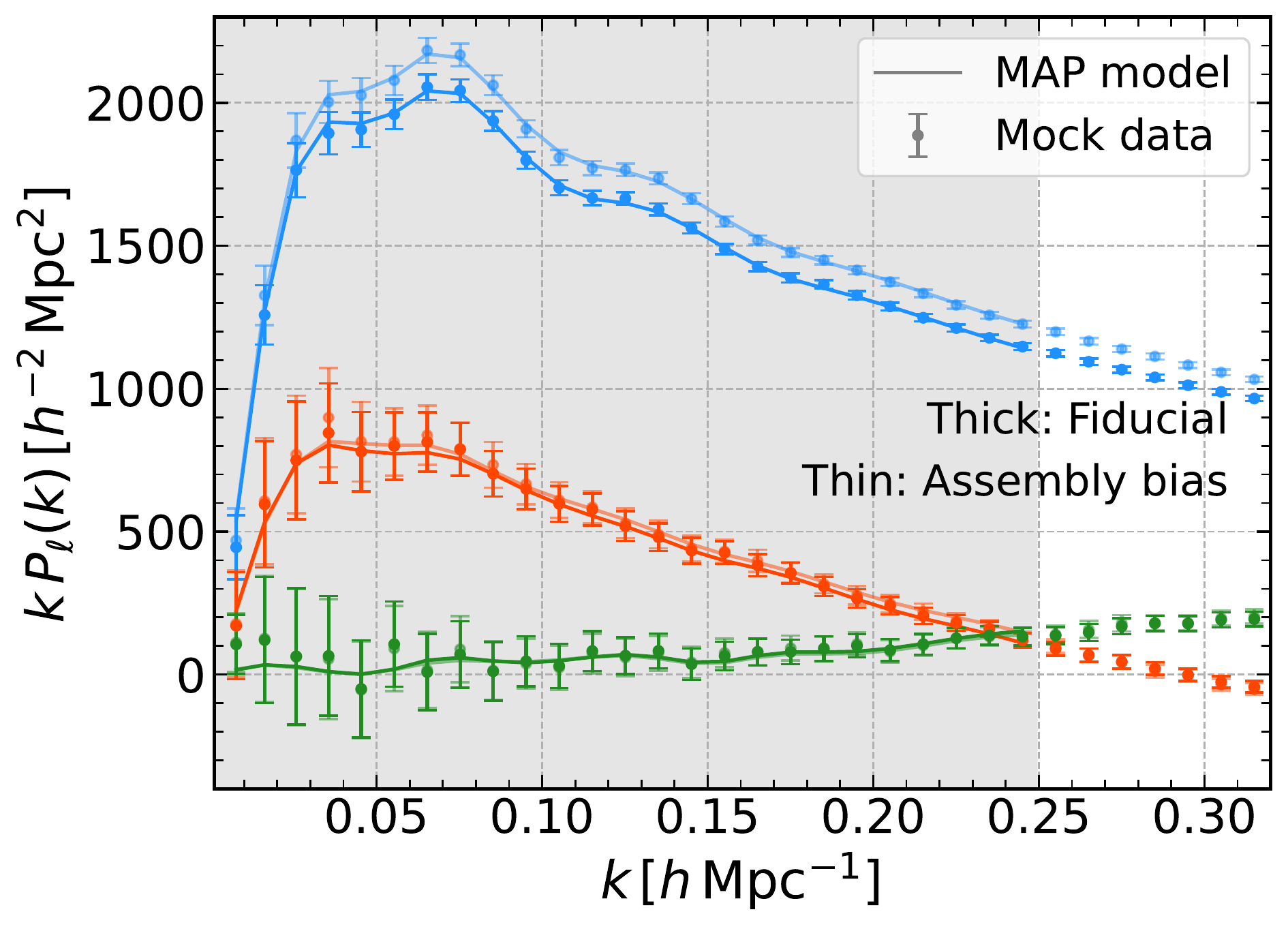}
\caption{
\ykrv{Comparison of the multipole moments between the measurements from the  mock catalog 
and the model predictions at MAP that are from the same analysis in Figure~\ref{fig:assemblybias_hod_posterior}. 
The gray shaded region is the $k$ range on which we use the data in the analysis, \textit{i.e.}, $k_\mathrm{max} = 0.25\,h\,\mathrm{Mpc}^{-1}$.}
}
\label{fig:assemblybias_map_mock}
\end{figure}

\subsection{Assembly bias}
\label{sec:assembly_bias}

Another concern of the halo model approach is the impact of the ``assembly bias'' effect; although the simple halo model assumes that the clustering amplitudes of halos (and galaxies) are determined by halo masses, they might depend on a secondary parameter, depending on the assembly history of halos/galaxies \citep{2006ApJ...652...71W,2008ApJ...687...12D}. 
Even in the presence of the assembly bias, the redshift-space distortion effect due to the peculiar velocities are unlikely to be affected by the assembly bias, because the peculiar velocities arise directly from the gravitational field \citep{PhysRevD.101.023510}. 
To test a possible effect of the assembly bias on cosmology analysis, we use the mock catalogs of galaxies including the assembly bias effect in Ref.~\cite{PhysRevD.101.023510}, where galaxies are populated preferentially into halos that have lower concentrations as a proxy of the assembly history \citep[see][for details]{PhysRevD.101.023510}. 
The mock catalogs including the assembly bias effect are generated from the same $N$-body simulations as those used in the right panel of Figure~\ref{fig:mock_z02_NS_p024_kmax0.25_bbn-planck-gauss-fsigv-Pshot_kmax0.2-0.3}; therefore the total volume is about $128\,(h^{-1}\,{\rm Gpc})^3$. 
The mock galaxies have about 30\% higher amplitudes in the real-space correlation function at large scales, compared to that of the mock galaxies without the assembly bias effect, which otherwise have the same HOD (see Figure~\ref{fig:assemblybias_map_mock}). 
We generated the mock signals for each of the BOSS-like galaxy samples, and applied the same pipeline of cosmology analysis to the mock signals.
Note that the assembly bias has not been detected with high significance from the BOSS galaxies \citep[\textit{e.g.}][]{2016ApJ...819..119L}. 

Figure~\ref{fig:dr11_assembly-bias-sigma0.10_z02_NS_p024_bbn-planck-gauss-fsigv-Pshot_kmax0.2-0.3} shows the cosmological parameters obtained from the mock catalogs including the assembly bias effect.
It can be found that the assembly bias does not cause a significant bias in the inferred cosmological parameters, $\Omega_\mathrm{m}, H_0$ and $\sigma_8$.
This is consistent with the Fisher forecast in the previous work \cite{PhysRevD.101.023510}, confirming that the BAO and RSD information are not affected by the assembly bias effect, even after marginalization over the galaxy-halo connection parameters. 
In other words, the cosmology analysis using the redshift-space power spectrum does not rely on the halo mass estimate.
This result is contrasted with that in Ref.~\cite{2021arXiv210100113M}: they found that the assembly bias causes a significant bias in the cosmological parameters, especially $\sigma_8$ and $\Omega_{\rm m}$, if a hypothetical joint-probe cosmology analysis using the galaxy-galaxy weak lensing and the projected correlation function is applied to 
the mock signals including the assembly bias effect.
In such a method, the galaxy-galaxy weak lensing plays an important role to constrain the average mass of host halos, which in turn helps determine the galaxy bias uncertainties to obtain the clustering amplitude of matter at large scales. 
However, the assembly bias disturbs the scaling relation of the large-scale bias amplitude with halo mass, and in turn leads to biases in the inferred cosmological parameters. 
On the other hand, our method using the redshift-space power spectrum does not rely on the halo mass estimate.
\ykrv{As can be found from Figure~\ref{fig:assemblybias_hod_posterior}, some of the galaxy-halo connection parameters are changed between the results for the fiducial and assembly bias mocks, suggesting that the assembly bias effect are absorbed by changes in the galaxy-halo connection parameters in our method. 
For example, Figure~\ref{fig:assemblybias_hod_posterior} shows the model which has a slightly smaller $\sigma_{\log M}$ than the input value is favored for the assembly bias mock, which leads to a larger bias amplitude on large scales as explicitly 
shown in Figure~\ref{fig:assemblybias_map_mock}. The figure clearly shows that the model well reproduces the mock measurements, especially the assembly bias effect that can be seen from the higher amplitude of the monopole moment
\citep[also see Figure~16 in Ref.][for similar discussion]{PhysRevD.101.023510}.}
Hence we conclude that our cosmological results for the BOSS galaxies are unlikely affected by the assembly bias, although the impact of more general assembly bias effects on cosmology inference with the redshift-space power spectrum further needs to be studied, \textit{e.g.}, using the results in cosmological hydrodynamical simulations \citep[\textit{e.g.}][]{2021MNRAS.501.1603H}.

\subsection{Impact of uncertainty in the galaxy-halo connection}
\label{sec:fixed_HOD}

Finally we comment on the impact of the galaxy-halo connection parameters on the cosmological parameters.
We have so far employed broad priors of these nuisance parameters.
Other observables such as galaxy-galaxy weak lensing \citep[][]{2015ApJ...806....2M,paszke2019pytorch} can be used to infer the HOD parameters. 
Throughout this paper, we have employed a fairly broad prior range for each of 7 galaxy-halo connection parameters. 
In this subsection, we study how the cosmological parameter estimation can be improved if we have some knowledge on the HOD parameters.

\begin{figure}
\center
\includegraphics[width=\columnwidth]{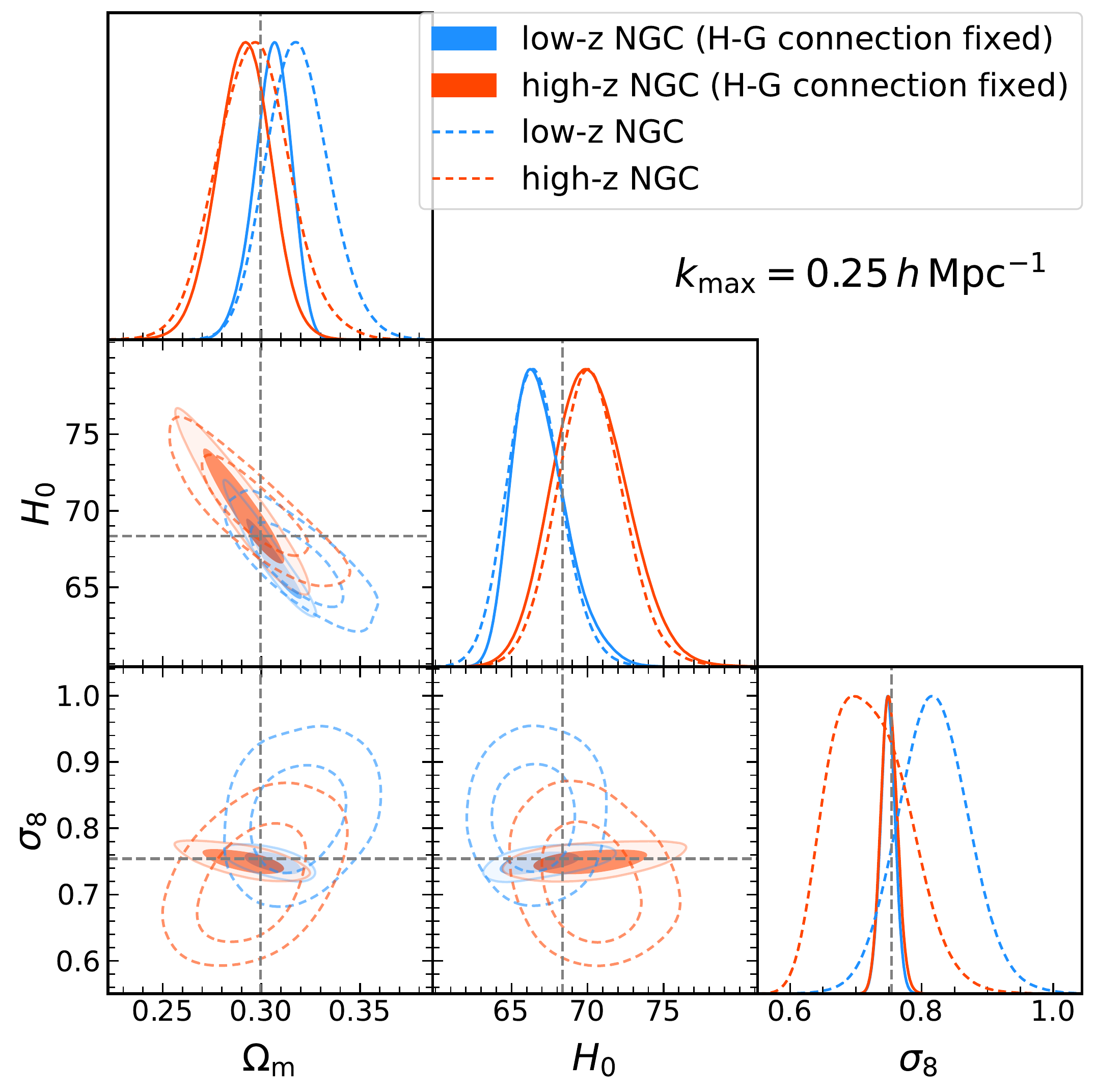}
\caption{
The posterior distributions of the cosmological parameters, obtained from the cosmology analysis of the real BOSS data for either of the low-$z$ or high-$z$ NGC sample, when fixing the 7 galaxy-halo connection parameters to their values at MAP in Figure~\ref{fig:real_p024_kmax0.25_bbn-planck-gauss-fsigv-Pshot_each-chunk}. 
The dashed-line contours and 1d posterior distribution are the same as in Figure~\ref{fig:real_p024_kmax0.25_bbn-planck-gauss-fsigv-Pshot_each-chunk}.}
\label{fig:impact_hod}
\end{figure}
As a working example, in Figure~\ref{fig:impact_hod} we show the posterior distributions of the cosmological parameters obtained from the BOSS data by fixing the galaxy-halo connection parameters to their values at MAP in our fiducial analysis (the results for Figure~\ref{fig:real_z02_NS_p024_bbn-planck-gauss-fsigv-Pshot_kmax0.25}). 
Therefore, this can be regarded as the best-case scenario where the galaxy-halo connection of the galaxy sample is perfectly known. 
The figure shows that the cosmological parameters are significantly improved.
It is interesting to observe that the level of improvement of the parameter constraint varies with the parameters. 
We can see that $\Omega_\mathrm{m}$ and $H_0$, which are mostly determined by the geometrical information through the BAO feature and also partly from the spectral shape information, are not significantly improved as compared to $\sigma_8$, which is an amplitude-related parameter. 
This is telling that the overall amplitude of the cosmological fluctuations is the hardest to interpret from observations due to the strong degeneracy with the galaxy bias uncertainty or the uncertainty in galaxy-halo connection. 

\section{Conclusion}
\label{sec:conclusion}

In this paper, we used an {\it emulator} of halo clustering statistics to estimate the cosmological parameters from the full-shape analysis of the redshift-space power spectra measured from the BOSS DR12 galaxy catalog over $0.2<z<0.75$.
Combining with the HOD model, the emulator allows us to compute the redshift-space power spectra of galaxies for a given cosmological model within the flat $\Lambda$CDM cosmology, in less than a CPU second. 
It enables the parameter inference of the BOSS spectra in a multi-dimensional parameter space (33 parameters in this study). 
We showed that the emulator model well reproduces the monopole, quadrupole and hexadecapole moments of the redshift-space power spectra simultaneously for all of the 4 galaxy subsamples. 
Our method yields stringent constraints on the cosmological parameters, $\Omega_\mathrm{m}$, $H_0$ and $\sigma_8$, even after marginalization over uncertainties in the nuisance parameters including the galaxy-halo connection parameters: 
more precisely, \ykrv{
$\Omega_\mathrm{m}=0.301^{+0.012}_{-0.011}$, $H_0=68.2 \pm 1.4~\mathrm{km\,s}^{-1}\mathrm{Mpc}^{-1}$, and $\sigma_8=0.786^{+0.036}_{-0.037}$
}, for the mode and the 68\% credible interval of the 1d posterior distribution (Figure~\ref{fig:real_z02_NS_p024_bbn-planck-gauss-fsigv-Pshot_kmax0.25} and Table~\ref{tab:constraint_result}).
The cosmological parameters we obtained are consistent with those from the independent studies for the same BOSS spectra using \ykrv{PT-based models as the theoretical template \citep{philcox2022boss,Chen_2022}}, even though the theoretical templates are totally different.
This shows the robustness of the redshift-space power spectrum method for estimating the cosmological parameters: the BAO, AP effects and RSD information can be robustly extracted as long as the uncertainties in the nonlinear effects and galaxy properties are marginalized over.
The statistical precision for each of the main parameters, $\Omega_\mathrm{m}$, $H_0$ and $\sigma_8$ is also comparable with that from the \ykrv{PT-based} methods.
One might think that the halo model could allow us to use the redshift-space power spectra down to a larger $k$ and therefore lead to a more stringent constraint on the parameters, compared to the \ykrv{PT-based} methods that treat all the galaxy bias parameters as nuisance parameters. 
However, this is not the case.
We think there are a few reasons for this result.
First, we employed quite broad priors for each of the HOD parameters and did not include any information on the abundance of galaxies.
Hence our analysis might be considered a conservative approach. 
Second, the power spectrum information at $k \gtrsim 0.2\,h\,\mathrm{Mpc}^{-1}$ is in the shot noise dominated regime, and the cosmological parameters are not improved even if including the power spectrum information on $k \gtrsim 0.2\,h\,\mathrm{Mpc}^{-1}$ (Figure~\ref{fig:real_z02_NS_p024_bbn-planck-gauss-fsigv-Pshot_kmax0.2-0.3}). 
Our results are also in good agreement with the {\it Planck} 2018 results \citep{planck_collaboration_2020} for $\Omega_\mathrm{m}$ and $H_0$, but indicate a slight tension for $\sigma_8$ similarly to those reported by the weak lensing analyses \citep{2019PASJ...71...43H,2018PhRvD..98d3526A,2021A&A...646A.140H}.

There is a promising route to improving our cosmological constraints. 
It is a joint-probes cosmology: although we use the redshift-space power spectrum as the data vector in this paper, there is another observable available in the BOSS footprint. 
The promising one is the galaxy-galaxy weak lensing, which can be obtained by cross-correlating the positions of BOSS galaxies with shapes of background galaxies, where the background galaxies can be taken from an imaging survey overlapping with some portion of the BOSS footprint. 
For this, the Subaru HSC and some parts of the DES and KiDS surveys have an overlapping region with the BOSS footprints, and the galaxy-galaxy weak lensing can be measured from the joint analysis of BOSS and these imaging surveys. 
Since the galaxy-galaxy weak lensing can measure the average mass distribution around the BOSS galaxies, it helps observationally disentangle the galaxy bias uncertainty or yields stringent constraints on the HOD parameters in the halo model picture. 
In fact, Ref.~\cite{2021arXiv211102419M} combined the galaxy-galaxy weak lensing, measured from the HSC data of small area coverage 140\,sq. degrees, with the {\it projected} correlation function of BOSS galaxies in the cosmology analysis and then obtained the accuracy of $\sigma(S_8) \simeq 0.05$, compared to our constraint of $\sigma(S_8) \sim 0.04$ although $S_8$ is not a best parameter to which the redshift-space power spectrum is sensitive.
The galaxy-galaxy weak lensing arises mainly from Fourier modes in the two-dimensional space perpendicular to the line-of-sight direction, and it was shown that it carries almost independent information from the redshift-space power spectrum that arises from Fourier modes in the three-dimensional space \cite{2013arXiv1308.6070D}. 
Hence, combining the redshift-space power spectrum with the galaxy-galaxy weak lensing helps disentangle the degeneracy between the galaxy bias uncertainty (the galaxy-halo connection) and the cosmological parameters, leading to an improved estimation of the cosmological parameters (see Figure~\ref{fig:impact_hod} for a possible improvement in the best-case scenario).
To do this, our emulator based method easily enables to jointly combine the two observables in the same halo model framework for parameter inference. 
This is definitely an interesting direction, and will be our future work.

\acknowledgments
We thank 
Mikhail~Ivanov, 
Oliver~Philcox,
Shun~Saito, 
Masato~Shirasaki,
Marko~Simonovi\'c,
Naonori~S.~Sugiyama, 
Sunao~Sugiyama,
Ryuichi~Takahashi,
Digvijay~Wadekar,
and
Matias~Zaldarriaga
for useful discussions.
\ykrv{After submission of this paper to arXiv, we noticed the full-shape cosmology analysis of BOSS power spectrum  using the EFTofLSS method in Ref.~\cite{2021arXiv211005530C}. The authors, Stephen~Chen and Martin~White, kindly informed us of the updated measurements of BOSS power spectrum done in Ref.~\cite{Beutler_McDonald_2021}. We would like to thank them for useful discussion.}
Y.K. also thanks to the Yukawa Institute for Theoretical Physics, Kyoto University for the warm hospitality where this work was partly done.
This work was supported in part by World Premier International Research Center Initiative, MEXT, Japan, and JSPS KAKENHI Grant No.~JP15H03654,
JP15H05887, JP15H05893, JP15K21733, JP17H01131, JP17K14273, JP19H00677, JP20H01932, JP20H04723, JP20H05850, JP20H05861, JP20H05855, and JP21H01081, by Japan Science and Technology Agency (JST) CREST JPMHCR1414, by JST AIP Acceleration Research Grant No.~JP20317829, Japan, and by Basic Research Grant (Super AI) of Institute for AI and Beyond of the University of Tokyo.
Y.K. was also supported by the Advanced Leading Graduate Course for Photon Science at the University of Tokyo.
The $N$-body simulations and subsequent halo-catalog creation in the \textsc{Dark Quest} simulation suite used in this work were carried out on Cray XC50 at Center for Computational Astrophysics, National Astronomical Observatory of Japan. 

\appendix

\section{Posterior distributions of the galaxy-halo connection parameters in the validation test}
\label{sec:hod_parameters_mockchallenges}

\begin{figure*}
\centering
\includegraphics[width=0.9\textwidth]{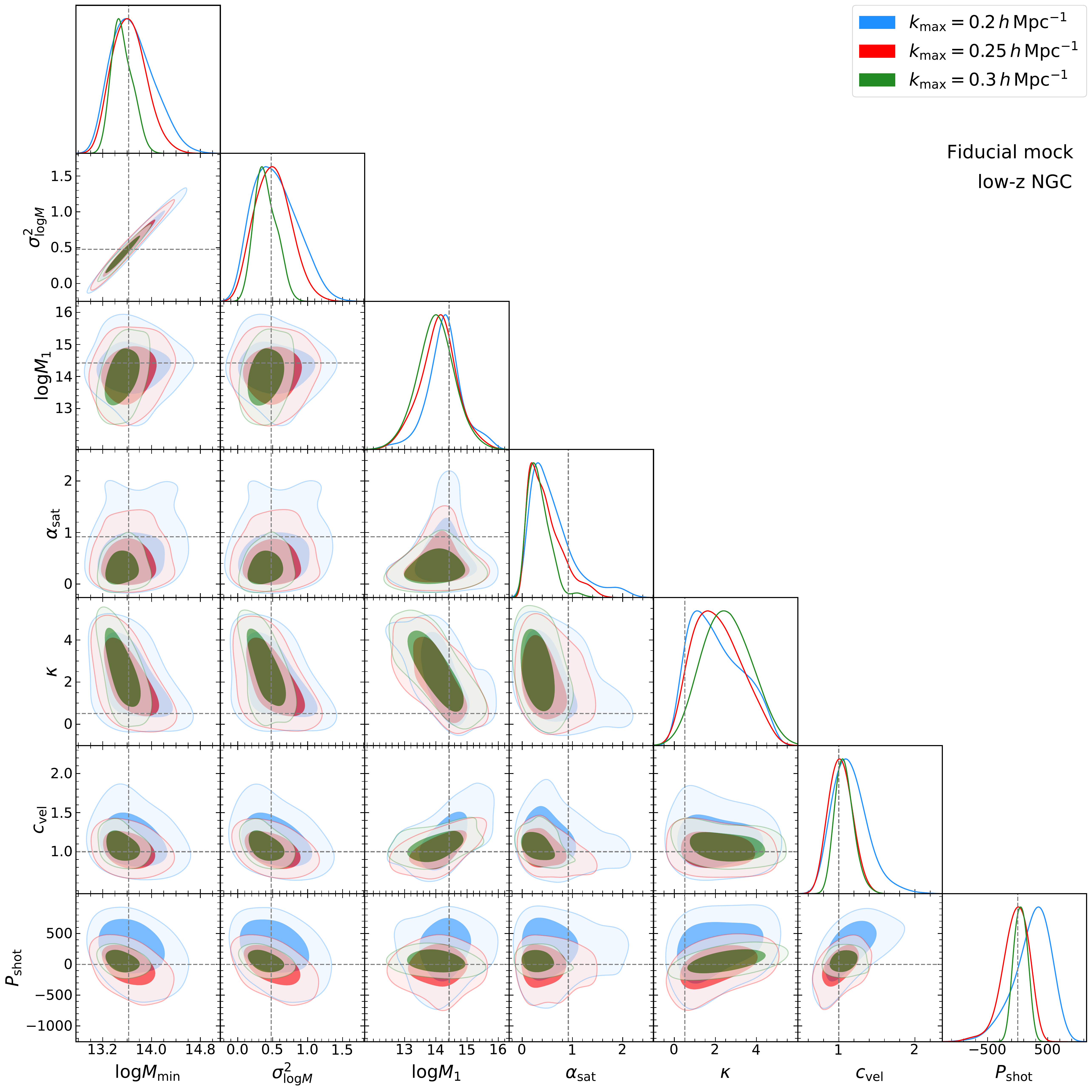}
\caption{
The posterior distributions of the galaxy-halo connection parameters, for the low-$z$ NGC sample as an example, obtained from the combined analysis using the mock catalogs that are generated using the same HOD model as in our theoretical template (the right panel of Figure~\ref{fig:mock_z02_NS_p024_kmax0.25_bbn-planck-gauss-fsigv-Pshot_kmax0.2-0.3}). 
The dashed lines in each plot denote the input value of each parameter that is used when generating the mock galaxy catalogs.
}
\label{fig:posterior_galaxy-halo_parameters_mockchallenge}
\end{figure*}
Figure~\ref{fig:posterior_galaxy-halo_parameters_mockchallenge} 
show the posterior distributions of the galaxy-halo connection parameters, obtained from the cosmology analysis of the mock data vector that are generated from the mock catalogs of BOSS-like galaxies. 
The mock galaxy catalog is the same as that used in the right panel of Figure~\ref{fig:mock_z02_NS_p024_kmax0.25_bbn-planck-gauss-fsigv-Pshot_kmax0.2-0.3}. 
The figure shows that some of the galaxy-halo connection parameters are not necessarily well recovered, although the cosmological parameters are fairly well recovered as shown in
Figure~\ref{fig:mock_z02_NS_p024_kmax0.25_bbn-planck-gauss-fsigv-Pshot_kmax0.2-0.3}. 
The size of the posterior contours for the galaxy-halo connection parameters indeed shrinks with including the power spectrum information on the higher $k$.

\section{Posterior distributions in full parameter space}
\label{sec:posterior_full_parameters}

\begin{figure*}
\centering
\includegraphics[width=0.9\textwidth]{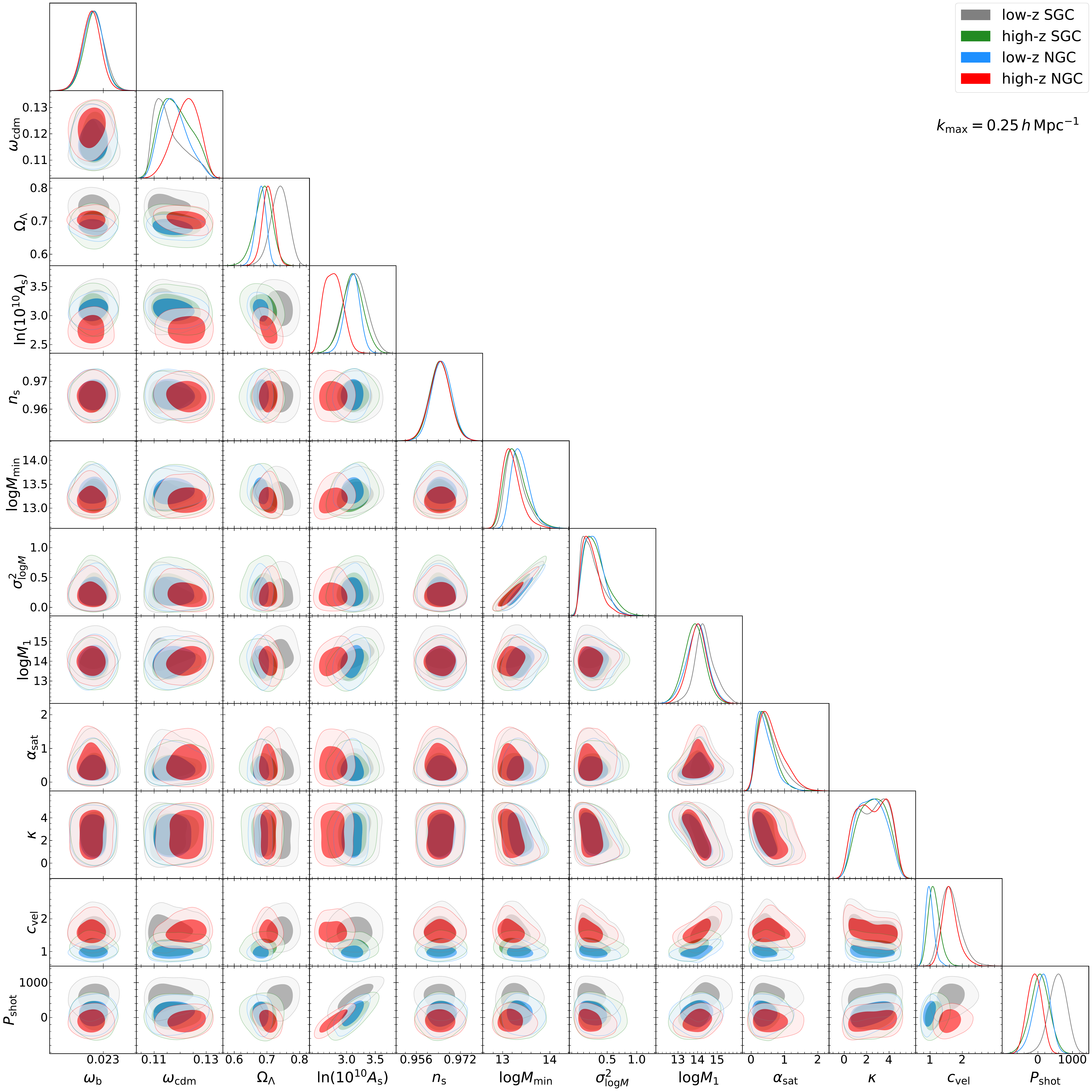}
\caption{
The posterior distributions in each 2d subspace of the full parameters for each of the 4 galaxy samples, for the cosmology analysis in Figure~\ref{fig:real_z02_NS_p024_bbn-planck-gauss-fsigv-Pshot_kmax0.25}: the low-$z$ NGC, low-$z$ SGC, high-$z$ NGC and high-$z$ SGC samples, respectively. 
The posterior distributions are from the joint parameter inference of these parameters (33 parameters in total as given in Table~\ref{tab:param_table}): 
5 cosmological parameters (while we impose tight Gaussian priors to $\omega_\mathrm{b}$ and $n_\mathrm{s}$) and each sample is characterized by 7 nuisance parameters (5 HOD parameters, the virial velocity parameter, and the residual shot noise parameter).
}
\label{fig:real_p024_kmax0.25_bbn-planck-gauss-fsigv-Pshot_each-chunk}
\end{figure*}

For comprehensiveness of our discussion, in Figure~\ref{fig:real_p024_kmax0.25_bbn-planck-gauss-fsigv-Pshot_each-chunk} we show the posterior distributions in each 2d subspace of the full parameters for the $\Lambda$CDM model for each galaxy sample, based on our baseline analysis setup (Table~\ref{tab:param_table}).

\section{Tests of the fiber collision effect and the 
minimum wavenumber $k_\mathrm{min}$}
\label{sec:fibercollision_kmin}

Next we investigate the impact of the fiber collision on the cosmological parameter inference from the redshift-space galaxy power spectrum. 
The fiber collision, which occurs due to the inability of the two adjacent optical fibers to be closer than some finite separation angle, is a potential systematic effect on the cosmological parameter inference.
The likelihood of the fiber collision depends on the number density of target galaxies on the celestial sphere, and it leads to an anisotropic effect in the measured power spectrum. 
Ref.~\cite{Hahn_2017} shows that the correction to the fiber collision effect on the galaxy power spectrum depends on the true power spectrum itself, and suggests the way for the correction. 
In this work, instead of implementing the correction in the model prediction, we examine the extent to which the fiber collisions affect the cosmology inference. 

\begin{figure}
\centering
\includegraphics[width=0.45\textwidth]{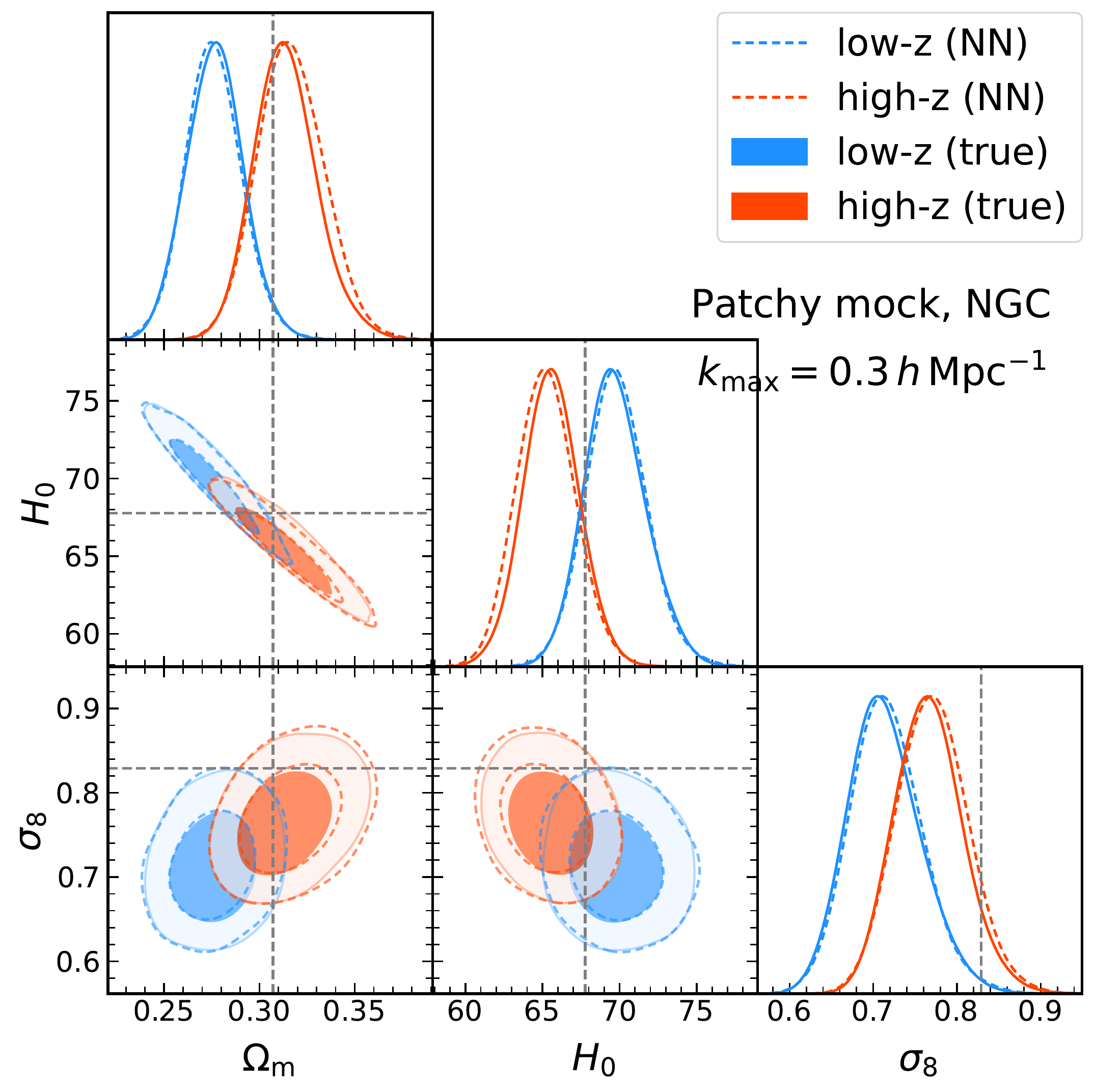}
\caption{
The comparison of the cosmological parameter inferences between the \textsc{Patchy} mock power spectra with and without the fiber collisions.
We show the cases of the low-z (blue) and high-z (red) NGCs and $k_\mathrm{max} = 0.3\,h\,\mathrm{Mpc}^{-1}$.
The empty dashed-line contours are the results for the \textsc{Patchy} mocks where we include the fiber collision weights.
The filled solid-line contours are those for the same mocks where we assume all of the fiber collision weights are unity, \textit{i.e.}, no fiber collisions.
The gray horizontal and vertical dashed lines indicate the input cosmological parameter values used in the simulations from which the \textsc{Patchy} mocks are created.
}
\label{fig:fiber_collision_patchy10_p024_bbn-planck-gauss-fsigv-Pshot_NGC}
\end{figure}
Figure~\ref{fig:fiber_collision_patchy10_p024_bbn-planck-gauss-fsigv-Pshot_NGC} shows a comparison of the parameter inference between the cases of with/without the fiber collisions. 
To investigate the influence of the fiber collisions on the cosmological parameter inference, we use the \textsc{Patchy} mocks which have the fiber collision weights. 
The fiber collision weights in the \textsc{Patchy} mocks are assigned following Ref.~\cite{Rodriguez-Torres_2016}, which reflects the nearest neighbor (NN) method.
In this figure we see that there are only marginal differences in the parameter posteriors between the case with the fiber collisions corrected by the NN weights (empty, dashed line contours) and that of the true power spectrum with no fiber collisions (filled, solid line contours), for both the low-$z$ and high-$z$ NGCs, up to $k_\mathrm{max} = 0.3\,h\,\mathrm{Mpc}^{-1}$. 

\begin{figure}
\centering
\includegraphics[width=0.45\textwidth]{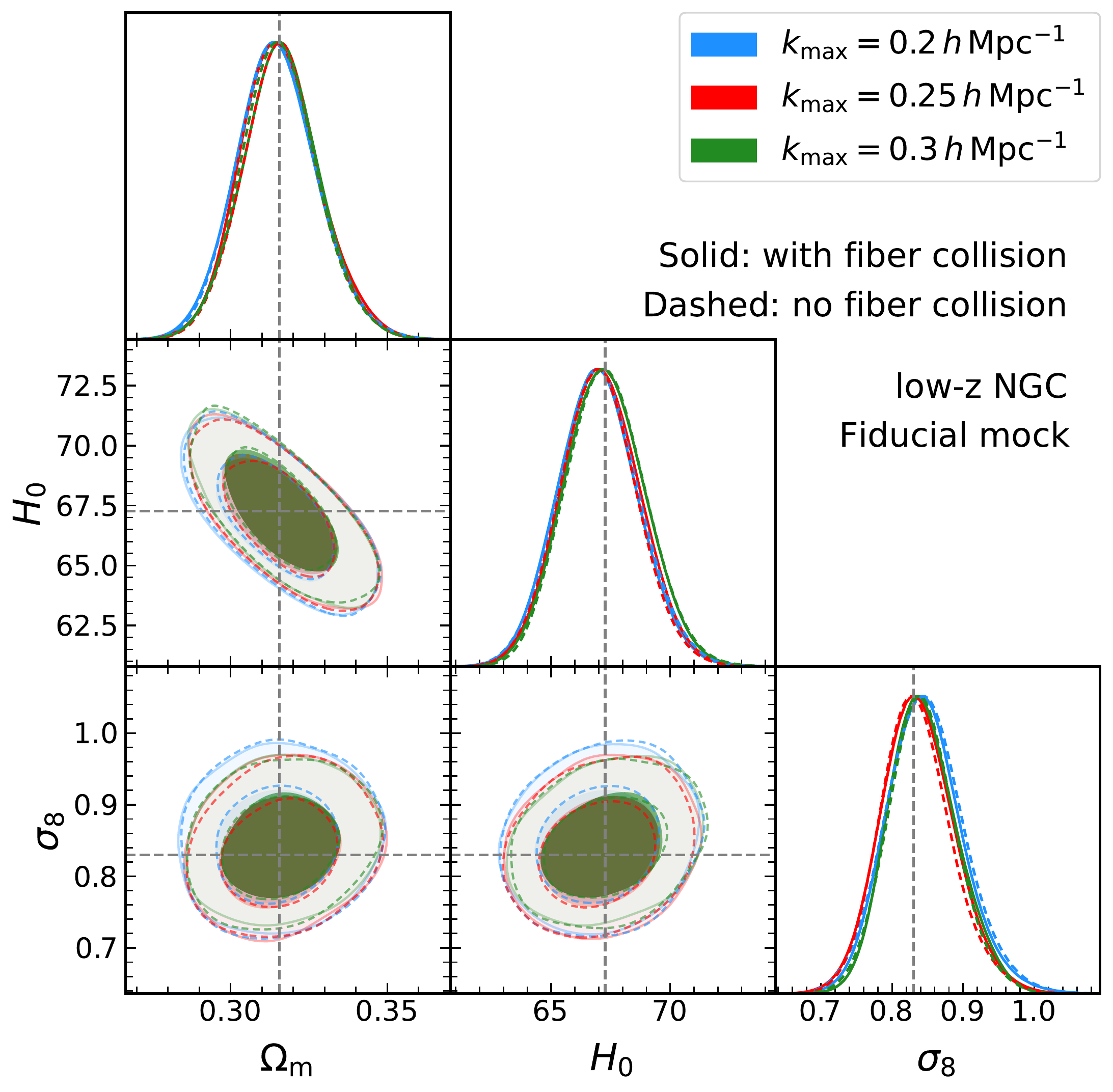}
\caption{
\ykrv{The comparison of the cosmological parameter inferences between the fiducial mock power spectrum with and without the fiber collision effect.
For mock signals with the fiber collisions, we add the fiber collision effect estimated by Ref.~\cite{Hahn_2017} (see their Figure 3) to the power spectrum monopole and quadrupole of the fiducial mock signals.}
}
\label{fig:fiducial_fcHahn17Fig3}
\end{figure}
\ykrv{
Figure~\ref{fig:fiducial_fcHahn17Fig3} shows another test based on mock power spectra. 
We quote the shifts of the power spectrum due to the fiber collisions \footnote{We obtain this shifts using WebPlotDigitizer (\url{https://automeris.io/WebPlotDigitizer/})} in the \textsc{Nseries} mocks shown in the left panel of Figure~3 of Ref.~\cite{Hahn_2017}, and add them to our fiducial mock signals after spline interpolation to reduce the noise contributions.
This mock fiber collision effect causes almost no change to the posterior distribution of the cosmological parameters, even in the case of $k_\mathrm{max} = 0.3\,h\,\mathrm{Mpc}^{-1}$.
From these studies, we conclude that the fiber collision effect has no significant impact on the cosmology analysis. 
Therefore, we conclude that the systematic shift in the $\sigma_8$ posterior in our real data analysis when we increase $k_\mathrm{max}$ (shown in Figure~\ref{fig:real_z02_NS_p024_bbn-planck-gauss-fsigv-Pshot_kmax0.2-0.3}) is not due to the fiber collisions.
}

\begin{figure}
\centering
\includegraphics[width=0.45\textwidth]{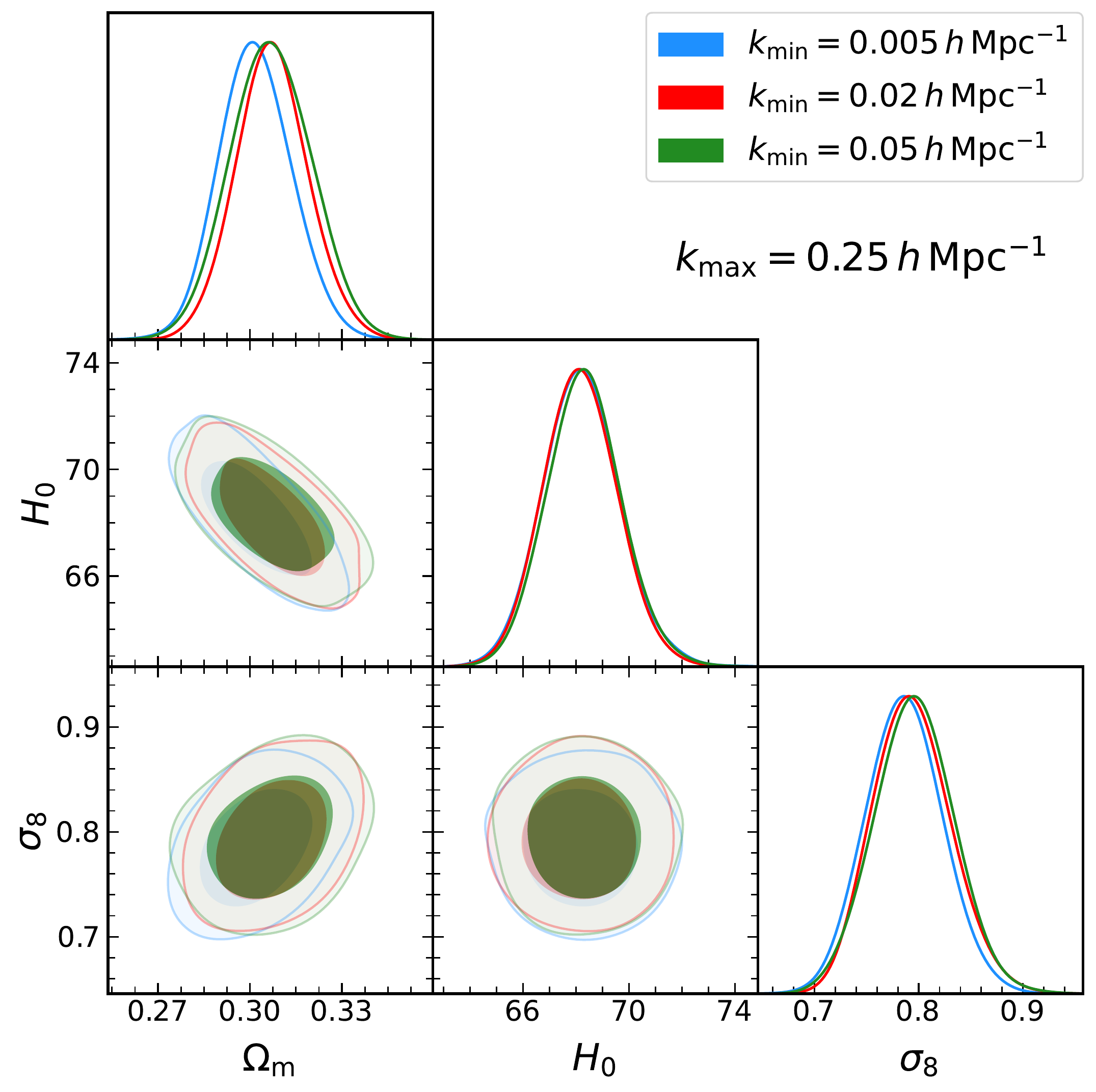}
\caption{
The dependence of the cosmological parameter constraints on the minimum wavenumber $k_\mathrm{min}$ used in 
the analysis.
We tested the cases of $k_\mathrm{min} = 0.005$ (our fiducial setting, blue), 0.02 (red) and $0.05\,h\,\mathrm{Mpc}^{-1}$ (green), while fixing $k_\mathrm{max} = 0.25 \, h\, \mathrm{Mpc}^{-1}$. 
}
\label{fig:real_z2_N_p024_kmin0-0.05}
\end{figure}

We mention the influence of the minimum wavenumber $k_\mathrm{min}$ of the power spectrum signals we use in the cosmological inference.
Figure~\ref{fig:real_z2_N_p024_kmin0-0.05} shows the parameter inference results for different values of $k_\mathrm{min} = 0.005, 0.02$, and $0.05 \, h\, \mathrm{Mpc}^{-1}$ while we keep $k_\mathrm{max} = 0.25\,h\,\mathrm{Mpc}^{-1}$. 
Here $k_\mathrm{min}=0.005\,h\mathrm{Mpc}^{-1}$ is our default choice throughout this paper.
It shows that the choice of $k_\mathrm{min}$ has no significant effect on the cosmological parameter inference.

\bibliographystyle{unsrt}
\bibliography{lssref}

\begin{thebibliography}{10}

\bibitem{cole:2005aa}
Shaun {Cole}, Will~J. {Percival}, John~A. {Peacock}, Peder {Norberg},
  Carlton~M. {Baugh}, Carlos~S. {Frenk}, Ivan {Baldry}, Joss {Bland-Hawthorn},
  Terry {Bridges}, Russell {Cannon}, Matthew {Colless}, Chris {Collins},
  Warrick {Couch}, Nicholas J.~G. {Cross}, Gavin {Dalton}, Vincent~R. {Eke},
  Roberto {De Propris}, Simon~P. {Driver}, George {Efstathiou}, Richard~S.
  {Ellis}, Karl {Glazebrook}, Carole {Jackson}, Adrian {Jenkins}, Ofer {Lahav},
  Ian {Lewis}, Stuart {Lumsden}, Steve {Maddox}, Darren {Madgwick}, Bruce~A.
  {Peterson}, Will {Sutherland}, and Keith {Taylor}.
\newblock {The 2dF Galaxy Redshift Survey: power-spectrum analysis of the final
  data set and cosmological implications}.
\newblock {\em \mnras}, 362(2):505--534, September 2005.

\bibitem{Eisenstein:2005su}
Daniel~J. {Eisenstein}, Idit {Zehavi}, David~W. {Hogg}, Roman {Scoccimarro},
  Michael~R. {Blanton}, Robert~C. {Nichol}, Ryan {Scranton}, Hee-Jong {Seo},
  Max {Tegmark}, Zheng {Zheng}, Scott~F. {Anderson}, Jim {Annis}, Neta
  {Bahcall}, Jon {Brinkmann}, Scott {Burles}, Francisco~J. {Castander}, Andrew
  {Connolly}, Istvan {Csabai}, Mamoru {Doi}, Masataka {Fukugita}, Joshua~A.
  {Frieman}, Karl {Glazebrook}, James~E. {Gunn}, John~S. {Hendry}, Gregory
  {Hennessy}, Zeljko {Ivezi{\'c}}, Stephen {Kent}, Gillian~R. {Knapp}, Huan
  {Lin}, Yeong-Shang {Loh}, Robert~H. {Lupton}, Bruce {Margon}, Timothy~A.
  {McKay}, Avery {Meiksin}, Jeffery~A. {Munn}, Adrian {Pope}, Michael~W.
  {Richmond}, David {Schlegel}, Donald~P. {Schneider}, Kazuhiro {Shimasaku},
  Christopher {Stoughton}, Michael~A. {Strauss}, Mark {SubbaRao}, Alexander~S.
  {Szalay}, Istv{\'a}n {Szapudi}, Douglas~L. {Tucker}, Brian {Yanny}, and
  Donald~G. {York}.
\newblock {Detection of the Baryon Acoustic Peak in the Large-Scale Correlation
  Function of SDSS Luminous Red Galaxies}.
\newblock {\em \apj}, 633(2):560--574, November 2005.

\bibitem{okumura08}
Teppei {Okumura}, Takahiko {Matsubara}, Daniel~J. {Eisenstein}, Issha {Kayo},
  Chiaki {Hikage}, Alexander~S. {Szalay}, and Donald~P. {Schneider}.
\newblock {Large-Scale Anisotropic Correlation Function of SDSS Luminous Red
  Galaxies}.
\newblock {\em \apj}, 676(2):889--898, April 2008.

\bibitem{parkinson12}
David {Parkinson}, Signe {Riemer-S{\o}rensen}, Chris {Blake}, Gregory~B.
  {Poole}, Tamara~M. {Davis}, Sarah {Brough}, Matthew {Colless}, Carlos
  {Contreras}, Warrick {Couch}, Scott {Croom}, Darren {Croton}, Michael~J.
  {Drinkwater}, Karl {Forster}, David {Gilbank}, Mike {Gladders}, Karl
  {Glazebrook}, Ben {Jelliffe}, Russell~J. {Jurek}, I.~hui {Li}, Barry
  {Madore}, D.~Christopher {Martin}, Kevin {Pimbblet}, Michael {Pracy}, Rob
  {Sharp}, Emily {Wisnioski}, David {Woods}, Ted~K. {Wyder}, and H.~K.~C.
  {Yee}.
\newblock {The WiggleZ Dark Energy Survey: Final data release and cosmological
  results}.
\newblock {\em \prd}, 86(10):103518, November 2012.

\bibitem{2017MNRAS.470.2617A}
Shadab {Alam}, Metin {Ata}, Stephen {Bailey}, Florian {Beutler}, Dmitry
  {Bizyaev}, Jonathan~A. {Blazek}, Adam~S. {Bolton}, Joel~R. {Brownstein},
  Angela {Burden}, Chia-Hsun {Chuang}, Johan {Comparat}, Antonio~J. {Cuesta},
  Kyle~S. {Dawson}, Daniel~J. {Eisenstein}, Stephanie {Escoffier}, H{\'e}ctor
  {Gil-Mar{\'\i}n}, Jan~Niklas {Grieb}, Nick {Hand}, Shirley {Ho}, Karen
  {Kinemuchi}, David {Kirkby}, Francisco {Kitaura}, Elena {Malanushenko},
  Viktor {Malanushenko}, Claudia {Maraston}, Cameron~K. {McBride}, Robert~C.
  {Nichol}, Matthew~D. {Olmstead}, Daniel {Oravetz}, Nikhil {Padmanabhan},
  Nathalie {Palanque-Delabrouille}, Kaike {Pan}, Marcos {Pellejero-Ibanez},
  Will~J. {Percival}, Patrick {Petitjean}, Francisco {Prada}, Adrian~M.
  {Price-Whelan}, Beth~A. {Reid}, Sergio~A. {Rodr{\'\i}guez-Torres}, Natalie~A.
  {Roe}, Ashley~J. {Ross}, Nicholas~P. {Ross}, Graziano {Rossi}, Jose~Alberto
  {Rubi{\~n}o-Mart{\'\i}n}, Shun {Saito}, Salvador {Salazar-Albornoz}, Lado
  {Samushia}, Ariel~G. {S{\'a}nchez}, Siddharth {Satpathy}, David~J.
  {Schlegel}, Donald~P. {Schneider}, Claudia~G. {Sc{\'o}ccola}, Hee-Jong {Seo},
  Erin~S. {Sheldon}, Audrey {Simmons}, An{\v{z}}e {Slosar}, Michael~A.
  {Strauss}, Molly E.~C. {Swanson}, Daniel {Thomas}, Jeremy~L. {Tinker}, Rita
  {Tojeiro}, Mariana~Vargas {Maga{\~n}a}, Jose~Alberto {Vazquez}, Licia
  {Verde}, David~A. {Wake}, Yuting {Wang}, David~H. {Weinberg}, Martin {White},
  W.~Michael {Wood-Vasey}, Christophe {Y{\`e}che}, Idit {Zehavi}, Zhongxu
  {Zhai}, and Gong-Bo {Zhao}.
\newblock {The clustering of galaxies in the completed SDSS-III Baryon
  Oscillation Spectroscopic Survey: cosmological analysis of the DR12 galaxy
  sample}.
\newblock {\em \mnras}, 470(3):2617--2652, Sep 2017.

\bibitem{2016PASJ...68...38O}
Teppei {Okumura}, Chiaki {Hikage}, Tomonori {Totani}, Motonari {Tonegawa},
  Hiroyuki {Okada}, Karl {Glazebrook}, Chris {Blake}, Pedro~G. {Ferreira},
  Surhud {More}, Atsushi {Taruya}, Shinji {Tsujikawa}, Masayuki {Akiyama},
  Gavin {Dalton}, Tomotsugu {Goto}, Takashi {Ishikawa}, Fumihide {Iwamuro},
  Takahiko {Matsubara}, Takahiro {Nishimichi}, Kouji {Ohta}, Ikkoh {Shimizu},
  Ryuichi {Takahashi}, Naruhisa {Takato}, Naoyuki {Tamura}, Kiyoto {Yabe}, and
  Naoki {Yoshida}.
\newblock {The Subaru FMOS galaxy redshift survey (FastSound). IV. New
  constraint on gravity theory from redshift space distortions at z\~1.4}.
\newblock {\em \pasj}, 68(3):38, June 2016.

\bibitem{2021PhRvD.103h3533A}
Shadab {Alam}, Marie {Aubert}, Santiago {Avila}, Christophe {Balland},
  Julian~E. {Bautista}, Matthew~A. {Bershady}, Dmitry {Bizyaev}, Michael~R.
  {Blanton}, Adam~S. {Bolton}, Jo~{Bovy}, Jonathan {Brinkmann}, Joel~R.
  {Brownstein}, Etienne {Burtin}, Sol{\`e}ne {Chabanier}, Michael~J. {Chapman},
  Peter~Doohyun {Choi}, Chia-Hsun {Chuang}, Johan {Comparat}, Marie-Claude
  {Cousinou}, Andrei {Cuceu}, Kyle~S. {Dawson}, Sylvain {de la Torre}, Arnaud
  {de Mattia}, Victoria de~Sainte {Agathe}, H{\'e}lion du~Mas {des Bourboux},
  Stephanie {Escoffier}, Thomas {Etourneau}, James {Farr}, Andreu
  {Font-Ribera}, Peter~M. {Frinchaboy}, Sebastien {Fromenteau}, H{\'e}ctor
  {Gil-Mar{\'\i}n}, Jean-Marc {Le Goff}, Alma~X. {Gonzalez-Morales}, Violeta
  {Gonzalez-Perez}, Kathleen {Grabowski}, Julien {Guy}, Adam~J. {Hawken},
  Jiamin {Hou}, Hui {Kong}, James {Parker}, Mark {Klaene}, Jean-Paul {Kneib},
  Sicheng {Lin}, Daniel {Long}, Brad~W. {Lyke}, Axel {de la Macorra}, Paul
  {Martini}, Karen {Masters}, Faizan~G. {Mohammad}, Jeongin {Moon}, Eva-Maria
  {Mueller}, Andrea {Mu{\~n}oz-Guti{\'e}rrez}, Adam~D. {Myers}, Seshadri
  {Nadathur}, Richard {Neveux}, Jeffrey~A. {Newman}, Pasquier {Noterdaeme},
  Audrey {Oravetz}, Daniel {Oravetz}, Nathalie {Palanque-Delabrouille}, Kaike
  {Pan}, Romain {Paviot}, Will~J. {Percival}, Ignasi {P{\'e}rez-R{\`a}fols},
  Patrick {Petitjean}, Matthew~M. {Pieri}, Abhishek {Prakash}, Anand
  {Raichoor}, Corentin {Ravoux}, Mehdi {Rezaie}, James {Rich}, Ashley~J.
  {Ross}, Graziano {Rossi}, Rossana {Ruggeri}, Vanina {Ruhlmann-Kleider},
  Ariel~G. {S{\'a}nchez}, F.~Javier {S{\'a}nchez}, Jos{\'e}~R.
  {S{\'a}nchez-Gallego}, Conor {Sayres}, Donald~P. {Schneider}, Hee-Jong {Seo},
  Arman {Shafieloo}, An{\v{z}}e {Slosar}, Alex {Smith}, Julianna {Stermer},
  Amelie {Tamone}, Jeremy~L. {Tinker}, Rita {Tojeiro}, Mariana
  {Vargas-Maga{\~n}a}, Andrei {Variu}, Yuting {Wang}, Benjamin~A. {Weaver},
  Anne-Marie {Weijmans}, Christophe {Y{\`e}che}, Pauline {Zarrouk}, Cheng
  {Zhao}, Gong-Bo {Zhao}, and Zheng {Zheng}.
\newblock {Completed SDSS-IV extended Baryon Oscillation Spectroscopic Survey:
  Cosmological implications from two decades of spectroscopic surveys at the
  Apache Point Observatory}.
\newblock {\em \prd}, 103(8):083533, April 2021.

\bibitem{2013AJ....145...10D}
Kyle~S. {Dawson}, David~J. {Schlegel}, Christopher~P. {Ahn}, Scott~F.
  {Anderson}, {\'E}ric {Aubourg}, Stephen {Bailey}, Robert~H. {Barkhouser},
  Julian~E. {Bautista}, Alessand~ra {Beifiori}, Andreas~A. {Berlind}, Vaishali
  {Bhardwaj}, Dmitry {Bizyaev}, Cullen~H. {Blake}, Michael~R. {Blanton},
  Michael {Blomqvist}, Adam~S. {Bolton}, Arnaud {Borde}, Jo~{Bovy}, W.~N.
  {Brandt}, Howard {Brewington}, Jon {Brinkmann}, Peter~J. {Brown}, Joel~R.
  {Brownstein}, Kevin {Bundy}, N.~G. {Busca}, William {Carithers}, Aurelio~R.
  {Carnero}, Michael~A. {Carr}, Yanmei {Chen}, Johan {Comparat}, Natalia
  {Connolly}, Frances {Cope}, Rupert A.~C. {Croft}, Antonio~J. {Cuesta},
  Luiz~N. {da Costa}, James R.~A. {Davenport}, Timoth{\'e}e {Delubac}, Roland
  {de Putter}, Saurav {Dhital}, Anne {Ealet}, Garrett~L. {Ebelke}, Daniel~J.
  {Eisenstein}, S.~{Escoffier}, Xiaohui {Fan}, N.~{Filiz Ak}, Hayley {Finley},
  Andreu {Font-Ribera}, R.~{G{\'e}nova-Santos}, James~E. {Gunn}, Hong {Guo},
  Daryl {Haggard}, Patrick~B. {Hall}, Jean-Christophe {Hamilton}, Ben {Harris},
  David~W. {Harris}, Shirley {Ho}, David~W. {Hogg}, Diana {Holder}, Klaus
  {Honscheid}, Joe {Huehnerhoff}, Beatrice {Jordan}, Wendell~P. {Jordan},
  Guinevere {Kauffmann}, Eyal~A. {Kazin}, David {Kirkby}, Mark~A. {Klaene},
  Jean-Paul {Kneib}, Jean-Marc {Le Goff}, Khee-Gan {Lee}, Daniel~C. {Long},
  Craig~P. {Loomis}, Britt {Lundgren}, Robert~H. {Lupton}, Marcio A.~G. {Maia},
  Martin {Makler}, Elena {Malanushenko}, Viktor {Malanushenko}, Rachel
  {Mandelbaum}, Marc {Manera}, Claudia {Maraston}, Daniel {Margala}, Karen~L.
  {Masters}, Cameron~K. {McBride}, Patrick {McDonald}, Ian~D. {McGreer},
  Richard~G. {McMahon}, Olga {Mena}, Jordi {Miralda-Escud{\'e}}, Antonio~D.
  {Montero-Dorta}, Francesco {Montesano}, Demitri {Muna}, Adam~D. {Myers},
  Tracy {Naugle}, Robert~C. {Nichol}, Pasquier {Noterdaeme}, Sebasti{\'a}n~E.
  {Nuza}, Matthew~D. {Olmstead}, Audrey {Oravetz}, Daniel~J. {Oravetz}, Russell
  {Owen}, Nikhil {Padmanabhan}, Nathalie {Palanque-Delabrouille}, Kaike {Pan},
  John~K. {Parejko}, Isabelle {P{\^a}ris}, Will~J. {Percival}, Ismael
  {P{\'e}rez-Fournon}, Ignasi {P{\'e}rez-R{\`a}fols}, Patrick {Petitjean},
  Robert {Pfaffenberger}, Janine {Pforr}, Matthew~M. {Pieri}, Francisco
  {Prada}, Adrian~M. {Price-Whelan}, M.~Jordan {Raddick}, Rafael {Rebolo},
  James {Rich}, Gordon~T. {Richards}, Constance~M. {Rockosi}, Natalie~A. {Roe},
  Ashley~J. {Ross}, Nicholas~P. {Ross}, Graziano {Rossi}, J.~A.
  {Rubi{\~n}o-Martin}, Lado {Samushia}, Ariel~G. {S{\'a}nchez}, Conor {Sayres},
  Sarah~J. {Schmidt}, Donald~P. {Schneider}, C.~G. {Sc{\'o}ccola}, Hee-Jong
  {Seo}, Alaina {Shelden}, Erin {Sheldon}, Yue {Shen}, Yiping {Shu}, An{\v{z}}e
  {Slosar}, Stephen~A. {Smee}, Stephanie~A. {Snedden}, Fritz {Stauffer}, Oliver
  {Steele}, Michael~A. {Strauss}, Alina {Streblyanska}, Nao {Suzuki}, Molly
  E.~C. {Swanson}, Tomer {Tal}, Masayuki {Tanaka}, Daniel {Thomas}, Jeremy~L.
  {Tinker}, Rita {Tojeiro}, Christy~A. {Tremonti}, M.~{Vargas Maga{\~n}a},
  Licia {Verde}, Matteo {Viel}, David~A. {Wake}, Mike {Watson}, Benjamin~A.
  {Weaver}, David~H. {Weinberg}, Benjamin~J. {Weiner}, Andrew~A. {West}, Martin
  {White}, W.~M. {Wood-Vasey}, Christophe {Yeche}, Idit {Zehavi}, Gong-Bo
  {Zhao}, and Zheng {Zheng}.
\newblock {The Baryon Oscillation Spectroscopic Survey of SDSS-III}.
\newblock {\em \aj}, 145:10, Jan 2013.

\bibitem{2016AJ....151...44D}
Kyle~S. {Dawson}, Jean-Paul {Kneib}, Will~J. {Percival}, Shadab {Alam},
  Franco~D. {Albareti}, Scott~F. {Anderson}, Eric {Armengaud}, {\'E}ric
  {Aubourg}, Stephen {Bailey}, Julian~E. {Bautista}, Andreas~A. {Berlind},
  Matthew~A. {Bershady}, Florian {Beutler}, Dmitry {Bizyaev}, Michael~R.
  {Blanton}, Michael {Blomqvist}, Adam~S. {Bolton}, Jo~{Bovy}, W.~N. {Brandt},
  Jon {Brinkmann}, Joel~R. {Brownstein}, Etienne {Burtin}, N.~G. {Busca}, Zheng
  {Cai}, Chia-Hsun {Chuang}, Nicolas {Clerc}, Johan {Comparat}, Frances {Cope},
  Rupert A.~C. {Croft}, Irene {Cruz-Gonzalez}, Luiz~N. {da Costa}, Marie-Claude
  {Cousinou}, Jeremy {Darling}, Axel {de la Macorra}, Sylvain {de la Torre},
  Timoth{\'e}e {Delubac}, H{\'e}lion {du Mas des Bourboux}, Tom {Dwelly}, Anne
  {Ealet}, Daniel~J. {Eisenstein}, Michael {Eracleous}, S.~{Escoffier}, Xiaohui
  {Fan}, Alexis {Finoguenov}, Andreu {Font-Ribera}, Peter {Frinchaboy}, Patrick
  {Gaulme}, Antonis {Georgakakis}, Paul {Green}, Hong {Guo}, Julien {Guy},
  Shirley {Ho}, Diana {Holder}, Joe {Huehnerhoff}, Timothy {Hutchinson}, Yipeng
  {Jing}, Eric {Jullo}, Vikrant {Kamble}, Karen {Kinemuchi}, David {Kirkby},
  Francisco-Shu {Kitaura}, Mark~A. {Klaene}, Russ~R. {Laher}, Dustin {Lang},
  Pierre {Laurent}, Jean-Marc {Le Goff}, Cheng {Li}, Yu~{Liang}, Marcos {Lima},
  Qiufan {Lin}, Weipeng {Lin}, Yen-Ting {Lin}, Daniel~C. {Long}, Britt
  {Lundgren}, Nicholas {MacDonald}, Marcio~Antonio {Geimba Maia}, Elena
  {Malanushenko}, Viktor {Malanushenko}, Vivek {Mariappan}, Cameron~K.
  {McBride}, Ian~D. {McGreer}, Brice {M{\'e}nard}, Andrea {Merloni}, Andres
  {Meza}, Antonio~D. {Montero-Dorta}, Demitri {Muna}, Adam~D. {Myers}, Kirpal
  {Nandra}, Tracy {Naugle}, Jeffrey~A. {Newman}, Pasquier {Noterdaeme}, Peter
  {Nugent}, Ricardo {Ogando}, Matthew~D. {Olmstead}, Audrey {Oravetz},
  Daniel~J. {Oravetz}, Nikhil {Padmanabhan}, Nathalie {Palanque-Delabrouille},
  Kaike {Pan}, John~K. {Parejko}, Isabelle {P{\^a}ris}, John~A. {Peacock},
  Patrick {Petitjean}, Matthew~M. {Pieri}, Alice {Pisani}, Francisco {Prada},
  Abhishek {Prakash}, Anand {Raichoor}, Beth {Reid}, James {Rich}, Jethro
  {Ridl}, Sergio {Rodriguez-Torres}, Aurelio {Carnero Rosell}, Ashley~J.
  {Ross}, Graziano {Rossi}, John {Ruan}, Mara {Salvato}, Conor {Sayres},
  Donald~P. {Schneider}, David~J. {Schlegel}, Uros {Seljak}, Hee-Jong {Seo},
  Branimir {Sesar}, Sarah {Shandera}, Yiping {Shu}, An{\v{z}}e {Slosar}, Flavia
  {Sobreira}, Alina {Streblyanska}, Nao {Suzuki}, Donna {Taylor}, Charling
  {Tao}, Jeremy~L. {Tinker}, Rita {Tojeiro}, Mariana {Vargas-Maga{\~n}a},
  Yuting {Wang}, Benjamin~A. {Weaver}, David~H. {Weinberg}, Martin {White},
  W.~M. {Wood-Vasey}, Christophe {Yeche}, Zhongxu {Zhai}, Cheng {Zhao}, Gong-bo
  {Zhao}, Zheng {Zheng}, Guangtun {Ben Zhu}, and Hu~{Zou}.
\newblock {The SDSS-IV Extended Baryon Oscillation Spectroscopic Survey:
  Overview and Early Data}.
\newblock {\em \aj}, 151(2):44, Feb 2016.

\bibitem{2014PASJ...66R...1T}
Masahiro {Takada}, Richard~S. {Ellis}, Masashi {Chiba}, Jenny~E. {Greene},
  Hiroaki {Aihara}, Nobuo {Arimoto}, Kevin {Bundy}, Judith {Cohen}, Olivier
  {Dor{\'e}}, Genevieve {Graves}, James~E. {Gunn}, Timothy {Heckman},
  Christopher~M. {Hirata}, Paul {Ho}, Jean-Paul {Kneib}, Olivier {Le
  F{\`e}vre}, Lihwai {Lin}, Surhud {More}, Hitoshi {Murayama}, Tohru {Nagao},
  Masami {Ouchi}, Michael {Seiffert}, John~D. {Silverman}, Laerte {Sodr{\'e}},
  David~N. {Spergel}, Michael~A. {Strauss}, Hajime {Sugai}, Yasushi {Suto},
  Hideki {Takami}, and Rosemary {Wyse}.
\newblock {Extragalactic science, cosmology, and Galactic archaeology with the
  Subaru Prime Focus Spectrograph}.
\newblock {\em \pasj}, 66(1):R1, February 2014.

\bibitem{Aghamousa:2016zmz}
Amir Aghamousa et~al.
\newblock {The DESI Experiment Part I: Science,Targeting, and Survey Design}.
\newblock 2016.

\bibitem{Laureijs:2011gra}
R.~{Laureijs}, J.~{Amiaux}, S.~{Arduini}, J.~L. {Augu{\`e}res},
  J.~{Brinchmann}, R.~{Cole}, M.~{Cropper}, C.~{Dabin}, L.~{Duvet}, A.~{Ealet},
  B.~{Garilli}, P.~{Gondoin}, L.~{Guzzo}, J.~{Hoar}, H.~{Hoekstra},
  R.~{Holmes}, T.~{Kitching}, T.~{Maciaszek}, Y.~{Mellier}, F.~{Pasian},
  W.~{Percival}, J.~{Rhodes}, G.~{Saavedra Criado}, M.~{Sauvage},
  R.~{Scaramella}, L.~{Valenziano}, S.~{Warren}, R.~{Bender}, F.~{Castander},
  A.~{Cimatti}, O.~{Le F{\`e}vre}, H.~{Kurki-Suonio}, M.~{Levi}, P.~{Lilje},
  G.~{Meylan}, R.~{Nichol}, K.~{Pedersen}, V.~{Popa}, R.~{Rebolo Lopez}, H.~W.
  {Rix}, H.~{Rottgering}, W.~{Zeilinger}, F.~{Grupp}, P.~{Hudelot},
  R.~{Massey}, M.~{Meneghetti}, L.~{Miller}, S.~{Paltani},
  S.~{Paulin-Henriksson}, S.~{Pires}, C.~{Saxton}, T.~{Schrabback},
  G.~{Seidel}, J.~{Walsh}, N.~{Aghanim}, L.~{Amendola}, J.~{Bartlett},
  C.~{Baccigalupi}, J.~P. {Beaulieu}, K.~{Benabed}, J.~G. {Cuby}, D.~{Elbaz},
  P.~{Fosalba}, G.~{Gavazzi}, A.~{Helmi}, I.~{Hook}, M.~{Irwin}, J.~P. {Kneib},
  M.~{Kunz}, F.~{Mannucci}, L.~{Moscardini}, C.~{Tao}, R.~{Teyssier},
  J.~{Weller}, G.~{Zamorani}, M.~R. {Zapatero Osorio}, O.~{Boulade}, J.~J.
  {Foumond}, A.~{Di Giorgio}, P.~{Guttridge}, A.~{James}, M.~{Kemp},
  J.~{Martignac}, A.~{Spencer}, D.~{Walton}, T.~{Bl{\"u}mchen}, C.~{Bonoli},
  F.~{Bortoletto}, C.~{Cerna}, L.~{Corcione}, C.~{Fabron}, K.~{Jahnke},
  S.~{Ligori}, F.~{Madrid}, L.~{Martin}, G.~{Morgante}, T.~{Pamplona},
  E.~{Prieto}, M.~{Riva}, R.~{Toledo}, M.~{Trifoglio}, F.~{Zerbi},
  F.~{Abdalla}, M.~{Douspis}, C.~{Grenet}, S.~{Borgani}, R.~{Bouwens},
  F.~{Courbin}, J.~M. {Delouis}, P.~{Dubath}, A.~{Fontana}, M.~{Frailis},
  A.~{Grazian}, J.~{Koppenh{\"o}fer}, O.~{Mansutti}, M.~{Melchior},
  M.~{Mignoli}, J.~{Mohr}, C.~{Neissner}, K.~{Noddle}, M.~{Poncet},
  M.~{Scodeggio}, S.~{Serrano}, N.~{Shane}, J.~L. {Starck}, C.~{Surace},
  A.~{Taylor}, G.~{Verdoes-Kleijn}, C.~{Vuerli}, O.~R. {Williams},
  A.~{Zacchei}, B.~{Altieri}, I.~{Escudero Sanz}, R.~{Kohley},
  T.~{Oosterbroek}, P.~{Astier}, D.~{Bacon}, S.~{Bardelli}, C.~{Baugh},
  F.~{Bellagamba}, C.~{Benoist}, D.~{Bianchi}, A.~{Biviano}, E.~{Branchini},
  C.~{Carbone}, V.~{Cardone}, D.~{Clements}, S.~{Colombi}, C.~{Conselice},
  G.~{Cresci}, N.~{Deacon}, J.~{Dunlop}, C.~{Fedeli}, F.~{Fontanot},
  P.~{Franzetti}, C.~{Giocoli}, J.~{Garcia-Bellido}, J.~{Gow}, A.~{Heavens},
  P.~{Hewett}, C.~{Heymans}, A.~{Holland}, Z.~{Huang}, O.~{Ilbert},
  B.~{Joachimi}, E.~{Jennins}, E.~{Kerins}, A.~{Kiessling}, D.~{Kirk},
  R.~{Kotak}, O.~{Krause}, O.~{Lahav}, F.~{van Leeuwen}, J.~{Lesgourgues},
  M.~{Lombardi}, M.~{Magliocchetti}, K.~{Maguire}, E.~{Majerotto}, R.~{Maoli},
  F.~{Marulli}, S.~{Maurogordato}, H.~{McCracken}, R.~{McLure},
  A.~{Melchiorri}, A.~{Merson}, M.~{Moresco}, M.~{Nonino}, P.~{Norberg},
  J.~{Peacock}, R.~{Pello}, M.~{Penny}, V.~{Pettorino}, C.~{Di Porto},
  L.~{Pozzetti}, C.~{Quercellini}, M.~{Radovich}, A.~{Rassat}, N.~{Roche},
  S.~{Ronayette}, E.~{Rossetti}, B.~{Sartoris}, P.~{Schneider}, E.~{Semboloni},
  S.~{Serjeant}, F.~{Simpson}, C.~{Skordis}, G.~{Smadja}, S.~{Smartt},
  P.~{Spano}, S.~{Spiro}, M.~{Sullivan}, A.~{Tilquin}, R.~{Trotta}, L.~{Verde},
  Y.~{Wang}, G.~{Williger}, G.~{Zhao}, J.~{Zoubian}, and E.~{Zucca}.
\newblock {Euclid Definition Study Report}.
\newblock {\em arXiv e-prints}, page arXiv:1110.3193, October 2011.

\bibitem{Gehrels:2014spa}
Neil Gehrels and David~N. Spergel.
\newblock {Wide-Field InfraRed Survey Telescope (WFIRST) Mission and Synergies
  with LISA and LIGO-Virgo}.
\newblock {\em Journal of Physics: Conference Series}, 610(1):012007, 2015.

\bibitem{kaiser87}
Nick {Kaiser}.
\newblock {Clustering in real space and in redshift space}.
\newblock {\em \mnras}, 227:1--21, July 1987.

\bibitem{Hamilton92}
A.~J.~S. {Hamilton}.
\newblock {Measuring Omega and the Real Correlation Function from the Redshift
  Correlation Function}.
\newblock {\em \apjl}, 385:L5, January 1992.

\bibitem{Hamilton_1998}
A.~J.~S. {Hamilton}.
\newblock {Linear Redshift Distortions: a Review}.
\newblock 231:185, January 1998.

\bibitem{PhysRevD.101.023510}
Yosuke Kobayashi, Takahiro Nishimichi, Masahiro Takada, and Ryuichi Takahashi.
\newblock Cosmological information content in redshift-space power spectrum of
  sdss-like galaxies in the quasinonlinear regime up to $k=0.3\text{ }\text{
  }h\text{ }{\mathrm{mpc}}^{\ensuremath{-}1}$.
\newblock {\em \prd}, 101:023510, Jan 2020.

\bibitem{2007PhRvL..99n1302Z}
Pengjie {Zhang}, Michele {Liguori}, Rachel {Bean}, and Scott {Dodelson}.
\newblock {Probing Gravity at Cosmological Scales by Measurements which Test
  the Relationship between Gravitational Lensing and Matter Overdensity}.
\newblock {\em \prl}, 99(14):141302, October 2007.

\bibitem{bernardeau02}
F.~{Bernardeau}, S.~{Colombi}, E.~{Gazta{\~n}aga}, and R.~{Scoccimarro}.
\newblock {Large-scale structure of the Universe and cosmological perturbation
  theory}.
\newblock {\em \physrep}, 367(1-3):1--248, September 2002.

\bibitem{Desjacques18}
Vincent {Desjacques}, Donghui {Jeong}, and Fabian {Schmidt}.
\newblock {Large-scale galaxy bias}.
\newblock {\em \physrep}, 733:1--193, February 2018.

\bibitem{taruya10}
Atsushi {Taruya}, Takahiro {Nishimichi}, and Shun {Saito}.
\newblock {Baryon acoustic oscillations in 2D: Modeling redshift-space power
  spectrum from perturbation theory}.
\newblock {\em \prd}, 82(6):063522, September 2010.

\bibitem{nishimichi11}
Takahiro {Nishimichi} and Atsushi {Taruya}.
\newblock {Baryon acoustic oscillations in 2D. II. Redshift-space halo
  clustering in N-body simulations}.
\newblock {\em \prd}, 84(4):043526, August 2011.

\bibitem{baumann12}
Daniel {Baumann}, Alberto {Nicolis}, Leonardo {Senatore}, and Matias
  {Zaldarriaga}.
\newblock {Cosmological non-linearities as an effective fluid}.
\newblock {\em \jcap}, 2012(7):051, July 2012.

\bibitem{2014MNRAS.444..476R}
Beth~A. {Reid}, Hee-Jong {Seo}, Alexie {Leauthaud}, Jeremy~L. {Tinker}, and
  Martin {White}.
\newblock {A 2.5 per cent measurement of the growth rate from small-scale
  redshift space clustering of SDSS-III CMASS galaxies}.
\newblock {\em \mnras}, 444(1):476--502, October 2014.

\bibitem{10.1093/mnras/stu1051}
Florian Beutler, Shun Saito, Hee-Jong Seo, Jon Brinkmann, Kyle~S. Dawson,
  Daniel~J. Eisenstein, Andreu Font-Ribera, Shirley Ho, Cameron~K. McBride,
  Francesco Montesano, Will~J. Percival, Ashley~J. Ross, Nicholas~P. Ross, Lado
  Samushia, David~J. Schlegel, Ariel~G. S^^c3^^a1nchez, Jeremy~L. Tinker, and
  Benjamin~A. Weaver.
\newblock {The clustering of galaxies in the SDSS-III Baryon Oscillation
  Spectroscopic Survey: testing gravity with redshift space distortions using
  the power spectrum multipoles}.
\newblock {\em Monthly Notices of the Royal Astronomical Society},
  443(2):1065--1089, 07 2014.

\bibitem{Beutler:2016arn}
Florian Beutler et~al.
\newblock {The clustering of galaxies in the completed SDSS-III Baryon
  Oscillation Spectroscopic Survey: Anisotropic galaxy clustering in
  Fourier-space}.
\newblock {\em \mnras}, 466(2):2242--2260, 2017.

\bibitem{Ivanov_2020}
Mikhail~M. Ivanov, Marko {Simonovi{\'c}}, and Matias Zaldarriaga.
\newblock Cosmological parameters from the boss galaxy power spectrum.
\newblock {\em Journal of Cosmology and Astroparticle Physics},
  2020(05):042^^e2^^80^^93042, May 2020.

\bibitem{d_Amico_2020}
Guido d’ Amico, J^^c3^^a9r^^c3^^b4me Gleyzes, Nickolas Kokron, Katarina
  Markovic, Leonardo Senatore, Pierre Zhang, Florian Beutler, and H^^c3^^a9ctor
  Gil-Mar^^c3^^adn.
\newblock The cosmological analysis of the sdss/boss data from the effective
  field theory of large-scale structure.
\newblock {\em Journal of Cosmology and Astroparticle Physics},
  2020(05):005^^e2^^80^^93005, May 2020.

\bibitem{2021arXiv211005530C}
Shi-Fan {Chen}, Zvonimir {Vlah}, and Martin {White}.
\newblock {A new analysis of the BOSS survey, including full-shape information
  and post-reconstruction BAO}.
\newblock {\em arXiv e-prints}, page arXiv:2110.05530, October 2021.

\bibitem{Pueblas_2009}
Sebasti^^c3^^a1n Pueblas and Rom^^c3^^a1n Scoccimarro.
\newblock Generation of vorticity and velocity dispersion by orbit crossing.
\newblock {\em Physical Review D}, 80(4), Aug 2009.

\bibitem{blas14}
Diego {Blas}, Mathias {Garny}, and Thomas {Konstandin}.
\newblock {Cosmological perturbation theory at three-loop order}.
\newblock {\em \jcap}, 2014(1):010, January 2014.

\bibitem{bernardeau:2014lr}
Francis {Bernardeau}, Atsushi {Taruya}, and Takahiro {Nishimichi}.
\newblock {Cosmic propagators at two-loop order}.
\newblock {\em \prd}, 89(2):023502, January 2014.

\bibitem{nishimichi2016}
Takahiro Nishimichi, Francis Bernardeau, and Atsushi Taruya.
\newblock Response function of the large-scale structure of the universe to the
  small scale inhomogeneities.
\newblock {\em Physics Letters B}, 762:247--252, November 2016.

\bibitem{taruya2017}
Atsushi Taruya and St{\'e}phane Colombi.
\newblock Post-collapse perturbation theory in {{1D}} cosmology -- beyond
  shell-crossing.
\newblock {\em Monthly Notices of the Royal Astronomical Society},
  470(4):4858--4884, October 2017.

\bibitem{saga2018}
Shohei Saga, Atsushi Taruya, and St{\'e}phane Colombi.
\newblock Lagrangian cosmological perturbation theory at shell-crossing.
\newblock {\em Physical Review Letters}, 121(24):241302, December 2018.

\bibitem{Halle_2020}
Ana^^c3^^ablle Halle, Takahiro Nishimichi, Atsushi Taruya, St^^c3^^a9phane
  Colombi, and Francis Bernardeau.
\newblock Power spectrum response of large-scale structure in 1d and in 3d:
  tests of prescriptions for post-collapse dynamics.
\newblock {\em Monthly Notices of the Royal Astronomical Society},
  499(2):1769^^e2^^80^^931787, Sep 2020.

\bibitem{2020arXiv200308277N}
Takahiro Nishimichi, Guido D’Amico, Mikhail~M. Ivanov, Leonardo Senatore,
  Marko {Simonovi{\'c}}, Masahiro Takada, Matias Zaldarriaga, and Pierre Zhang.
\newblock Blinded challenge for precision cosmology with large-scale structure:
  Results from effective field theory for the redshift-space galaxy power
  spectrum.
\newblock {\em Physical Review D}, 102(12), Dec 2020.

\bibitem{Kobayashi_2020}
Yosuke {Kobayashi}, Takahiro {Nishimichi}, Masahiro {Takada}, Ryuichi
  {Takahashi}, and Ken {Osato}.
\newblock {Accurate emulator for the redshift-space power spectrum of dark
  matter halos and its application to galaxy power spectrum}.
\newblock {\em \prd}, 102(6):063504, September 2020.

\bibitem{Cooray02}
Asantha {Cooray} and Ravi {Sheth}.
\newblock {Halo models of large scale structure}.
\newblock {\em \physrep}, 372(1):1--129, December 2002.

\bibitem{Nishimichi_2019}
Takahiro Nishimichi, Masahiro Takada, Ryuichi Takahashi, Ken Osato, Masato
  Shirasaki, Taira Oogi, Hironao Miyatake, Masamune Oguri, Ryoma Murata, Yosuke
  Kobayashi, and Naoki Yoshida.
\newblock Dark quest. i. fast and accurate emulation of halo clustering
  statistics and its application to galaxy clustering.
\newblock {\em The Astrophysical Journal}, 884(1):29, oct 2019.

\bibitem{1998ApJ...494....1J}
Y.~P. {Jing}, H.~J. {Mo}, and G.~{B{\"o}rner}.
\newblock {Spatial Correlation Function and Pairwise Velocity Dispersion of
  Galaxies: Cold Dark Matter Models versus the Las Campanas Survey}.
\newblock {\em \apj}, 494:1--12, February 1998.

\bibitem{seljak:2000uq}
Uro{\v{s}} {Seljak}.
\newblock {Analytic model for galaxy and dark matter clustering}.
\newblock {\em \mnras}, 318(1):203--213, October 2000.

\bibitem{peacock:2000qy}
J.~A. {Peacock} and R.~E. {Smith}.
\newblock {Halo occupation numbers and galaxy bias}.
\newblock {\em \mnras}, 318:1144--1156, November 2000.

\bibitem{scoccimarro:2001fj}
Rom{\'a}n {Scoccimarro}, Ravi~K. {Sheth}, Lam {Hui}, and Bhuvnesh {Jain}.
\newblock {How Many Galaxies Fit in a Halo? Constraints on Galaxy Formation
  Efficiency from Spatial Clustering}.
\newblock {\em \apj}, 546(1):20--34, January 2001.

\bibitem{2005ApJ...633..791Z}
Zheng {Zheng}, Andreas~A. {Berlind}, David~H. {Weinberg}, Andrew~J. {Benson},
  Carlton~M. {Baugh}, Shaun {Cole}, Romeel {Dav{\'e}}, Carlos~S. {Frenk}, Neal
  {Katz}, and Cedric~G. {Lacey}.
\newblock {Theoretical Models of the Halo Occupation Distribution: Separating
  Central and Satellite Galaxies}.
\newblock {\em \apj}, 633(2):791--809, November 2005.

\bibitem{zheng09}
Zheng {Zheng}, Idit {Zehavi}, Daniel~J. {Eisenstein}, David~H. {Weinberg}, and
  Y.~P. {Jing}.
\newblock {Halo Occupation Distribution Modeling of Clustering of Luminous Red
  Galaxies}.
\newblock {\em \apj}, 707(1):554--572, December 2009.

\bibitem{Note1}
\protect \url {https://data.sdss.org/sas/dr12/boss/lss/}.

\bibitem{Aver_2015}
Erik Aver, Keith~A. Olive, and Evan~D. Skillman.
\newblock The effects of he i $\lambda$10830 on helium abundance
  determinations.
\newblock {\em Journal of Cosmology and Astroparticle Physics},
  2015(07):011^^e2^^80^^93011, Jul 2015.

\bibitem{planck_collaboration_2020}
{Planck Collaboration}, N.~{Aghanim}, Y.~{Akrami}, M.~{Ashdown}, J.~{Aumont},
  C.~{Baccigalupi}, M.~{Ballardini}, A.~J. {Banday}, R.~B. {Barreiro},
  N.~{Bartolo}, S.~{Basak}, R.~{Battye}, K.~{Benabed}, J.~P. {Bernard},
  M.~{Bersanelli}, P.~{Bielewicz}, J.~J. {Bock}, J.~R. {Bond}, J.~{Borrill},
  F.~R. {Bouchet}, F.~{Boulanger}, M.~{Bucher}, C.~{Burigana}, R.~C. {Butler},
  E.~{Calabrese}, J.~F. {Cardoso}, J.~{Carron}, A.~{Challinor}, H.~C. {Chiang},
  J.~{Chluba}, L.~P.~L. {Colombo}, C.~{Combet}, D.~{Contreras}, B.~P. {Crill},
  F.~{Cuttaia}, P.~{de Bernardis}, G.~{de Zotti}, J.~{Delabrouille}, J.~M.
  {Delouis}, E.~{Di Valentino}, J.~M. {Diego}, O.~{Dor{\'e}}, M.~{Douspis},
  A.~{Ducout}, X.~{Dupac}, S.~{Dusini}, G.~{Efstathiou}, F.~{Elsner}, T.~A.
  {En{\ss}lin}, H.~K. {Eriksen}, Y.~{Fantaye}, M.~{Farhang}, J.~{Fergusson},
  R.~{Fernandez-Cobos}, F.~{Finelli}, F.~{Forastieri}, M.~{Frailis}, A.~A.
  {Fraisse}, E.~{Franceschi}, A.~{Frolov}, S.~{Galeotta}, S.~{Galli},
  K.~{Ganga}, R.~T. {G{\'e}nova-Santos}, M.~{Gerbino}, T.~{Ghosh},
  J.~{Gonz{\'a}lez-Nuevo}, K.~M. {G{\'o}rski}, S.~{Gratton}, A.~{Gruppuso},
  J.~E. {Gudmundsson}, J.~{Hamann}, W.~{Handley}, F.~K. {Hansen}, D.~{Herranz},
  S.~R. {Hildebrandt}, E.~{Hivon}, Z.~{Huang}, A.~H. {Jaffe}, W.~C. {Jones},
  A.~{Karakci}, E.~{Keih{\"a}nen}, R.~{Keskitalo}, K.~{Kiiveri}, J.~{Kim},
  T.~S. {Kisner}, L.~{Knox}, N.~{Krachmalnicoff}, M.~{Kunz}, H.~{Kurki-Suonio},
  G.~{Lagache}, J.~M. {Lamarre}, A.~{Lasenby}, M.~{Lattanzi}, C.~R. {Lawrence},
  M.~{Le Jeune}, P.~{Lemos}, J.~{Lesgourgues}, F.~{Levrier}, A.~{Lewis},
  M.~{Liguori}, P.~B. {Lilje}, M.~{Lilley}, V.~{Lindholm},
  M.~{L{\'o}pez-Caniego}, P.~M. {Lubin}, Y.~Z. {Ma}, J.~F.
  {Mac{\'\i}as-P{\'e}rez}, G.~{Maggio}, D.~{Maino}, N.~{Mandolesi},
  A.~{Mangilli}, A.~{Marcos-Caballero}, M.~{Maris}, P.~G. {Martin},
  M.~{Martinelli}, E.~{Mart{\'\i}nez-Gonz{\'a}lez}, S.~{Matarrese}, N.~{Mauri},
  J.~D. {McEwen}, P.~R. {Meinhold}, A.~{Melchiorri}, A.~{Mennella},
  M.~{Migliaccio}, M.~{Millea}, S.~{Mitra}, M.~A. {Miville-Desch{\^e}nes},
  D.~{Molinari}, L.~{Montier}, G.~{Morgante}, A.~{Moss}, P.~{Natoli}, H.~U.
  {N{\o}rgaard-Nielsen}, L.~{Pagano}, D.~{Paoletti}, B.~{Partridge},
  G.~{Patanchon}, H.~V. {Peiris}, F.~{Perrotta}, V.~{Pettorino},
  F.~{Piacentini}, L.~{Polastri}, G.~{Polenta}, J.~L. {Puget}, J.~P. {Rachen},
  M.~{Reinecke}, M.~{Remazeilles}, A.~{Renzi}, G.~{Rocha}, C.~{Rosset},
  G.~{Roudier}, J.~A. {Rubi{\~n}o-Mart{\'\i}n}, B.~{Ruiz-Granados},
  L.~{Salvati}, M.~{Sandri}, M.~{Savelainen}, D.~{Scott}, E.~P.~S. {Shellard},
  C.~{Sirignano}, G.~{Sirri}, L.~D. {Spencer}, R.~{Sunyaev}, A.~S. {Suur-Uski},
  J.~A. {Tauber}, D.~{Tavagnacco}, M.~{Tenti}, L.~{Toffolatti}, M.~{Tomasi},
  T.~{Trombetti}, L.~{Valenziano}, J.~{Valiviita}, B.~{Van Tent}, L.~{Vibert},
  P.~{Vielva}, F.~{Villa}, N.~{Vittorio}, B.~D. {Wandelt}, I.~K. {Wehus},
  M.~{White}, S.~D.~M. {White}, A.~{Zacchei}, and A.~{Zonca}.
\newblock {Planck 2018 results. VI. Cosmological parameters}.
\newblock {\em \aap}, 641:A6, September 2020.

\bibitem{philcox2022boss}
Oliver H.~E. Philcox and Mikhail~M. Ivanov.
\newblock Boss dr12 full-shape cosmology: $\lambda$cdm constraints from the
  large-scale galaxy power spectrum and bispectrum monopole.
\newblock {\em Physical Review D}, 105(4), Feb 2022.

\bibitem{Chen_2022}
Shi-Fan Chen, Zvonimir Vlah, and Martin White.
\newblock A new analysis of galaxy 2-point functions in the boss survey,
  including full-shape information and post-reconstruction bao.
\newblock {\em Journal of Cosmology and Astroparticle Physics}, 2022(02):008,
  Feb 2022.

\bibitem{Beutler_McDonald_2021}
Florian Beutler and Patrick McDonald.
\newblock Unified galaxy power spectrum measurements from 6dfgs, boss, and
  eboss.
\newblock {\em Journal of Cosmology and Astroparticle Physics}, 2021(11):031,
  Nov 2021.

\bibitem{Kitaura_2016}
Francisco-Shu Kitaura, Sergio Rodr^^c3^^adguez-Torres, Chia-Hsun Chuang, Cheng
  Zhao, Francisco Prada, H^^c3^^a9ctor Gil-Mar^^c3^^adn, Hong Guo, Gustavo
  Yepes, Anatoly Klypin, Claudia~G. Sc^^c3^^b3ccola, and et~al.
\newblock The clustering of galaxies in the sdss-iii baryon oscillation
  spectroscopic survey: mock galaxy catalogues for the boss final data release.
\newblock {\em Monthly Notices of the Royal Astronomical Society},
  456(4):4156^^e2^^80^^934173, Jan 2016.

\bibitem{Rodriguez-Torres_2016}
Sergio~A. Rodr^^c3^^adguez-Torres, Chia-Hsun Chuang, Francisco Prada, Hong Guo,
  Anatoly Klypin, Peter Behroozi, Chang~Hoon Hahn, Johan Comparat, Gustavo
  Yepes, Antonio~D. Montero-Dorta, Joel~R. Brownstein, Claudia Maraston,
  Cameron~K. McBride, Jeremy Tinker, Stefan Gottl^^c3^^b6ber, Ginevra Favole,
  Yiping Shu, Francisco-Shu Kitaura, Adam Bolton, Rom^^c3^^a1n Scoccimarro,
  Lado Samushia, David Schlegel, Donald~P. Schneider, and Daniel Thomas.
\newblock {The clustering of galaxies in the SDSS-III Baryon Oscillation
  Spectroscopic Survey: modelling the clustering and halo occupation
  distribution of BOSS CMASS galaxies in the Final Data Release}.
\newblock {\em Monthly Notices of the Royal Astronomical Society},
  460(2):1173--1187, 04 2016.

\bibitem{Klypin_2016}
Anatoly Klypin, Gustavo Yepes, Stefan Gottl^^c3^^b6ber, Francisco Prada, and
  Steffen He^^c3^^9f.
\newblock Multidark simulations: the story of dark matter halo concentrations
  and density profiles.
\newblock {\em Monthly Notices of the Royal Astronomical Society},
  457(4):4340^^e2^^80^^934359, Feb 2016.

\bibitem{Hartlap_2006}
J.~{Hartlap}, P.~{Simon}, and P.~{Schneider}.
\newblock {Why your model parameter confidences might be too optimistic.
  Unbiased estimation of the inverse covariance matrix}.
\newblock {\em \aap}, 464(1):399--404, March 2007.

\bibitem{2021arXiv211102419M}
Hironao {Miyatake}, Sunao {Sugiyama}, Masahiro {Takada}, Takahiro {Nishimichi},
  Masato {Shirasaki}, Yosuke {Kobayashi}, Rachel {Mandelbaum}, Surhud {More},
  Masamune {Oguri}, Ken {Osato}, Youngsoo {Park}, Ryuichi {Takahashi}, Jean
  {Coupon}, Chiaki {Hikage}, Bau-Ching {Hsieh}, Alexie {Leauthaud}, Xiangchong
  {Li}, Wentao {Luo}, Robert~H. {Lupton}, Satoshi {Miyazaki}, Hitoshi
  {Murayama}, Atsushi~J. {Nishizawa}, Paul~A. {Price}, Melanie {Simet},
  Joshua~S. {Speagle}, Michael~A. {Strauss}, Masayuki {Tanaka}, and Naoki
  {Yoshida}.
\newblock {Cosmological inference from the emulator based halo model II: Joint
  analysis of galaxy-galaxy weak lensing and galaxy clustering from HSC-Y1 and
  SDSS}.
\newblock {\em arXiv e-prints}, page arXiv:2111.02419, November 2021.

\bibitem{NFW}
Julio~F. {Navarro}, Carlos~S. {Frenk}, and Simon D.~M. {White}.
\newblock {The Structure of Cold Dark Matter Halos}.
\newblock {\em \apj}, 462:563, May 1996.

\bibitem{hikage12a}
Chiaki {Hikage}, Masahiro {Takada}, and David~N. {Spergel}.
\newblock {Using galaxy-galaxy weak lensing measurements to correct the finger
  of God}.
\newblock {\em \mnras}, 419(4):3457--3481, February 2012.

\bibitem{hikage:2013kx}
Chiaki {Hikage}, Rachel {Mandelbaum}, Masahiro {Takada}, and David~N.
  {Spergel}.
\newblock {Where are the Luminous Red Galaxies (LRGs)? Using correlation
  measurements and lensing to relate LRGs to dark matter haloes}.
\newblock {\em \mnras}, 435(3):2345--2370, November 2013.

\bibitem{2015ApJ...799..108D}
Benedikt {Diemer} and Andrey~V. {Kravtsov}.
\newblock {A Universal Model for Halo Concentrations}.
\newblock {\em \apj}, 799(1):108, January 2015.

\bibitem{Diemer_2019}
Benedikt Diemer and Michael Joyce.
\newblock An accurate physical model for halo concentrations.
\newblock {\em The Astrophysical Journal}, 871(2):168, Jan 2019.

\bibitem{Diemer_2018}
Benedikt Diemer.
\newblock Colossus: A python toolkit for cosmology, large-scale structure, and
  dark matter halos.
\newblock {\em The Astrophysical Journal Supplement Series}, 239(2):35, Dec
  2018.

\bibitem{alcock79}
C.~{Alcock} and B.~{Paczynski}.
\newblock {An evolution free test for non-zero cosmological constant}.
\newblock {\em \nat}, 281:358, October 1979.

\bibitem{Matsubara_1996}
Takahiko Matsubara and Yasushi Suto.
\newblock Cosmological redshift distortion of correlation functions as a probe
  of the density parameter and the cosmological constant.
\newblock {\em The Astrophysical Journal}, 470(1):L1^^e2^^80^^93L5, Oct 1996.

\bibitem{Cooke_2018}
Ryan~J. Cooke, Max Pettini, and Charles~C. Steidel.
\newblock One percent determination of the primordial deuterium abundance.
\newblock {\em The Astrophysical Journal}, 855(2):102, Mar 2018.

\bibitem{Sch_neberg_2019}
Nils Sch^^c3^^b6neberg, Julien Lesgourgues, and Deanna~C. Hooper.
\newblock The bao+bbn take on the hubble tension.
\newblock {\em Journal of Cosmology and Astroparticle Physics},
  2019(10):029^^e2^^80^^93029, Oct 2019.

\bibitem{metropolis_1949}
Nicholas Metropolis and S.~Ulam.
\newblock The monte carlo method.
\newblock {\em Journal of the American Statistical Association},
  44(247):335--341, 1949.
\newblock PMID: 18139350.

\bibitem{Feroz_2009}
F.~Feroz, M.~P. Hobson, and M.~Bridges.
\newblock Multinest: an efficient and robust bayesian inference tool for
  cosmology and particle physics.
\newblock {\em Monthly Notices of the Royal Astronomical Society},
  398(4):1601^^e2^^80^^931614, Oct 2009.

\bibitem{Buchner_2014}
J.~Buchner, A.~Georgakakis, K.~Nandra, L.~Hsu, C.~Rangel, M.~Brightman,
  A.~Merloni, M.~Salvato, J.~Donley, and D.~Kocevski.
\newblock X-ray spectral modelling of the agn obscuring region in the cdfs:
  Bayesian model selection and catalogue.
\newblock {\em Astronomy \& Astrophysics}, 564:A125, Apr 2014.

\bibitem{Vehtari_2021}
Aki Vehtari, Andrew Gelman, Daniel Simpson, Bob Carpenter, and Paul-Christian
  B^^c3^^bcrkner.
\newblock {Rank-Normalization, Folding, and Localization: An Improved
  $\widehat{R}$ for Assessing Convergence of MCMC (with Discussion)}.
\newblock {\em Bayesian Analysis}, 16(2):667 -- 718, 2021.

\bibitem{Gelman_Rubin_1992}
Andrew Gelman and Donald~B. Rubin.
\newblock {Inference from Iterative Simulation Using Multiple Sequences}.
\newblock {\em Statistical Science}, 7(4):457 -- 472, 1992.

\bibitem{Brooks_Gelman_1998}
Stephen~P. Brooks and Andrew Gelman.
\newblock General methods for monitoring convergence of iterative simulations.
\newblock {\em Journal of Computational and Graphical Statistics},
  7(4):434--455, 1998.

\bibitem{Gelman_2013}
Andrew Gelman, John~B. Carlin, Hal~S. Stern, David~B. Dunson, Aki Vehtari, and
  Donald~B. Rubin.
\newblock {\em Bayesian Data Analysis (3rd ed.)}.
\newblock CRC Press, 2013.

\bibitem{Lewis:2019xzd}
Antony Lewis.
\newblock {GetDist: a Python package for analysing Monte Carlo samples}.
\newblock 2019.

\bibitem{2020JCAP...05..032P}
Oliver H.~E. {Philcox}, Mikhail~M. {Ivanov}, Marko {Simonovi{\'c}}, and Matias
  {Zaldarriaga}.
\newblock {Combining full-shape and BAO analyses of galaxy power spectra: a
  1.6\% CMB-independent constraint on H$_{0}$}.
\newblock {\em \jcap}, 2020(5):032, May 2020.

\bibitem{2021arXiv210100113M}
Hironao {Miyatake}, Yosuke {Kobayashi}, Masahiro {Takada}, Takahiro
  {Nishimichi}, Masato {Shirasaki}, Sunao {Sugiyama}, Ryuichi {Takahashi}, Ken
  {Osato}, Surhud {More}, and Youngsoo {Park}.
\newblock {Cosmological inference from emulator based halo model I: Validation
  tests with HSC and SDSS mock catalogs}.
\newblock {\em arXiv e-prints}, page arXiv:2101.00113, December 2020.

\bibitem{Sugiyama_2020}
Sunao Sugiyama, Masahiro Takada, Yosuke Kobayashi, Hironao Miyatake, Masato
  Shirasaki, Takahiro Nishimichi, and Youngsoo Park.
\newblock Validating a minimal galaxy bias method for cosmological parameter
  inference using hsc-sdss mock catalogs.
\newblock {\em Physical Review D}, 102(8), Oct 2020.

\bibitem{Zhang_2022}
Pierre Zhang, Guido D’Amico, Leonardo Senatore, Cheng Zhao, and Yifu Cai.
\newblock Boss correlation function analysis from the effective field theory of
  large-scale structure.
\newblock {\em Journal of Cosmology and Astroparticle Physics}, 2022(02):036,
  Feb 2022.

\bibitem{2021arXiv211103156S}
Agne {Semenaite}, Ariel~G. {S{\'a}nchez}, Andrea {Pezzotta}, Jiamin {Hou},
  Roman {Scoccimarro}, Alexander {Eggemeier}, Martin {Crocce}, Chia-Hsun
  {Chuang}, Alexander {Smith}, Cheng {Zhao}, Joel~R. {Brownstein}, Graziano
  {Rossi}, and Donald~P. {Schneider}.
\newblock {Cosmological implications of the full shape of anisotropic
  clustering measurements in BOSS and eBOSS}.
\newblock {\em arXiv e-prints}, page arXiv:2111.03156, November 2021.

\bibitem{2021PhRvD.104l1301B}
Samuel {Brieden}, H{\'e}ctor {Gil-Mar{\'\i}n}, and Licia {Verde}.
\newblock {Model-independent versus model-dependent interpretation of the
  SDSS-III BOSS power spectrum: Bridging the divide}.
\newblock {\em \prd}, 104(12):L121301, December 2021.

\bibitem{2019PASJ...71...43H}
Chiaki {Hikage}, Masamune {Oguri}, Takashi {Hamana}, Surhud {More}, Rachel
  {Mandelbaum}, Masahiro {Takada}, Fabian {K{\"o}hlinger}, Hironao {Miyatake},
  Atsushi~J. {Nishizawa}, Hiroaki {Aihara}, Robert {Armstrong}, James {Bosch},
  Jean {Coupon}, Anne {Ducout}, Paul {Ho}, Bau-Ching {Hsieh}, Yutaka
  {Komiyama}, Fran{\c{c}}ois {Lanusse}, Alexie {Leauthaud}, Robert~H. {Lupton},
  Elinor {Medezinski}, Sogo {Mineo}, Shoken {Miyama}, Satoshi {Miyazaki}, Ryoma
  {Murata}, Hitoshi {Murayama}, Masato {Shirasaki}, Crist{\'o}bal {Sif{\'o}n},
  Melanie {Simet}, Joshua {Speagle}, David~N. {Spergel}, Michael~A. {Strauss},
  Naoshi {Sugiyama}, Masayuki {Tanaka}, Yousuke {Utsumi}, Shiang-Yu {Wang}, and
  Yoshihiko {Yamada}.
\newblock {Cosmology from cosmic shear power spectra with Subaru Hyper
  Suprime-Cam first-year data}.
\newblock {\em \pasj}, 71(2):43, April 2019.

\bibitem{reid09}
Beth~A. {Reid} and David~N. {Spergel}.
\newblock {Constraining the Luminous Red Galaxy Halo Occupation Distribution
  Using Counts-In-Cylinders}.
\newblock {\em \apj}, 698(1):143--154, June 2009.

\bibitem{2011ApJ...728..126W}
Martin {White}, M.~{Blanton}, A.~{Bolton}, D.~{Schlegel}, J.~{Tinker},
  A.~{Berlind}, L.~{da Costa}, E.~{Kazin}, Y.~T. {Lin}, M.~{Maia}, C.~K.
  {McBride}, N.~{Padmanabhan}, J.~{Parejko}, W.~{Percival}, F.~{Prada},
  B.~{Ramos}, E.~{Sheldon}, F.~{de Simoni}, R.~{Skibba}, D.~{Thomas},
  D.~{Wake}, I.~{Zehavi}, Z.~{Zheng}, R.~{Nichol}, Donald~P. {Schneider},
  Michael~A. {Strauss}, B.~A. {Weaver}, and David~H. {Weinberg}.
\newblock {The Clustering of Massive Galaxies at z
  \raisebox{-0.5ex}\textasciitilde 0.5 from the First Semester of BOSS Data}.
\newblock {\em \apj}, 728(2):126, February 2011.

\bibitem{2015ApJ...806....2M}
Surhud {More}, Hironao {Miyatake}, Rachel {Mandelbaum}, Masahiro {Takada},
  David~N. {Spergel}, Joel~R. {Brownstein}, and Donald~P. {Schneider}.
\newblock {The Weak Lensing Signal and the Clustering of BOSS Galaxies. II.
  Astrophysical and Cosmological Constraints}.
\newblock {\em \apj}, 806(1):2, June 2015.

\bibitem{hikage:2013lr}
Chiaki {Hikage} and Kazuhiro {Yamamoto}.
\newblock {Impacts of satellite galaxies on the redshift-space distortions}.
\newblock {\em \jcap}, 2013(8):019, August 2013.

\bibitem{2006ApJ...652...71W}
Risa~H. {Wechsler}, Andrew~R. {Zentner}, James~S. {Bullock}, Andrey~V.
  {Kravtsov}, and Brandon {Allgood}.
\newblock {The Dependence of Halo Clustering on Halo Formation History,
  Concentration, and Occupation}.
\newblock {\em \apj}, 652(1):71--84, Nov 2006.

\bibitem{2008ApJ...687...12D}
Neal {Dalal}, Martin {White}, J.~Richard {Bond}, and Alexander {Shirokov}.
\newblock {Halo Assembly Bias in Hierarchical Structure Formation}.
\newblock {\em \apj}, 687(1):12--21, November 2008.

\bibitem{2016ApJ...819..119L}
Yen-Ting {Lin}, Rachel {Mandelbaum}, Yun-Hsin {Huang}, Hung-Jin {Huang}, Neal
  {Dalal}, Benedikt {Diemer}, Hung-Yu {Jian}, and Andrey {Kravtsov}.
\newblock {On Detecting Halo Assembly Bias with Galaxy Populations}.
\newblock {\em \apj}, 819(2):119, March 2016.

\bibitem{2021MNRAS.501.1603H}
Boryana {Hadzhiyska}, Sownak {Bose}, Daniel {Eisenstein}, and Lars {Hernquist}.
\newblock {Extensions to models of the galaxy-halo connection}.
\newblock {\em \mnras}, 501(2):1603--1620, February 2021.

\bibitem{paszke2019pytorch}
Adam Paszke, Sam Gross, Francisco Massa, Adam Lerer, James Bradbury, Gregory
  Chanan, Trevor Killeen, Zeming Lin, Natalia Gimelshein, Luca Antiga, Alban
  Desmaison, Andreas K^^c3^^b6pf, Edward Yang, Zach DeVito, Martin Raison,
  Alykhan Tejani, Sasank Chilamkurthy, Benoit Steiner, Lu~Fang, Junjie Bai, and
  Soumith Chintala.
\newblock Pytorch: An imperative style, high-performance deep learning library,
  2019.

\bibitem{2018PhRvD..98d3526A}
T.~M.~C. {Abbott}, F.~B. {Abdalla}, A.~{Alarcon}, J.~{Aleksi{\'c}}, S.~{Allam},
  S.~{Allen}, A.~{Amara}, J.~{Annis}, J.~{Asorey}, S.~{Avila}, D.~{Bacon},
  E.~{Balbinot}, M.~{Banerji}, N.~{Banik}, W.~{Barkhouse}, M.~{Baumer},
  E.~{Baxter}, K.~{Bechtol}, M.~R. {Becker}, A.~{Benoit-L{\'e}vy}, B.~A.
  {Benson}, G.~M. {Bernstein}, E.~{Bertin}, J.~{Blazek}, S.~L. {Bridle},
  D.~{Brooks}, D.~{Brout}, E.~{Buckley-Geer}, D.~L. {Burke}, M.~T. {Busha},
  A.~{Campos}, D.~{Capozzi}, A.~{Carnero Rosell}, M.~{Carrasco Kind},
  J.~{Carretero}, F.~J. {Castander}, R.~{Cawthon}, C.~{Chang}, N.~{Chen},
  M.~{Childress}, A.~{Choi}, C.~{Conselice}, R.~{Crittenden}, M.~{Crocce},
  C.~E. {Cunha}, C.~B. {D'Andrea}, L.~N. {da Costa}, R.~{Das}, T.~M. {Davis},
  C.~{Davis}, J.~{De Vicente}, D.~L. {DePoy}, J.~{DeRose}, S.~{Desai}, H.~T.
  {Diehl}, J.~P. {Dietrich}, S.~{Dodelson}, P.~{Doel}, A.~{Drlica-Wagner},
  T.~F. {Eifler}, A.~E. {Elliott}, F.~{Elsner}, J.~{Elvin-Poole}, J.~{Estrada},
  A.~E. {Evrard}, Y.~{Fang}, E.~{Fernandez}, A.~{Fert{\'e}}, D.~A. {Finley},
  B.~{Flaugher}, P.~{Fosalba}, O.~{Friedrich}, J.~{Frieman},
  J.~{Garc{\'{\i}}a-Bellido}, M.~{Garcia-Fernandez}, M.~{Gatti},
  E.~{Gaztanaga}, D.~W. {Gerdes}, T.~{Giannantonio}, M.~S.~S. {Gill},
  K.~{Glazebrook}, D.~A. {Goldstein}, D.~{Gruen}, R.~A. {Gruendl},
  J.~{Gschwend}, G.~{Gutierrez}, S.~{Hamilton}, W.~G. {Hartley}, S.~R.
  {Hinton}, K.~{Honscheid}, B.~{Hoyle}, D.~{Huterer}, B.~{Jain}, D.~J. {James},
  M.~{Jarvis}, T.~{Jeltema}, M.~D. {Johnson}, M.~W.~G. {Johnson},
  T.~{Kacprzak}, S.~{Kent}, A.~G. {Kim}, A.~{King}, D.~{Kirk}, N.~{Kokron},
  A.~{Kovacs}, E.~{Krause}, C.~{Krawiec}, A.~{Kremin}, K.~{Kuehn},
  S.~{Kuhlmann}, N.~{Kuropatkin}, F.~{Lacasa}, O.~{Lahav}, T.~S. {Li}, A.~R.
  {Liddle}, C.~{Lidman}, M.~{Lima}, H.~{Lin}, N.~{MacCrann}, M.~A.~G. {Maia},
  M.~{Makler}, M.~{Manera}, M.~{March}, J.~L. {Marshall}, P.~{Martini}, R.~G.
  {McMahon}, P.~{Melchior}, F.~{Menanteau}, R.~{Miquel}, V.~{Miranda},
  D.~{Mudd}, J.~{Muir}, A.~{M{\"o}ller}, E.~{Neilsen}, R.~C. {Nichol},
  B.~{Nord}, P.~{Nugent}, R.~L.~C. {Ogando}, A.~{Palmese}, J.~{Peacock}, H.~V.
  {Peiris}, J.~{Peoples}, W.~J. {Percival}, D.~{Petravick}, A.~A. {Plazas},
  A.~{Porredon}, J.~{Prat}, A.~{Pujol}, M.~M. {Rau}, A.~{Refregier}, P.~M.
  {Ricker}, N.~{Roe}, R.~P. {Rollins}, A.~K. {Romer}, A.~{Roodman},
  R.~{Rosenfeld}, A.~J. {Ross}, E.~{Rozo}, E.~S. {Rykoff}, M.~{Sako}, A.~I.
  {Salvador}, S.~{Samuroff}, C.~{S{\'a}nchez}, E.~{Sanchez}, B.~{Santiago},
  V.~{Scarpine}, R.~{Schindler}, D.~{Scolnic}, L.~F. {Secco}, S.~{Serrano},
  I.~{Sevilla-Noarbe}, E.~{Sheldon}, R.~C. {Smith}, M.~{Smith}, J.~{Smith},
  M.~{Soares-Santos}, F.~{Sobreira}, E.~{Suchyta}, G.~{Tarle}, D.~{Thomas},
  M.~A. {Troxel}, D.~L. {Tucker}, B.~E. {Tucker}, S.~A. {Uddin}, T.~N. {Varga},
  P.~{Vielzeuf}, V.~{Vikram}, A.~K. {Vivas}, A.~R. {Walker}, M.~{Wang}, R.~H.
  {Wechsler}, J.~{Weller}, W.~{Wester}, R.~C. {Wolf}, B.~{Yanny}, F.~{Yuan},
  A.~{Zenteno}, B.~{Zhang}, Y.~{Zhang}, J.~{Zuntz}, and {Dark Energy Survey
  Collaboration}.
\newblock {Dark Energy Survey year 1 results: Cosmological constraints from
  galaxy clustering and weak lensing}.
\newblock {\em \prd}, 98(4):043526, August 2018.

\bibitem{2021A&A...646A.140H}
Catherine {Heymans}, Tilman {Tr{\"o}ster}, Marika {Asgari}, Chris {Blake},
  Hendrik {Hildebrandt}, Benjamin {Joachimi}, Konrad {Kuijken}, Chieh-An {Lin},
  Ariel~G. {S{\'a}nchez}, Jan~Luca {van den Busch}, Angus~H. {Wright},
  Alexandra {Amon}, Maciej {Bilicki}, Jelte {de Jong}, Martin {Crocce}, Andrej
  {Dvornik}, Thomas {Erben}, Maria~Cristina {Fortuna}, Fedor {Getman}, Benjamin
  {Giblin}, Karl {Glazebrook}, Henk {Hoekstra}, Shahab {Joudaki}, Arun
  {Kannawadi}, Fabian {K{\"o}hlinger}, Chris {Lidman}, Lance {Miller},
  Nicola~R. {Napolitano}, David {Parkinson}, Peter {Schneider}, HuanYuan
  {Shan}, Edwin~A. {Valentijn}, Gijs {Verdoes Kleijn}, and Christian {Wolf}.
\newblock {KiDS-1000 Cosmology: Multi-probe weak gravitational lensing and
  spectroscopic galaxy clustering constraints}.
\newblock {\em \aap}, 646:A140, February 2021.

\bibitem{2013arXiv1308.6070D}
Roland {de Putter}, Olivier {Dor{\'e}}, and Masahiro {Takada}.
\newblock {The Synergy between Weak Lensing and Galaxy Redshift Surveys}.
\newblock {\em arXiv e-prints}, page arXiv:1308.6070, August 2013.

\bibitem{Hahn_2017}
ChangHoon Hahn, Roman Scoccimarro, Michael~R. Blanton, Jeremy~L. Tinker, and
  Sergio Rodr^^c3^^adguez-Torres.
\newblock The effect of fiber collisions on the galaxy power spectrum
  multipoles.
\newblock {\em Monthly Notices of the Royal Astronomical Society}, page stx185,
  Jan 2017.

\bibitem{Note2}
We obtain this shifts using WebPlotDigitizer (\protect \url
  {https://automeris.io/WebPlotDigitizer/}).

\end{thebibliography}

\end{document}